\documentclass[%
reprint,
superscriptaddress,
showpacs,
nofootinbib,
 amsmath,amssymb,
 aps,
eqsecnum
]{revtex4-1}

\usepackage{graphicx}
\usepackage{dcolumn}
\usepackage{bm}

\begin{document}

\title{Gauged twistor formulation of a massive spinning particle in four dimensions}

\author{Shinichi Deguchi}
\email[E-mail: ]{deguchi@phys.cst.nihon-u.ac.jp}
\affiliation{Institute of Quantum Science, College of Science and Technology, 
Nihon University, Chiyoda-ku, Tokyo 101-8308, Japan}
\affiliation{Department of Quantum Science and Technology, Graduate School of Science and Technology, 
Nihon University, Chiyoda-ku, Tokyo 101-8308, Japan}
\author{Satoshi Okano}
\email[E-mail: ]{okano@phys.cst.nihon-u.ac.jp}
\affiliation{Department of Quantum Science and Technology, Graduate School of Science and Technology, 
Nihon University, Chiyoda-ku, Tokyo 101-8308, Japan}


\date{\today}

\begin{abstract}
We present a gauged twistor model of a free massive spinning particle in four-dimensional Minkowski space. 
This model is governed by an action, referred to here as the gauged generalized Shirafuji (GGS) action, 
that consists of twistor variables, auxiliary variables, and $U(1)$ and $SU(2)$ gauge fields on 
the one-dimensional parameter space of a particle's worldline. 
The GGS action remains invariant under reparametrization and the local $U(1)$ and $SU(2)$ transformations 
of the relevant variables, although the $SU(2)$ symmetry is nonlinearly realized. 
We consider the canonical Hamiltonian formalism based on the GGS action in the unitary gauge 
by following Dirac's recipe for constrained Hamiltonian systems. 
It is shown that just sufficient constraints for the twistor variables are consistently derived 
by virtue of the gauge symmetries of the GGS action. 
In the subsequent quantization procedure, these constraints turn into 
simultaneous differential equations for a twistor function. 
We perform the Penrose transform of this twistor function to define a massive spinor field of arbitrary rank,  
demonstrating that the spinor field satisfies generalized Dirac-Fierz-Pauli equations with $SU(2)$ indices.  
We also investigate the rank-one spinor fields in detail to clarify the physical meanings of 
the $U(1)$ and $SU(2)$ symmetries.  
\end{abstract}


\maketitle

\section{Introduction}

Twistor theory is basically appropriate for describing massless systems 
with conformal symmetry \cite{PenMac, PenRin, HugTod}. 
Nevertheless, there have been some approaches to formulating massive particle systems in terms of twistors 
\cite{Penrose, Perjes1, Perjes2, Perjes3, Hughston, 
FedZim, BALE, AFLM, FFLM, AIL, MRT, FedLuk, AFIL, MRT2}. 
For describing a massive particle, it is common to use two or more independent twistors. 
In fact, introducing two twistors has been considered until recently   
\cite{FedZim, BALE, AFLM, FFLM, AIL, MRT, FedLuk, AFIL, MRT2}, 
and introducing more than two twistors was considered in some earlier studies  
\cite{Penrose, Perjes1, Perjes2, Perjes3, Hughston}.  
By virtue of using two or more independent twistors, 
an extra symmetry between the twistors occurs naturally in the system.  
Penrose, Perj\'{e}s, and Hughston proposed the idea of identifying this symmetry 
with an internal symmetry in particle physics, such as weak isospin or flavor, 
toward explaining internal symmetries of elementary particles on the basis of twistor theory 
\cite{Penrose, Perjes1, Perjes2, Perjes3, Hughston}. 
Although this idea is quite interesting, it seems that its detailed investigations have been made 
from neither a mechanical point of view nor a dynamical point of view. 
Therefore we would have to say that the idea is still poorly understood.

Lagrangian mechanics of a massive spinning particle 
formulated in terms of two twistors has been studied in Refs.  
\cite{FedZim, BALE, AFLM, FFLM, AIL, MRT, FedLuk, AFIL, MRT2}. 
Most of these papers begin with generalization of the Shirafuji action that describes  
a free massless spinning particle in four dimensions in terms of a twistor \cite{Shirafuji}. 
In fact, various generalizations of the Shirafuji action have been presented to specify twistorial models 
of massive spinning particles. 
The generalized Shirafuji actions are constructed by incorporating a mass-shell condition of a particle 
and certain other conditions for the twistor variables. 
The canonical formalism based on each generalized Shirafuji action and its subsequent quantization 
were also studied in Refs. \cite{FedZim, AFLM, FFLM, MRT, FedLuk, AFIL, MRT2}. 
It was shown that the canonical quantization of each twistorial model leads to generalized Dirac equations   
or the Dirac-Fierz-Pauli (DFP) equations for massive spinor fields of arbitrary rank \cite{Dirac1, Fierz, FiePau}. 
(The supersymmetric Shirafuji action \cite{Shirafuji} that is written in terms of a supertwistor 
and describes a massless superparticle in four dimensions has been generalized to the twistorial actions 
for massive superparticles in four dimensions \cite{AIL, MRT, MRT2}. 
In this paper, however, we are not concerned with the supersymmetric cases.)

In this paper, we consider an alternative generalization of the Shirafuji action to define 
a new twistor model of a free massive spinning particle in four dimensions by using two twistors.  
Our formulation is precisely a non-Abelian extension of the gauged twistor formulation of  
a free massless spinning particle in four dimensions \cite{BarPic, DEN, DNOS}. 
In the gauged twistor formulation, the Shirafuji action is modified in accordance with the gauge principle  
so that it can become invariant under the local $U(1)$ (phase) transformation of twistor variables.   
Here ^^ ^^ local" means that the transformation parameter depends on a worldline parameter 
along the particle's worldline.  
This modification is accomplished by gauging the Shirafuji action with the aid of a $U(1)$ gauge field 
on the one-dimensional (1D) parameter space of the worldline and by adding the 1D Chern-Simons term 
consisting of the $U(1)$ gauge field. 
The modified action, named the {\em gauged} Shirafuju action, includes a helicity constraint term due to the modification. 
Hence it follows that this action describes a free massless spinning particle with a fixed value of helicity. 
Remarkably, the gauged Shirafuji action is equivalent to the action for a massless particle with rigidity,    
at least at the classical mechanical level \cite{DegSuz}.  
The Shirafuji action can furthermore be modified so as to be invariant under the local scale transformation of  
twistor variables with the aid of another gauge field on the 1D parameter space. 
From the point of view of twistor theory, it is desirable that the modified action remains invariant under 
the combination of the local $U(1)$ and local scale transformations, 
which is referred to in Refs. \cite{DEN, DNOS} as the complexified local scale transformation.  
In actuality, the gauge field for the local scale transformation can be gauged away 
by a scaling of the twistor variables. 
Therefore it turns out that only the local $U(1)$ transformation is essential 
and one does not need to consider the local scale transformation in practice.

In the next section, we begin with setting up a generalized Shirafuji action that consists of two twistors 
and involves a mass-shell condition. 
Here, for convenience, 
we exploit the mass-shell condition with a complexified mass parameter introduced in Refs. \cite{FedLuk, AFIL}.   
The generalized Shirafuji action remains invariant under the global $U(1)$ transformation 
of twistor variables supplemented with that of auxiliary fields on the 1D parameter space. 
In addition, the generalized Shirafuji action remains invariant under the global $SU(2)$ transformation 
defined for a doublet of twistors. 
In accordance with the gauge principle, we modify the generalized Shirafuji action in such a way that 
the modified action remains invariant under the local $U(1)$ and $SU(2)$ transformations of twistor variables.  
The modification is performed by gauging  
the generalized Shirafuji action with the aid of $U(1)$ and $SU(2)$ gauge fields 
on the 1D parameter space and by adding the 1D $U(1)$ and $SU(2)$ Chern-Simons terms. 
The 1D $SU(2)$ Chern-Simons term, however, vanishes owing to the traceless property of the $SU(2)$ gauge field.  
For this reason, the variation of the modified action with respect to the $SU(2)$ gauge field yields too strong constraints 
that, after quantizing the model, permit us to have only massive spinless fields in four dimensions.  
A similar consequence has been found by Fedoruk and Lukierski in their twistorial model of a massive particle \cite{FedLuk}. 
To overcome such an undesirable situation, they modified the model by incorporating 
the Souriau-Wess-Zumino term, following the successful argument for a twistorial model of 
a massive spinning particle in three dimensions \cite{MRT}.  
In the present paper, we consider an alternative approach based on a nonlinear realization of $SU(2)$ 
to eventually obtain massive spinor fields of arbitrary rank.  
This approach makes it possible to define the 1D $U(1)$ Chern-Simons term 
consisting of the third (or diagonal) component of the $SU(2)$ gauge field in a particular gauge. 
In addition, this approach can provide a novel gauge-invariant term consisting of the first and second 
(or off-diagonal) components of the same $SU(2)$ gauge field.  
With the new terms, we furthermore modify the generalized Shirafuji action 
by adding these terms to the modified action mentioned above. 
The completely modified action is thus the sum of the gauged twistorial part, the two 1D $U(1)$ 
Chern-Simons terms, and the novel term. 
This action, hereafter referred to as the {\em gauged} generalized Shirafuji (GGS) action,  
remains invariant under reparametrization of the worldline parameter and under 
the local $U(1)$ and $SU(2)$ transformations. 
The GGS action yields just sufficient constraints for the twistor variables 
in a systematic and consistent manner. 
All the constraints except for the mass-shell condition are derived on the basis of the gauge symmetry.    
This is an advantage of our gauged twistor model.

Having obtained the GGS action, we study the canonical Hamiltonian formalism based on it 
by completely following the Dirac algorithm for Hamiltonian systems with constraints
\cite{Dirac2, HRT, HenTei}. 
We see that most of the Dirac brackets between the twistor variables take on complicated forms. 
Fortunately, these Dirac brackets can be reduced to simple Dirac brackets for  
new twistor variables that are in one-to-one correspondence with the old ones. 
Also, all the constraints for the (old) twistor variables can be written completely in terms of the new twistor variables.  
The canonical quantization of the twistor model governed by the GGS action is performed with the commutation relations 
between the operators that correspond to the new twistor variables or the other canonical variables. 
Some of the first-class constraints eventually turn into 
simultaneous differential equations for a holomorphic function of half the new twistor variables. 
Each solution of the simultaneous differential equations, referred to here as a twistor function, 
is characterized by the three quantum numbers that originate from  
the $U(1)$ and $SU(2)$ symmetries inherent in the GGS action.

We also consider the Penrose transform of the twistor function to define a four-dimensional spinor field of arbitrary rank.  
The spinor field defined in this manner has extra upper and lower $SU(2)$ indices 
in addition to dotted and undotted spinor indices.  
Because of the structure of the Penrose transform, the number of upper (lower) 
$SU(2)$ indices is equal to the number of undotted (dotted) spinor indices. 
We demonstrate that the present spinor field satisfies generalized DFP equations with $SU(2)$ indices. 
In the simplest case, the generalized DFP equations reduce to the ordinary Dirac equations for particle and 
antiparticle spinor fields. Investigating properties of these fields, 
we clarify the physical meanings of the $U(1)$ and $SU(2)$ symmetries; 
ultimately, we see that the $U(1)$ symmetry is a gauge symmetry concerning the chiralities  
of the particle and antiparticle spinor fields, while the $SU(2)$ symmetry is a gauge symmetry 
realized in a doublet consisting of the particle and antiparticle spinor fields. 
Therefore it turns out that the idea proposed by Penrose, Perj\'{e}s, and Hughston, 
in which the $SU(2)$ symmetry is identified with the weak isospin symmetry, is not valid in our gauged twistor formulation.

This paper is organized as follows. 
In Sec. II, we elaborate the GGS action, after making some preliminary arrangements.  
The canonical Hamiltonian formalism based on the GGS action is studied in Sec. III, 
and the subsequent canonical quantization is performed in Sec. IV.  
In Sec. V, we define a massive spinor field of arbitrary rank by the Penrose transform of a twistor function 
and demonstrate that this spinor field satisfies the generalized DFP equations. 
In Sec. VI, we particularly investigate the rank-one spinor fields to clarify the physical meanings of 
the $U(1)$ and $SU(2)$ symmetries.  
Section VII is devoted to a summary and discussion. 
In the Appendix, we treat the Pauli-Lubanski pseudovector written in terms of the twistor variables.

\section{Construction of the GGS action}

In this section, we construct the GGS action for a free massive spinning  particle 
in four-dimensional Minkowski space.

In order to describe a massive particle in terms of twistors, 
we introduce two twistors   
$Z_{i}^{A}=(\omega_{i}^{\alpha}, \pi_{i \dot{\alpha}})$ 
$(A=0,1,2,3 \:\! ;\:\! \alpha=0,1;\:\! \dot{\alpha}=\dot{0}, \dot{1} )$
distinguished by the extra index $i$ $(i=1, 2)$   
and their dual twistors  
$\bar{Z}^{i}_{A}=(\bar{\pi}^{i}_{\alpha}, \bar{\omega}{}^{i\dot{\alpha}})$.  
Here, $\bar{\pi}^{i}_{\alpha}$ and $\bar{\omega}{}^{i\dot{\alpha}}$ denote the complex conjugates 
of $\pi_{i \dot{\alpha}}$ and $\omega_{i}^{\alpha}$, respectively: 
$\bar{\pi}^{i}_{\alpha}:=\overline{\pi_{i \dot{\alpha}}}\:\!$,  
$\;\!\bar{\omega}{}^{i\dot{\alpha}}:=\overline{\omega_{i}^{\alpha}}\:\!$.  
It is assumed that $Z_{1}^{A}$ and $Z_{2}^{A}$ are not proportional to each other:  
$Z_{1}^{A} \neq c^{\:\!} Z_{2}^{A}$ ($c \in \Bbb{C}$), so that 
$\bar{Z}^{1}_{A} \neq \bar{c}^{\:\!} \bar{Z}^{2}_{A}$.  
The 2-component spinors $\omega_{i}^{\alpha}$ and $\pi_{i \dot{\alpha}}$ 
are related by  
\begin{align}
\omega_{i}^{\alpha}=i z^{\alpha \dot{\alpha}} \pi_{i \dot{\alpha}} \,, 
\label{2.1}
\end{align}
where $z^{\alpha \dot{\alpha}}$ are coordinates of a point in complexified 
Minkowski space, $\Bbb{C}\mathbf{M}$, with  
the metric tensor $\eta_{\mu\nu}=\mathrm{diag}(1,-1,-1,-1)$. 
As can be seen in the literature on twistor theory 
\cite{Penrose, Perjes1, Perjes2, Perjes3, Hughston, 
FedZim, BALE, AFLM, FFLM, AIL, MRT, FedLuk, AFIL}, 
the four-momentum of a massive particle is expressed as 
$p_{\alpha\dot{\alpha}}=\bar{\pi}^{1}_{\alpha} \pi_{1 \dot{\alpha}} +\bar{\pi}^{2}_{\alpha} \pi_{2 \dot{\alpha}} 
\equiv \bar{\pi}^{i}_{\alpha} \pi_{i \dot{\alpha}}$. 
(For this reason, $\pi_{i \dot{\alpha}}$ and $\bar{\pi}^{i}_{\alpha}$ are named as momentum spinors.) 
The squared norm of $p_{\alpha\dot{\alpha}}$ remains nonvanishing 
even after using the formula 
$\pi_{i \dot{\alpha}} \pi_{i}^{\dot{\alpha}}
=\epsilon^{\dot{\alpha} \dot{\beta}} \pi_{i \dot{\alpha}} \pi_{i \dot{\beta}}
=0$ 
(no sum with respect to $i^{}$) and its complex conjugate,\footnote{The two-dimensional Levi-Civita symbols 
$\epsilon^{\alpha\beta}$, $\epsilon_{\alpha\beta}$, $\epsilon^{\dot{\alpha}\dot{\beta}}$, 
$\epsilon_{\dot{\alpha}\dot{\beta}}$, $\epsilon^{ij}$, and $\epsilon_{ij}$ 
are defined as $\epsilon^{01}=\epsilon_{01}=\epsilon^{\dot{0}\dot{1}}
=\epsilon_{\dot{0}\dot{1}}=\epsilon^{12}=\epsilon_{12}=1$ and conform to the rules 
$\overline{\epsilon^{\alpha\beta}}=\epsilon^{\dot{\alpha}\dot{\beta}}$, 
$\overline{\epsilon_{\alpha\beta}}=\epsilon_{\dot{\alpha}\dot{\beta}}\:\!$, 
$\overline{\epsilon^{ij}}=\epsilon_{ij}$,  and $\overline{\epsilon_{ij}}=\epsilon^{ij}$. 
The contravariant spinors $\pi_{i}^{\dot{\alpha}}$ and $\bar{\pi}^{i \alpha}$ are defined by   
$\pi_{i}^{\dot{\alpha}}=\epsilon^{\dot{\alpha} \dot{\beta}} \pi_{i \dot{\beta}}$ and 
$\bar{\pi}^{i \alpha}=\epsilon^{\alpha\beta} \bar{\pi}^{i}_{\beta \:\!}$, respectively. 
These relations can be expressed as 
$\pi_{i \dot{\alpha}}=\pi_{i}^{\dot{\beta}} \epsilon_{\dot{\beta}\dot{\alpha}}$ and 
$\bar{\pi}^{i}_{\alpha}=\bar{\pi}^{i \beta} \epsilon_{\beta\alpha}$. 
}   
because the cross terms provided from different twistors still survive:   
$p_{\alpha\dot{\alpha}} p^{\alpha\dot{\alpha}}
=\bar{\pi}^{i}_{\alpha} \pi_{i \dot{\alpha}} \bar{\pi}^{j \alpha} \pi_{j}^{\dot{\alpha}} 
=2 \big| \pi_{1 \dot{\alpha}} \pi_{2}^{\dot{\alpha}} \big|{}^{2}. $
Thus the mass-shell condition $p_{\alpha\dot{\alpha}} p^{\alpha\dot{\alpha}}=m^{2}$ 
with a mass parameter $m$ can be written as 
\begin{align}
\bar{\pi}^{i}_{\alpha} \pi_{i \dot{\alpha}} \bar{\pi}^{j \alpha} \pi_{j}^{\dot{\alpha}} =m^2 .
\label{2.2}
\end{align}
It is easy to see that this condition is equivalent to 
\begin{subequations}
\label{2.3}
\begin{align}
\epsilon^{ij} \pi_{i \dot{\alpha}} \pi_{j}^{\dot{\alpha}} -\sqrt{2} \:\! m e^{i\varphi} &=0 \,, 
\label{2.3a}
\\
\epsilon_{ij} \bar{\pi}^{i}_{\alpha} \bar{\pi}^{j \alpha} -\sqrt{2} \:\! m e^{-i\varphi} &=0 \,, 
\label{2.3b}
\end{align}
\end{subequations}
where $\varphi$ is a real parameter.    
These equations have been incorporated in twistorial models of massive spinning particles \cite{FedLuk, AFIL}, 
in which $m e^{i\varphi}/\sqrt{2}$ is called a complexified mass parameter. 
In this paper, we also adopt the equation pair Eqs. (\ref{2.3a}) and (\ref{2.3b}) as the mass-shell condition 
because of the convenience for our formulation.

The Shirafuji action of a free massless spinning particle\footnote{With a twistor $Z^{A}$ and 
its dual twistor $\bar{Z}_{A}$, the Shirafuji action is defined by \cite{Shirafuji} 
$$S_{0} = \int_{\tau_0}^{\tau_1} d\tau  
\:\! \frac{i}{2} \Big(\bar{Z}_{A} \dot{Z}^{A}
-Z^{A} \dot{\bar{Z}}_{A} \Big) \,.$$} 
can be generalized to 
describe a free spinning particle of mass $m$ propagating 
in four-dimensional Minkowski space $\mathbf{M}$.  
A generalized Shirafuji action is indeed given by 
\begin{align}
S_{m} &= \int_{\tau_0}^{\tau_1} d\tau 
\bigg[\;\! 
\frac{i}{2} \Big( \bar{Z}_{A}^{i} \dot{Z}^{A}_{i} 
-Z^{A}_{i} \dot{\bar{Z}}_{A}^{i} \Big) 
\nonumber
\\
& \quad \,
+h \Big( \epsilon^{ij} \pi_{i \dot{\alpha}} \pi_{j}^{\dot{\alpha}} -\sqrt{2} \:\! m e^{i\varphi} \Big) 
\nonumber
\\
& \quad \,
+\bar{h} \Big( \epsilon_{ij} \bar{\pi}^{i}_{\alpha} \bar{\pi}^{j \alpha} -\sqrt{2} \:\! m e^{-i\varphi} \Big)
\bigg] \,,
\label{2.4}
\end{align}
where $Z_{i}^{A}=Z_{i}^{A}(\tau)$ and $\bar{Z}_{A}^{i}=\bar{Z}_{A}^{i}(\tau)$ are understood 
as complex scalar fields on the one-dimensional parameter space 
$\mathcal{T}:=\{  \tau\;\! |\, \tau_{0} \leq \tau \leq \tau_{1} \}$ of a particle's worldline, 
and $h=h(\tau)$ is treated as a complex scalar-density field of weight 1 on $\mathcal{T}$. 
[${}^{\:\!}$That is, $h$ transforms as 
$h(\tau) \rightarrow h^{\prime}(\tau^{\prime})=(d\tau/d\tau^{\prime}) h(\tau)$ 
under the proper reparametrization $\tau \rightarrow \tau^{\prime}=\tau^{\prime} (\tau)$ ($d\tau^{\prime}/d\tau>0$).] 
The exponent $\varphi$ is now considered a real scalar field on $\mathcal{T}$ and hence 
is treated as a real function $\varphi=\varphi(\tau)$. 
This setting is different from that in Refs. \cite{FedLuk, AFIL}, in which the complexified mass parameter 
is regarded as a constant.   
A dot over a variable denotes its derivative with respect to $\tau$. 
The variation of $S_{m}$ with respect to $h$ and $\bar{h}$ yields the mass-shell condition (\ref{2.3}).\footnote{Instead of 
the action $S_{m}$, we can consider an alternative action 
\begin{align}
S^{\prime}_{m} &= \int_{\tau_0}^{\tau_1} d\tau 
\bigg[\;\! 
\frac{i}{2} \Big( \bar{Z}_{A}^{i} \dot{Z}^{A}_{i} 
-Z^{A}_{i} \dot{\bar{Z}}_{A}^{i} \Big) 
\nonumber
\\
& \quad \,
+ \frac{1}{2} f \big( 
\bar{\pi}^{i}_{\alpha} \pi_{i \dot{\alpha}} \bar{\pi}^{j \alpha} \pi_{j}^{\dot{\alpha}}
-m^2 \big) 
\bigg] \,,
\notag
\end{align}
where $f=f(\tau)$ is a real scalar-density field of weight 1 on $\mathcal{T}$. 
The variation of $S^{\prime}_{m}$ with respect to $f$ yields the mass-shell condition (\ref{2.2}).}

The generalized Shirafuji action $S_{m}$ remains invariant under the reparametrization 
$\tau \rightarrow \tau^{\prime}=\tau^{\prime} (\tau)$. 
In addition, $S_{m}$ remains invariant under the global $U(1)$ transformation 
\begin{subequations}
\label{2.5}
\begin{alignat}{3} 
Z_{i}^{A} &\rightarrow Z_{i}^{\prime A} =e^{i\theta} Z_{i}^{A} \,, 
&\quad \;
\bar{Z}{}^{i}_{A} &\rightarrow \bar{Z}^{\prime\:\! i}_{A} =e^{-i\theta} \bar{Z}^{i}_{A} \,, 
\label{2.5a}
\\
h &\rightarrow h^{\prime} =e^{-2i\theta} h \,, 
&\quad \;
\bar{h} &\rightarrow \bar{h}^{\prime} =e^{2i\theta} \bar{h} \,, 
\label{2.5b}
\\
\varphi &\rightarrow \varphi^{\prime} =\varphi+2\theta \,, 
&\quad \; 
~& 
\end{alignat}
\end{subequations}
with a real constant parameter $\theta$ and under the global $SU(2)$ transformation 
\begin{subequations}
\label{2.6}
\begin{alignat}{3} 
Z_{i}^{A} &\rightarrow Z_{i}^{\prime A} =U_{i}{}^{j} Z^{A}_{j} \,, 
&\quad \;
\bar{Z}{}^{i}_{A} &\rightarrow \bar{Z}{}^{\prime\:\! i}_{A} =\bar{Z}{}^{j}_{A} U^{\dagger}{}_{j}{}^{i} \,, 
\label{2.6a}
\\
h &\rightarrow h^{\prime} =h \,, 
&\quad \;
\bar{h} &\rightarrow \bar{h}^{\prime} =\bar{h} \,, 
\label{2.6b}
\\
\varphi &\rightarrow \varphi^{\prime} =\varphi \,, 
&\quad \; 
~& 
\end{alignat}
\end{subequations}
with a constant matrix $U$ belonging to $SU(2)$. 
The $SU(2)$ invariance of $S_{m}$ can be verified using 
$\epsilon^{ij} U_{i}{}^{k} U_{j}{}^{l}=\epsilon^{kl}$ and 
$\epsilon_{ij} U^{\dagger}{}_{k}{}^{i} U^{\dagger}{}_{l}{}^{j}=\epsilon_{kl}$  
together with the unitarity property of $U$. 
We thus see that $S_{m}$ possesses two global internal symmetries specified by $U(1)$ and $SU(2)$.  
We also see that the two terms $\bar{Z}_{A}^{i} \dot{Z}^{A}_{i}$ and 
$Z^{A}_{i} \dot{\bar{Z}}_{A}^{i}$ in Eq. (\ref{2.4}) are invariant under 
the global $SU(2,2)$ transformation  (or more simply, the global conformal transformation)  
$Z_{i}^{A} \rightarrow Z_{i}^{\prime A} =\mathcal{U}^{A}{}_{B} Z^{B}_{i} , 
\;\! \bar{Z}{}^{i}_{A} \rightarrow \bar{Z}{}^{\prime\:\! i}_{A} 
=\bar{Z}{}^{i}_{B\;\!} \mathcal{U}^{\dagger B}{}_{A \:\!}$, 
with a constant matrix $\mathcal{U}$ belonging to $SU(2,2)$. 
In contrast, the two terms $\epsilon^{ij} \pi_{i \dot{\alpha}} \pi_{j}^{\dot{\alpha}}$ and 
$\epsilon_{ij} \bar{\pi}^{i}_{\alpha} \bar{\pi}^{j \alpha}$ in Eq. (\ref{2.4}) are 
invariant only under 
the global $SL(2, \Bbb{C}) \ltimes \Bbb{R}^{1,3}$ transformation  
(or more simply, the global Poincar\'{e} transformation).  
Hence it turns out that the symmetry reduction from $SU(2,2)$ 
to $SL(2, \Bbb{C}) \ltimes \Bbb{R}^{1,3}$ occurs in $S_{m}$ as a result of  
adding the term proportional to $h$ and its complex conjugate term.

Now, we perform a gauging of the global $U(1)$ and $SU(2)$ symmetries 
in such a way that the gauged action remains invariant under the local $U(1)$ and 
$SU(2)$ transformations that depend on $\tau$. 
That is, we consider a $U(1) \times SU(2)$ gauge theory on the parameter space $\mathcal{T}$. 
To this end, in accordance with the gauge principle, we introduce a $U(1)$ gauge field, 
$a=a(\tau)$, and an $SU(2)$ gauge field, $b=b(\tau)$. 
The field $a$ is assumed to be a real scalar-density field of weight 1 on $\mathcal{T}$, 
while $b$ is assumed to be a $2\times2$ traceless Hermitian matrix  
that behaves as a scalar-density field of weight 1 on $\mathcal{T}$. 
The field $b$ can be represented as $(b_{i}{}^{j})$ with its matrix elements $b_{i}{}^{j}$ and 
can be expanded in terms of the Pauli matrices $\sigma_{r}$ ($r=1,2,3$), 
satisfying $[^{\:\!} \sigma_{r}, \sigma_{s}]=2i\epsilon_{rst} \sigma_{t}$, as $b=b^{r}\sigma_{r}$. 
Here, $b^{r}=b^{r}(\tau)$ are real scalar-density fields of weight 1 on $\mathcal{T}$.  
The (primitive) gauged action, $S_{m \mathrm{g}}$, can be obtained by replacing $d/d\tau$ in $S_{m}$ 
with a covariant derivative operator as follows: 
\begin{align} 
S_{m \mathrm{g}} &= \int_{\tau_0}^{\tau_1} d\tau 
\bigg[\;\! 
\frac{i}{2} \big(\bar{Z}_{A}^{i} DZ^{A}_{i} 
-Z^{A}_{i} \bar{D}\bar{Z}_{A}^{i} \big) 
\nonumber
\\
& \quad \,
+h \Big( \epsilon^{ij} \pi_{i \dot{\alpha}} \pi_{j}^{\dot{\alpha}} -\sqrt{2} \:\! m e^{i\varphi} \Big) 
\nonumber
\\
& \quad \,
+\bar{h} \Big( \epsilon_{ij} \bar{\pi}^{i}_{\alpha} \bar{\pi}^{j \alpha} -\sqrt{2} \:\! m e^{-i\varphi} \Big)
\bigg] \,,
\label{2.7}
\end{align}
where 
\begin{subequations}
\label{2.8}
\begin{align}
DZ^{A}_{i} &:=\dot{Z}^{A}_{i} -iaZ^{A}_{i} -ib_{i}{}^{j} Z^{A}_{j} \,, 
\label{2.8a}
\\
\bar{D}\bar{Z}_{A}^{i} &:=\dot{\bar{Z}}_{A}^{i} +ia\bar{Z}_{A}^{i} +i\bar{Z}_{A}^{j} b_{j}{}^{i} \,.
\label{2.8b}
\end{align}
\end{subequations}
We see that the action $S_{m \mathrm{g}}$ is reparametrization invariant. 
It can easily be verified that $S_{m \mathrm{g}}$ remains invariant 
under the local $U(1)$ transformation 
\begin{subequations}
\label{2.9}
\begin{align}
Z_{i}^{A} &\rightarrow Z_{i}^{\prime A} =e^{i\theta(\tau)} Z_{i}^{A} \,, 
\label{2.9a}
\\
\bar{Z}{}^{i}_{A} &\rightarrow \bar{Z}^{\prime\:\! i}_{A} =e^{-i\theta(\tau)} \bar{Z}^{i}_{A} \,, 
\label{2.9b}
\\
h &\rightarrow h^{\prime} =e^{-2i\theta(\tau)} h \,, 
\label{2.9c}
\\
\bar{h} &\rightarrow \bar{h}^{\prime} =e^{2i\theta(\tau)} \bar{h} \,, 
\label{2.9d}
\\
\varphi &\rightarrow \varphi^{\prime} =\varphi+2\theta(\tau) \,, 
\label{2.9e}
\\
a &\rightarrow a^{\prime}=a+\dot{\theta} \,, 
\label{2.9f}
\\
b &\rightarrow b^{\prime}=b \,, 
\label{2.9g}
\end{align}
\end{subequations}
with a real gauge function $\theta=\theta(\tau)$ and under the local $SU(2)$ transformation 
\begin{subequations}
\label{2.10}
\begin{align}
Z_{i}^{A} &\rightarrow Z_{i}^{\prime A} =U_{i}{}^{j}(\tau) Z^{A}_{j} \,, 
\label{2.10a}
\\
\bar{Z}{}^{i}_{A} &\rightarrow \bar{Z}{}^{\prime\:\! i}_{A} =\bar{Z}{}^{j}_{A} U^{\dagger}{}_{j}{}^{i}(\tau) \,, 
\label{2.10b}
\\
h &\rightarrow h^{\prime} =h \,, 
\label{2.10c}
\\
\bar{h} &\rightarrow \bar{h}^{\prime} =\bar{h} \,, 
\label{2.10d}
\\
\varphi &\rightarrow \varphi^{\prime} =\varphi \,, 
\label{2.10e}
\\
a & \rightarrow a^{\prime}=a \,, 
\label{2.10f}
\\
b &\rightarrow b^{\prime}=UbU^{\dagger}-i \dot{U} U^{\dagger} , 
\label{2.10g}
\end{align}
\end{subequations}
with a gauge function $U=U(\tau)$ taking its value in $SU(2)$. 
Because each of $a$ and $b^{r}$ is a single-component gauge field associated with $d/d\tau$, 
we cannot define their field strengths. 
For this reason, there exists neither the Maxwell action for $a$ nor the Yang-Mills action for $b$. 
As for $a$, it is possible to define the (nonvanishing) 1D $U(1)$ Chern-Simons term 
\begin{align}
S_{a}=-2s \int_{\tau_0}^{\tau_1} d\tau a \,, 
\label{2.11}
\end{align}
where $s$ is a real constant. 
The 1D $SU(2)$ Chern-Simons term for $b$, 
i.e., $S_{b}=-2t \int_{\tau_0}^{\tau_1} d\tau \mathrm{Tr} {}^{\:\!} b$ vanishes 
by the reason of $\mathrm{Tr} {}^{\:\!} b=0$.  
Since $a$ is a scalar-density field of weight 1, $S_{a}$ is reparametrization invariant. 
Also, $S_{a}$ remains invariant under the gauge transformation (\ref{2.9f}), 
provided that $\theta$ satisfies an appropriate boundary condition such as $\theta(\tau_1)=\theta(\tau_0)$. 
The $SU(2)$ invariance of $S_{a}$ is evident from Eq. (\ref{2.10f}). 
Therefore we can consider the reparametrization-invariant and gauge-invariant action 
$\tilde{S}_{m \mathrm{g}}:=S_{m \mathrm{g}}+S_{a}$.\footnote{The action 
$\tilde{S}_{m \mathrm{g}}$ is a simple and natural generalization of 
the {\em gauged} Shirafuji action (without invariance under the local scale transformation of 
$Z^{A}$ and $\bar{Z}_{A}$) 
$$\tilde{S}_{0 \mathrm{g}} =\int_{\tau_0}^{\tau_1} d\tau 
\bigg[\;\! \frac{i}{2} \big(\bar{Z}_{A} DZ^{A} -Z^{A} \bar{D}\bar{Z}_{A} \big) -2sa\bigg] \,,$$
where $D:=d/d\tau-ia$. This action describes a free massless spinning particle of helicity $s$ 
\cite{BarPic, DEN, DNOS} and is equivalent to the action for a massless particle with rigidity       
at least at the classical mechanical level \cite{DegSuz}.} 
However, $\tilde{S}_{m \mathrm{g}}$ eventually turns out to govern only massive {\em spinless} fields in four dimensions 
owing to the too strong constraints $\bar{Z}_{A}^{i} \sigma_{ri}{}^{j} Z^{A}_{j}=0$ ($r=1,2,3$) 
that are derived by varying $\tilde{S}_{m \mathrm{g}}$ with respect to $b^{r}$.\footnote{From the action 
$\tilde{S}_{m \mathrm{g}}$, the Pauli-Lubanski spin vector $W^{\alpha\dot{\alpha}}$ is found to be 
\begin{align*}
W^{\alpha\dot{\alpha}}=T_{r} \sigma_{ri}{}^{j} \bar{\pi}{}^{i \alpha} \pi_{j}^{\dot{\alpha}} \,, 
\quad \;  
T_{r}:=\frac{1}{2} \bar{Z}{}^{i}_{A} \sigma_{ri}{}^{j} Z^{A}_{j} 
\end{align*}
(see Appendix). 
Using the mass-shell condition (\ref{2.3}), we can show that 
$W_{\alpha\dot{\alpha}} W^{\alpha\dot{\alpha}} =-m^{2} T_{r} T_{r}$.  
Obviously, $T_{r}=0$ ($r=1,2,3$) leads to $W_{\alpha\dot{\alpha}} W^{\alpha\dot{\alpha}}=0$. 
Hence, it follows that only massive {\em spinless} particles are admissible in the model 
defined by $\tilde{S}_{m \mathrm{g}}$. 
Accordingly, it turns out that only massive {\em spinless} fields are provided after quantizing the model. 
}
$[^{\:\!}$Here, $\sigma_{rj}{}^{k}$ denotes the $(j,k)$ entry of the Pauli matrix $\sigma_{r}$.]  
To avoid such an undesirable situation, next we perform a modification of $\tilde{S}_{m \mathrm{g}}$ 
with the aid of a nonlinear realization of $SU(2)$.

Let us now consider the coset space $SU(2)/U(1)(\;\!\cong \mathbb{C}\mathbf{P}^{1})$ 
and representative elements, $V(\xi, \bar{\xi}\;\!)$ ($V\in SU(2)$, $\xi \in \mathbb{C}$), 
that are chosen one by one from each left coset of $U(1)$ in $SU(2)$.  
Here, $\xi$ labels the cosets in a way of one-to-one correspondence
and can be regarded as an inhomogeneous coordinate of a point on $SU(2)/U(1)$. 
$[^{\:\!}$To completely coordinatize $SU(2)/U(1)$, 
it is necessary to use $\xi^{-1}$ in addition to $\xi$.]  
The representative elements $V(\xi, \bar{\xi}^{\;\!})$ are assumed to constitute a smooth function of 
$\xi$ and $\bar{\xi}$ so that we can simply treat $V(\xi, \bar{\xi}^{\;\!})$  
as an $SU(2)$-valued smooth function.  
We consider $\xi$ to be a complex scalar field $\xi=\xi(\tau)$ on $\mathcal{T}$.  
The left action of $U$ on $V(\xi, \bar{\xi}\;\!)$ generates a nonlinear transformation 
$\xi \rightarrow \xi^{\prime}=\xi^{\prime}(\xi)$ in accordance with 
\begin{align}
V(\xi, \bar{\xi}\;\!) \rightarrow 
V(\xi^{\prime}, \bar{\xi}^{\prime}\:\!) 
&=U(\tau) V(\xi, \bar{\xi}\;\!) \varTheta^{-1}(\tau) \,, 
\label{2.12}
\end{align}
where $\varTheta(\tau):=\exp\{ i\vartheta(\tau) \sigma_3 \}$, 
and $\vartheta=\vartheta(\tau)$ is a real gauge function \cite{CWZ, SalStr, Nieuwenhuizen}. 
Note here that $\vartheta$ is determined depending on $(\xi, \bar{\xi}\;\!)$ as well as $U$. 
Using $V=V(\xi, \bar{\xi}\;\!)$, we define the following new fields on $\mathcal{T}$: 
\begin{subequations}
\label{2.13}
\begin{align}
\mathsf{Z}_{i}^{A} &:=V^{\dagger}{}_{i}{}^{j} Z^{A}_{j} \,, 
\quad\;  
\bar{\mathsf{Z}}{}^{i}_{A} :=\bar{Z}{}^{j}_{A} V_{j}{}^{i} \,, 
\label{2.13a}
\\
\mathsf{b} &:=V^{\dagger} bV-i \dot{V}^{\dagger} V \,.
\label{2.13b}
\end{align}
\end{subequations}
The field $\mathsf{b}$ can be expanded as $\mathsf{b}=\mathsf{b}^{r} \sigma_{r}$, 
where $\mathsf{b}^{r}=\mathsf{b}^{r}(\tau)$ are real fields. 
Clearly, $\mathsf{b}^{r}$ behave as scalar-density fields of weight 1 on $\mathcal{T}$.  
With the new fields, the local $U(1)$ transformation (\ref{2.9}) reads 
\begin{subequations}
\label{2.14}
\begin{align}
\mathsf{Z}_{i}^{A} &\rightarrow \mathsf{Z}_{i}^{\prime A} =e^{i\theta(\tau)} \mathsf{Z}_{i}^{A} \,, 
\label{2.14a}
\\
\bar{\mathsf{Z}}{}^{i}_{A} &\rightarrow \bar{\mathsf{Z}}^{\prime\:\! i}_{A} 
=e^{-i\theta(\tau)} \bar{\mathsf{Z}}^{i}_{A} \,, 
\label{2.14b}
\\
h &\rightarrow h^{\prime} =e^{-2i\theta(\tau)} h \,, 
\label{2.14c}
\\
\bar{h} &\rightarrow \bar{h}^{\prime} =e^{2i\theta(\tau)} \bar{h} \,, 
\label{2.14d}
\\
\varphi &\rightarrow \varphi^{\prime} =\varphi+2\theta(\tau) \,, 
\label{2.14e}
\\
a &\rightarrow a^{\prime}=a+\dot{\theta} \,, 
\label{2.14f}
\\
\mathsf{b} &\rightarrow \mathsf{b}^{\prime}=\mathsf{b} \,.  
\label{2.14g}
\end{align}
\end{subequations}
On the other hand, from Eqs. (\ref{2.10}) and (\ref{2.12}), we have 
\begin{subequations}
\label{2.15}
\begin{align}
\mathsf{Z}_{i}^{A} &\rightarrow \mathsf{Z}_{i}^{\prime A} =\varTheta_{i}{}^{j}(\tau) \mathsf{Z}^{A}_{j} \,, 
\label{2.15a}
\\
\bar{\mathsf{Z}}{}^{i}_{A} &\rightarrow \bar{\mathsf{Z}}{}^{\prime\:\! i}_{A} 
=\bar{\mathsf{Z}}{}^{j}_{A} \varTheta^{\dagger}{}_{j}{}^{i}(\tau) \,, 
\label{2.15b}
\\
h &\rightarrow h^{\prime} =h \,, 
\label{2.15c}
\\
\bar{h} &\rightarrow \bar{h}^{\prime} =\bar{h} \,, 
\label{2.15d}
\\
\varphi &\rightarrow \varphi^{\prime} =\varphi \,, 
\label{2.15e}
\\
a & \rightarrow a^{\prime}=a \,, 
\label{2.15f}
\\
\mathsf{b} &\rightarrow \mathsf{b}^{\prime}=\varTheta \mathsf{b}\varTheta^{\dagger} +\dot{\vartheta} \sigma_{3} \,.  
\label{2.15g}
\end{align}
\end{subequations}
Equation (\ref{2.15}) is precisely a local $U(1)$ transformation. 
Hereafter, we refer to the local $U(1)$ transformation specified by 
Eq. (\ref{2.9}), or Eq. (\ref{2.14}), as 
the $U(1)_{a}$ transformation and refer to that specified by Eq. (\ref{2.15}) as 
the $U(1)_{\mathsf{b}}$ transformation. 
Their corresponding gauge groups are simply denoted as $U(1)_{a}$ and $U(1)_{\mathsf{b}}$. 
The local SU(2) transformation is not manifestly seen in Eq. (\ref{2.15}); 
instead, it is realized as a nonlinear transformation of $\xi$. 
We may say that the function $V$ converts the local $SU(2)$ transformation into 
the $U(1)_{\mathsf{b}}$ transformation while $\xi$ undergoes a nonlinear transformation. 
Equation (\ref{2.15g}) defines the transformation rules of the fields $\mathsf{b}^{r}$,  
\begin{subequations}
\label{2.16}
\begin{align}
\mathsf{b}^{1} & \rightarrow \mathsf{b}^{\prime 1}
=\mathsf{b}^{1} \cos 2\vartheta +\mathsf{b}^{2} \sin 2\vartheta \,,
\label{2.16a}
\\
\mathsf{b}^{2} & \rightarrow \mathsf{b}^{\prime 2}
=-\mathsf{b}^{1} \sin 2\vartheta +\mathsf{b}^{2} \cos 2\vartheta \,, 
\label{2.16b}
\\
\mathsf{b}^{3} & \rightarrow \mathsf{b}^{\prime 3}=\mathsf{b}^{3} +\dot{\vartheta} \,. 
\label{2.16c}
\end{align}
\end{subequations}
We see that $\mathsf{b}^{\hat{\imath}}$ $(\hat{\imath}=1,2)$ transform homogeneously, obeying 
together an $SO(2)$ rotation, while $\mathsf{b}^{3}$ 
transforms inhomogeneously as a $U(1)$ gauge field.

Now, we can provide the following two terms: 
\begin{align}
S_{\mathsf{b}12}=-k \int_{\tau_0}^{\tau_1} d\tau 
\sqrt{\mathsf{b}^{\hat{\imath}} \mathsf{b}^{\hat{\imath}} } \,, 
\label{2.17}
\end{align}
with $\mathsf{b}^{\hat{\imath}} \mathsf{b}^{\hat{\imath}} :=
(\mathsf{b}^{1})^{2} +(\mathsf{b}^{2})^{2}$, and  
\begin{align}
S_{\mathsf{b}3}=-2t \int_{\tau_0}^{\tau_1} d\tau \:\! \mathsf{b}^{3} \,. 
\label{2.18}
\end{align}
Here, $k$ is a positive constant and $t$ is a  real constant. 
Since $\mathsf{b}^{r}$ are scalar-density fields of weight 1 on $\mathcal{T}$, 
both $S_{\mathsf{b}12}$ and $S_{\mathsf{b}3}$ are reparametrization invariant. 
It is obvious that $S_{\mathsf{b}12}$ remains invariant under 
the $SO(2)$ rotation defined by Eqs. (\ref{2.16a}) and (\ref{2.16b}). 
Also, $S_{\mathsf{b}3}$, which is the 1D Chern-Simons term for $\mathsf{b}^{3}$, 
remains invariant under the gauge transformation (\ref{2.16c}), 
provided that $\vartheta$ satisfies an appropriate boundary condition such as 
$\vartheta(\tau_1)=\vartheta(\tau_0)$. 
We thus see that both $S_{\mathsf{b}12}$ and $S_{\mathsf{b}3}$ possess the $U(1)_{\mathsf{b}}$ symmetry. 
The $U(1)_{a}$ invariance of  $S_{\mathsf{b}12}$ and $S_{\mathsf{b}3}$ is evident from Eq. (\ref{2.14g}). 
For our investigation, it is convenient to express $S_{\mathsf{b}12}$ as  
\begin{align}
S_{\mathsf{be}}=-\int_{\tau_0}^{\tau_1} d\tau 
\! \left( \frac{1}{2\mathsf{e}} \mathsf{b}^{\hat{\imath}} \mathsf{b}^{\hat{\imath}} 
+\frac{k^2}{2} \mathsf{e} \right) 
\label{2.19}
\end{align}
with the aid of $\mathsf{e}=\mathsf{e}(\tau)$ being a positive scalar-density field of weight 1 on $\mathcal{T}$. 
It is assumed that $\mathsf{e}$ does not change under the $U(1)_{a}$ and $U(1)_{\mathsf{b}}$ transformations.  
(At this stage, we should include the transformation rule $\mathsf{e} \rightarrow \mathsf{e}^{\prime}=\mathsf{e}$ 
in each of Eqs. (\ref{2.14}) and (\ref{2.15}).)  
The action $S_{m \mathrm{g}}$ can be rewritten in terms of  
$\mathsf{Z}_{i}^{A}$, $\bar{\mathsf{Z}}{}^{i}_{A}$, $h$, $\bar{h}$, $\varphi$, $a$, and $\mathsf{b}$. 
The resulting rewritten expression of $S_{m \mathrm{g}}$ is precisely what is obtained by replacing 
$Z_{i}^{A}$, $\bar{Z}{}^{i}_{A}$, and $b$ in Eq. (\ref{2.7}) 
with $\mathsf{Z}_{i}^{A}$, $\bar{\mathsf{Z}}{}^{i}_{A}$, and $\mathsf{b}$, respectively. 
With this expression, we modify $\tilde{S}_{m \mathrm{g}}=S_{m \mathrm{g}}+S_{a}$ 
by adding $S_{\mathsf{be}}$ and $S_{\mathsf{b}3}$ to it. 
That is, we consider the modified action 
${S}:=S_{m \mathrm{g}}+S_{a}+S_{\mathsf{be}}+S_{\mathsf{b}3}$, or more precisely,  
\begin{align}
{S} &= \int_{\tau_0}^{\tau_1} d\tau 
\bigg[\;\! 
\frac{i}{2} \big(\bar{\mathsf{Z}}_{A}^{i} \mathsf{D} \mathsf{Z}^{A}_{i} 
-\mathsf{Z}^{A}_{i} \bar{\mathsf{D}} \bar{\mathsf{Z}}_{A}^{i} \big) -2sa -2t\mathsf{b}^{3}  
\nonumber
\\
& \quad \,
-\frac{1}{2\mathsf{e}} \mathsf{b}^{\hat{\imath}} \mathsf{b}^{\hat{\imath}} 
-\frac{k^2}{2} \mathsf{e}
+h \Big(\epsilon^{ij} \varpi_{i \dot{\alpha}} \varpi_{j}^{\dot{\alpha}} -\sqrt{2}\:\! m e^{i\varphi} \Big) 
\nonumber 
\\
& \quad \,
+\bar{h} \Big(\epsilon_{ij} \bar{\varpi}^{i}_{\alpha} \bar{\varpi}^{j \alpha} -\sqrt{2}\:\! m e^{-i\varphi} \Big) 
\bigg] \,,
\label{2.20}
\end{align}
with 
\begin{subequations}
\label{2.21}
\begin{align}
\mathsf{D} \mathsf{Z}^{A}_{i} &:=\dot{\mathsf{Z}}^{A}_{i} -ia\mathsf{Z}^{A}_{i} -i\mathsf{b}_{i}{}^{j} \mathsf{Z}^{A}_{j} \,, 
\label{2.21a}
\\
\bar{\mathsf{D}}\bar{\mathsf{Z}}_{A}^{i} &:=\dot{\bar{\mathsf{Z}}}_{A}^{i} 
+ia\bar{\mathsf{Z}}_{A}^{i} +i\bar{\mathsf{Z}}_{A}^{j} \mathsf{b}_{j}{}^{i} \,.
\label{2.21b}
\end{align}
\end{subequations}
Here, $\varpi_{i \dot{\alpha}}$ and $\bar{\varpi}^{i}_{\alpha}$ are 
momentum-spinor components of the twistors 
$\mathsf{Z}_{i}^{A}=(\varrho_{i}^{\alpha}, \varpi_{i \dot{\alpha}})$ and 
$\bar{\mathsf{Z}}^{i}_{A}=(\bar{\varpi}^{i}_{\alpha}, \bar{\varrho}{}^{i\dot{\alpha}})$, respectively. 
We refer to $S$ as the GGS action. 
From Eq. (\ref{2.13a}), it follows that  
$\varpi_{i \dot{\alpha}}=V^{\dagger}{}_{i}{}^{j} \pi_{j \dot{\alpha}}$ and 
$\bar{\varpi}^{i}_{\alpha}=\bar{\pi}^{j}_{\alpha} V_{j}{}^{i}$. 
The other components are given by 
$\varrho_{i}^{\alpha}=V^{\dagger}{}_{i}{}^{j} \omega_{j}^{\alpha}$ and 
$\bar{\varrho}{}^{i\dot{\alpha}}=\bar{\omega}^{j \dot{\alpha}} V_{j}{}^{i}$. 
It is now obvious that $\overline{\varpi_{i \dot{\alpha}}}=\bar{\varpi}^{i}_{\alpha}$ and 
$\overline{\varrho_{i}^{\alpha}}=\bar{\varrho}{}^{i\dot{\alpha}}$.  
In terms of $\mathsf{Z}_{i}^{A}$, Eq. (\ref{2.1}) can be written as  
\begin{align}
\varrho_{i}^{\alpha}=i z^{\alpha \dot{\alpha}} \varpi_{i \dot{\alpha}} \,.  
\label{2.22}
\end{align}
It is clear from (\ref{2.20}) that the GGS action $S$ remains invariant under the reparametrization and 
the $U(1)_{a}$ and $U(1)_{\mathsf{b}}$ transformations. 
However, in actuality, ${S}$ remains invariant under the reparametrization and   
the $U(1)_{a}$ and local $SU(2)$ transformations, 
because the $U(1)_{\mathsf{b}}$ transformation is induced by the local $SU(2)$ transformation 
in accordance with Eq. (\ref{2.12}). 
In fact, we can express ${S}$ in a manifestly $SU(2)$ invariant form as follows: 
\begin{align}
{S} &= \int_{\tau_0}^{\tau_1} d\tau 
\bigg[\;\! 
\frac{i}{2} \big(\bar{Z}_{A}^{i} DZ^{A}_{i} 
-Z^{A}_{i} \bar{D}\bar{Z}_{A}^{i} \big)  
\nonumber
\\
& \quad \,
-2sa -2t \Big( b^{r} \mathcal{V}_{r}{}^{3} -\dot{\xi} e_{\xi}{}^{3} -\Dot{\Bar{\xi}} e_{\bar{\xi}}{}^{3} \Big)
\nonumber
\\
& \quad \, 
-\frac{1}{\mathsf{e}} g_{\xi\bar{\xi}} D\xi D\bar{\xi} -\frac{k^2}{2} \mathsf{e}
+h \Big( \epsilon^{ij} \pi_{i \dot{\alpha}} \pi_{j}^{\dot{\alpha}} -\sqrt{2} \:\! m e^{i\varphi} \Big) 
\nonumber
\\
& \quad \,
+\bar{h} \Big( \epsilon_{ij} \bar{\pi}^{i}_{\alpha} \bar{\pi}^{j \alpha} -\sqrt{2} \:\! m e^{-i\varphi} \Big)
\bigg] \,,
\label{2.23}
\end{align}
with $g_{\xi\bar{\xi}}:=e_{\xi}{}^{\hat{\imath}} e_{\bar{\xi}}{}^{\hat{\imath}}$, 
$D\xi :=\dot{\xi}-b^{r} K_{r}{}^{\xi}$, and 
$D\bar{\xi} :=\dot{\bar{\xi}}-b^{r} K_{r}{}^{\bar{\xi}}$. 
Here, $g_{\xi\bar{\xi}}$ is a metric on $SU(2)/U(1)$, 
$( K_{r}{}^{\xi}, K_{r}{}^{\bar{\xi}\;\!} )$ $(r=1,2,3)$ are the $SU(2)$ Killing vectors 
on this coset space, 
and $e_{\xi}{}^{r}$ and $e_{\bar{\xi}}{}^{r}$ $(r={\hat{\imath}}, 3)$ are defined by 
$e_{\xi}{}^{r} \sigma_{r}=-iV^{\dagger} (\partial V/{\partial \xi})$ and 
$e_{\bar{\xi}}{}^{r} \sigma_{r}=-iV^{\dagger} (\partial V/{\partial \bar{\xi}}^{\;\!})$, respectively. 
Also, $\mathcal{V}_{r}{}^{3}$ is defined according to  
$V^{\dagger} \sigma_{r} V
=\mathcal{V}_{r}{}^{\hat{\imath}} \sigma_{\hat{\imath}} +\mathcal{V}_{r}{}^{3} \sigma_{3}$. 
Using the transformation rule (\ref{2.12}), we can show that 
$\mathcal{V}_{r}{}^{\hat{\imath}} 
=K_{r}{}^{\xi} e_{\xi}{}^{\hat{\imath}} +K_{r}{}^{\bar{\xi}} e_{\bar{\xi}}{}^{\hat{\imath}}$. 
In addition, it can be verified that 
$K_{r}:=K_{r}{}^{\xi} \partial/\partial \xi +K_{r}{}^{\bar{\xi}} \partial/\partial \bar{\xi}$ 
satisfy the $SU(2)$ commutation relations. 
In the expression (\ref{2.20}), we should understand that the local $SU(2)$ symmetry of ${S}$ 
is hidden rather than is broken, because no symmetry breaking mechanisms are incorporated in the model. 
The action (\ref{2.20}) can be regarded as the action (\ref{2.23}) in a particular gauge $\xi(\tau)=\xi_{0}$, 
where $\xi_{0}$ is a constant such that $V(\xi_{0}, \bar{\xi}_{0})=1$.  
We term this gauge the unitary gauge, 
because it corresponds to the so-called unitary gauge in massive Yang-Mills theory 
\cite{tHooft, Weinberg}.  
Then $\mathsf{b}$ can be said to be the $SU(2)$ gauge field in the unitary gauge. 
The action (\ref{2.20}) can be written as 
\begin{align}
{S} &= \int_{\tau_0}^{\tau_1} d\tau 
\bigg[\;\! 
\frac{i}{2} \Big( \bar{\mathsf{Z}}_{A}^{i} \dot{\mathsf{Z}}{}^{A}_{i} 
-\mathsf{Z}^{A}_{i} \dot{\bar{\mathsf{Z}}}{}_{A}^{i} \Big) 
+a \big(\bar{\mathsf{Z}}_{A}^{i} \mathsf{Z}^{A}_{i} -2s \big) 
\nonumber
\\
& \quad \,
+\mathsf{b}^{3} \big( \bar{\mathsf{Z}}_{A}^{j} \sigma_{3j}{}^{k} \mathsf{Z}^{A}_{k} -2t \big) 
+\mathsf{b}^{\hat{\imath}} \bar{\mathsf{Z}}_{A}^{j} \sigma_{\hat{\imath} j}{}^{k} \mathsf{Z}^{A}_{k} 
\nonumber
\\
& \quad \,
-\frac{1}{2\mathsf{e}} \mathsf{b}^{\hat{\imath}} \mathsf{b}^{\hat{\imath}} -\frac{k^2}{2} \mathsf{e}
+h \Big(\epsilon^{ij} \varpi_{i \dot{\alpha}} \varpi_{j}^{\dot{\alpha}} -\sqrt{2}\:\! m e^{i\varphi} \Big) 
\nonumber 
\\
& \quad \,
+\bar{h} \Big(\epsilon_{ij} \bar{\varpi}^{i}_{\alpha} \bar{\varpi}^{j \alpha} -\sqrt{2}\:\! m e^{-i\varphi} \Big) 
\bigg] \,. 
\label{2.24}
\end{align}

\section{Canonical Formalism} 

In this section, we study the canonical Hamiltonian formalism of the model governed by the GGS action  
in the unitary gauge. 

Let $L$ be the Lagrangian defined in Eq. (\ref{2.24}) as the integrand of the GGS action $S$. 
We treat the variables  
$\left( \mathsf{Z}^{A}_{i}, \bar{\mathsf{Z}}_{A}^{i}, a, \mathsf{b}^{r},  \mathsf{e}, h, \bar{h}, \varphi \right)$ 
as canonical coordinates. 
Their canonical conjugate momenta are found to be 
\begin{subequations}
\label{3.1}
\begin{align}
P_{A}^{i}
&:= \frac{\partial L}{\partial \dot{\mathsf{Z}}^{A}_{i}}
=\frac{i}{2} \bar{\mathsf{Z}}_{A}^{i} \,, 
\label{3.1a}
\\
\bar{P}^{A}_{i}
&:= \frac{\partial L}{\partial \dot{\bar{\mathsf{Z}}}_{A}^{i}}
=-\frac{i}{2} \mathsf{Z}^{A}_{i} , 
\label{3.1b}
\\
P^{(a)}
&:= \frac{\partial L}{\partial \dot{a}}
= 0 \,, 
\label{3.1c}
\\
P^{(\mathsf{b})}_{r}
&:= \frac{\partial L}{\partial \dot{\mathsf{b}}{}^{r}}
= 0 \,, 
\label{3.1d}
\\
P^{(\mathsf{e})}
&:= \frac{\partial L}{\partial \dot{\mathsf{e}}}
= 0 \,, 
\label{3.1e}
\\
P^{(h)}
&:= \frac{\partial L}{\partial \dot{h}}
= 0 \,, 
\label{3.1f}
\\
P^{(\bar{h})}
&:= \frac{\partial L}{\partial \dot{\bar{h}}}
= 0 \,, 
\label{3.1g} 
\\
P^{(\varphi)}
&:= \frac{\partial L}{\partial \dot{\varphi}}
= 0 \,. 
\label{3.1h}
\end{align}
\end{subequations}
The canonical Hamiltonian corresponding to $L$ is defined by the Legendre transform of $L$, 
\begin{align}
H_{\rm{C}} &:=\dot{\mathsf{Z}}^{A}_{i} P_{A}^{i} 
+\dot{\bar{\mathsf{Z}}}_{A}^{i} \bar{P}^{A}_{i} 
+\dot{a} P^{(a)} +\dot{\mathsf{b}}{}^{r} P^{(\mathsf{b})}_{r} 
\notag
\\
&\quad\:\, +\dot{\mathsf{e}} P^{(\mathsf{e})} +\dot{h} P^{(h)} +\dot{\bar{h}} P^{(\bar{h})} +\dot{\varphi} P^{(\varphi)} -L
\notag 
\\
&\;=-a \big(\bar{\mathsf{Z}}_{A}^{i} \mathsf{Z}^{A}_{i} -2s \big) 
-\mathsf{b}^{3} \big( \bar{\mathsf{Z}}_{A}^{j} \sigma_{3j}{}^{k} \mathsf{Z}^{A}_{k} -2t \big) 
\notag
\\
&\quad\:\, -\mathsf{b}^{\hat{\imath}} \bar{\mathsf{Z}}_{A}^{j} \sigma_{\hat{\imath} j}{}^{k} \mathsf{Z}^{A}_{k} 
+\frac{1}{2\mathsf{e}} \mathsf{b}^{\hat{\imath}} \mathsf{b}^{\hat{\imath}} +\frac{k^2}{2} \mathsf{e}
\notag
\\
& \quad\:\,
-h \Big(\epsilon^{ij} \varpi_{i \dot{\alpha}} \varpi_{j}^{\dot{\alpha}} -\sqrt{2}\:\! m e^{i\varphi} \Big) 
\nonumber 
\\
& \quad\:\,
-\bar{h} \Big(\epsilon_{ij} \bar{\varpi}^{i}_{\alpha} \bar{\varpi}^{j \alpha} -\sqrt{2}\:\! m e^{-i\varphi} \Big) . 
\label{3.2}
\end{align}
The equal-time Poisson brackets between the canonical variables are given by 
\begin{alignat}{3}
\Big\{ \mathsf{Z}^{A}_{i} , P_{B}^{j} \Big\} &= \delta_{i}^{j} \delta_{B}^{A}  \,, 
&\quad 
\Big\{ \bar{\mathsf{Z}}_{A}^{i} , \bar{P}^{B}_{j} \Big\} &= \delta^{i}_{j} \delta_{A}^{B} \,, 
\notag
\\
\left\{ a \:\!, P^{(a)} \right\} &= 1 \,, 
&\quad 
\left\{ \mathsf{b}^{r} , P^{(\mathsf{b})}_{s} \right\} &= \delta_{s}^{r} \,, 
\notag
\\
\left\{ \mathsf{e} \:\!, P^{(\mathsf{e})} \right\} &= 1 \,, 
&\quad 
\left\{ h \:\!, P^{(h)} \right\} &= 1 \,, 
\notag
\\
\left\{ \bar{h} \:\!, P^{(\bar{h})} \right\} &= 1 \,, 
&\quad
\left\{ \varphi \:\!, P^{(\varphi)} \right\} &= 1 \,, 
\notag
\\
\mbox{all others} &=0 \,, 
\label{3.3}
\end{alignat}
which can be used for calculating the Poisson bracket between two arbitrary analytic functions 
of the canonical variables.

Equations (\ref{3.1a})--(\ref{3.1h}) are read as the primary constraints 
\begin{subequations}
\label{3.4}
\begin{align}
\phi_{A}^{i} &:=P_{A}^{i} -\frac{i}{2} \bar{\mathsf{Z}}_{A}^{i} \approx 0 \,, 
\label{3.4a}
\\
\bar{\phi}^{A}_{i} &:=\bar{P}^{A}_{i} +\frac{i}{2} \mathsf{Z}^{A}_{i} \approx 0 \,, 
\label{3.4b}
\\
\phi^{(a)}	&:=P^{(a)} \approx 0 \,, 
\label{3.4c}
\\
\phi^{(\mathsf{b})}_{r} &:=P^{(\mathsf{b})}_{r} \approx 0 \,, 
\label{3.4d}
\\
\phi^{(\mathsf{e})} &:=P^{(\mathsf{e})} \approx 0 \,, 
\label{3.4e}
\\
\phi^{(h)}	&:=P^{(h)} \approx 0 \,, 
\label{3.4f}
\\
\phi^{(\bar{h})}	&:=P^{(\bar{h})} \approx 0 \,, 
\label{3.4g}
\\
\phi^{(\varphi)} &:=P^{(\varphi)} \approx 0 \,, 
\label{3.4h}
\end{align}
\end{subequations}
where the symbol ^^ ^^ $\approx$" denotes the weak equality. 
Now, we follow the Dirac algorithm for constrained Hamiltonian systems \cite{Dirac2, HRT, HenTei} 
to establish the canonical formalism of the present model. 
We see that the Poisson brackets between the primary constraint functions $\phi$'s 
are summarized in 
\begin{align}
\big\{ \phi^{i}_{A} , \bar{\phi}{}_{j}^{B} \big\} =-i\delta^{i}_{j} \delta_{A}^{B} \,,  
\quad 
\mbox{all others} =0 \,. 
\label{3.5}
\end{align}
The Poisson brackets between $H_{\rm{C}}$ and the primary constraint functions are found to be 
\begin{subequations}
\label{3.6}
\begin{align}
\left\{ \phi^{i}_{A} , H_{\rm{C}} \right\}
&= a\bar{\mathsf{Z}}{}^{i}_{A} +\mathsf{b}^{r} \sigma_{rj}{}^{i} \bar{\mathsf{Z}}{}^{j}_{A}
+2h\epsilon^{ij} I_{AB} \mathsf{Z}^{B}_{j} ,
\label{3.6a}
\\
\left\{ \bar{\phi}_{i}^{A} , H_{\rm{C}} \right\}
&= a\mathsf{Z}_{i}^{A} +\mathsf{b}^{r} \sigma_{ri}{}^{j} \mathsf{Z}_{j}^{A} 
+2\bar{h}\epsilon_{ij} I^{AB} \bar{\mathsf{Z}}_{B}^{j} \,, 
\label{3.6b}
\\
\left\{ \phi^{(a)} , H_{\rm{C}} \right\}
&=\bar{\mathsf{Z}}_{A}^{i} \mathsf{Z}^{A}_{i} -2s \,, 
\label{3.6c}
\\
\left\{ \phi^{(\mathsf{b})}_{\hat{\imath}} , H_{\rm{C}} \right\}
&=\bar{\mathsf{Z}}_{A}^{j} \sigma_{\hat{\imath} j}{}^{k} \mathsf{Z}^{A}_{k} 
-\frac{1}{\mathsf{e}} \mathsf{b}_{\hat{\imath}} \,, 
\label{3.6d}
\\
\left\{ \phi^{(\mathsf{b})}_{3} , H_{\rm{C}} \right\}
&=\bar{\mathsf{Z}}_{A}^{j} \sigma_{3j}{}^{k} \mathsf{Z}^{A}_{k} -2t \,, 
\label{3.6e}
\\
\left\{ \phi^{(\mathsf{e})} , H_{\rm{C}} \right\}
&=\frac{1}{2\mathsf{e}^{2}} 
\big( \mathsf{b}^{\hat{\imath}} \mathsf{b}^{\hat{\imath}} -k^{2} \mathsf{e}^{2} \big) \,, 
\label{3.6f}
\\
\left\{ \phi^{(h)} , H_{\rm{C}} \right\}
&=\epsilon^{ij} \varpi_{i \dot{\alpha}} \varpi_{j}^{\dot{\alpha}} -\sqrt{2}\:\! m e^{i\varphi} ,
\label{3.6g}
\\
\left\{ \phi^{(\bar{h})} , H_{\rm{C}} \right\}
&=\epsilon_{ij} \bar{\varpi}^{i}_{\alpha} \bar{\varpi}^{j \alpha} -\sqrt{2}\:\! m e^{-i\varphi} , 
\label{3.6h}
\\
\left\{ \phi^{(\varphi)} , H_{\rm{C}} \right\}
&=-i\sqrt{2}\:\! m \big( h e^{i\varphi} -\bar{h} e^{-i\varphi} \big) \,, 
\label{3.6i}
\end{align}
\end{subequations}
where $I_{AB}$ and $I^{AB}$ are the so-called infinity twistors \cite{PenMac, PenRin, Hughston}, 
defined by 
\begin{align}
I_{AB}:= \left(\,
\begin{array}{cc}
0 & 0 \\
0 & \, \epsilon^{\dot{\alpha} \dot{\beta}} 
\end{array}
\! \right) , 
\quad 
I^{AB}:= \left(
\begin{array}{cc}
\epsilon^{\alpha\beta} &\:\! 0 \\
0 & \:\! 0 
\end{array}
\:\! \right) .
\notag
\end{align}
With $H_{\rm{C}}$ and the primary constraint functions, we define the total Hamiltonian 
\begin{align}
H_{\rm{T}} &:=H_{\rm{C}} +u_{i}^{A} \phi^{i}_{A} +\bar{u}^{i}_{A} \bar{\phi}_{i}^{A} 
+u_{(a)} \phi^{(a)} +u_{(\mathsf{b})}^{r} \phi^{(\mathsf{b})}_{r} 
\notag
\\
&\quad\:\, 
+u_{(\mathsf{e})} \phi^{(\mathsf{e})} 
+u_{(h)} \phi^{(h)} +u_{(\bar{h})} \phi^{(\bar{h})} +u_{(\varphi)} \phi^{(\varphi)} , 
\label{3.7}
\end{align}
where $u_{i}^{A}$, $\bar{u}^{i}_{A}$, $u_{(a)}$, $u_{(\mathsf{b})}^{r}$, $u_{(\mathsf{e})}$, 
$u_{(h)}$, $u_{(\bar{h})}$, and $u_{(\varphi)}$ are Lagrange multipliers.  
The time evolution of a function $f$ of the canonical variables is governed by 
the canonical equation 
\begin{align}
\dot{f}=\{ f, H_{\rm{T}} \}\,. 
\label{3.8}
\end{align}
Using this equation together with Eqs. (\ref{3.4})--(\ref{3.7}), 
we can evaluate the time evolution of the primary constraint functions. 
Because the primary constraints (\ref{3.4a})--(\ref{3.4h}) are valid at any time, 
they must be preserved in time. This fact leads to the consistency conditions 
\begin{subequations}
\label{3.9}
\begin{align}
\dot{\phi}{}^{i}_{A} &
=\left\{ \phi^{i}_{A} , H_{\rm{T}} \right\}
\notag
\\
& \approx a\bar{\mathsf{Z}}{}^{i}_{A} 
+\mathsf{b}^{r} \sigma_{rj}{}^{i} \bar{\mathsf{Z}}{}^{j}_{A} 
+2h\epsilon^{ij} I_{AB} \mathsf{Z}^{B}_{j} 
-i\bar{u}^{i}_{A} 
\approx 0 \,,
\label{3.9a}
\\
\Dot{\Bar{\phi}}{}_{i}^{A} &
=\left\{ \bar{\phi}_{i}^{A} , H_{\rm{T}} \right\} 
\notag
\\
& \approx a\mathsf{Z}_{i}^{A} +\mathsf{b}^{r} \sigma_{ri}{}^{j} \mathsf{Z}_{j}^{A} 
+2\bar{h}\epsilon_{ij} I^{AB} \bar{\mathsf{Z}}_{B}^{j}
+iu_{i}^{A} 
\approx 0 \,, 
\label{3.9b}
\\
\dot{\phi}{}^{(a)} &
=\Big\{ \phi^{(a)} , H_{\rm{T}} \Big\}
\approx \bar{\mathsf{Z}}_{A}^{i} \mathsf{Z}^{A}_{i} -2s 
\approx 0 \,,
\label{3.9c}
\\
\dot{\phi}{}^{(\mathsf{b})}_{\hat{\imath}} &
=\Big\{ \phi^{(\mathsf{b})}_{\hat{\imath}} , H_{\rm{T}} \Big\}
\approx 
\bar{\mathsf{Z}}_{A}^{j} \sigma_{\hat{\imath} j}{}^{k} \mathsf{Z}^{A}_{k} 
-\frac{1}{\mathsf{e}} \mathsf{b}_{\hat{\imath}}
\approx 0 \,,
\label{3.9d}
\\
\dot{\phi}{}^{(\mathsf{b})}_{3} &
=\Big\{ \phi^{(\mathsf{b})}_{3} , H_{\rm{T}} \Big\}
\approx 
\bar{\mathsf{Z}}_{A}^{j} \sigma_{3j}{}^{k} \mathsf{Z}^{A}_{k} -2t 
\approx 0 \,,
\label{3.9e}
\\
\dot{\phi}{}^{(\mathsf{e})} &
=\Big\{ \phi^{(\mathsf{e})} , H_{\rm{T}} \Big\}
\approx 
\frac{1}{2\mathsf{e}^{2}} 
\big( \mathsf{b}^{\hat{\imath}} \mathsf{b}^{\hat{\imath}} -k^{2} \mathsf{e}^{2} \big) 
\approx 0 \,,
\label{3.9f}
\\
\dot{\phi}{}^{(h)} &
=\Big\{ \phi^{(h)} , H_{\rm{T}} \Big\}
\approx 
\epsilon^{ij} \varpi_{i \dot{\alpha}} \varpi_{j}^{\dot{\alpha}} -\sqrt{2}\:\! m e^{i\varphi} 
\approx 0 \,,
\label{3.9g}
\\
\dot{\phi}{}^{(\bar{h})} &
=\Big\{ \phi^{(\bar{h})} , H_{\rm{T}} \Big\}
\approx 
\epsilon_{ij} \bar{\varpi}^{i}_{\alpha} \bar{\varpi}^{j \alpha} -\sqrt{2}\:\! m e^{-i\varphi} 
\approx 0 \,,
\label{3.9h}
\\
\dot{\phi}{}^{(\varphi)} &
=\Big\{ \phi^{(\varphi)} , H_{\rm{T}} \Big\}
\approx 
-i\sqrt{2}\:\! m \big( h e^{i\varphi} -\bar{h} e^{-i\varphi} \big) 
\approx 0 \,.
\label{3.9i}
\end{align}
\end{subequations}
Equations (\ref{3.9a}) and (\ref{3.9b}) determine $\bar{u}^{i}_{A}$ and $u_{i}^{A}$, respectively, 
as follows: 
\begin{subequations}
\label{3.10}
\begin{align}
\bar{u}^{i}_{A}&= -ia\bar{\mathsf{Z}}{}^{i}_{A} -i\mathsf{b}^{r} \sigma_{rj}{}^{i} \bar{\mathsf{Z}}{}^{j}_{A} 
-2i h\epsilon^{ij} I_{AB} \mathsf{Z}^{B}_{j} \,, 
\label{3.10a}
\\
u_{i}^{A}&= ia\mathsf{Z}_{i}^{A} +i\mathsf{b}^{r} \sigma_{ri}{}^{j} \mathsf{Z}_{j}^{A} 
+2i \bar{h}\epsilon_{ij} I^{AB} \bar{\mathsf{Z}}_{B}^{j} \,.  
\label{3.10b}
\end{align}
\end{subequations}
In contrast, Eqs. (\ref{3.9c})--(\ref{3.9i}) give rise to the secondary constraints 
\begin{subequations}
\label{3.11}
\begin{align}
\chi^{(a)} &:= \bar{\mathsf{Z}}_{A}^{i} \mathsf{Z}^{A}_{i} -2s 
\approx 0 \,,
\label{3.11a}
\\
\chi^{(\mathsf{b})}_{\hat{\imath}} &:= 
\bar{\mathsf{Z}}_{A}^{j} \sigma_{\hat{\imath} j}{}^{k} \mathsf{Z}^{A}_{k} 
-\frac{1}{\mathsf{e}} \mathsf{b}_{\hat{\imath}}
\approx 0 \,,
\label{3.11b}
\\
\chi^{(\mathsf{b})}_{3} &:= \bar{\mathsf{Z}}_{A}^{j} \sigma_{3j}{}^{k} \mathsf{Z}^{A}_{k} -2t 
\approx 0 \,,
\label{3.11c}
\\
\chi^{(\mathsf{e})} &:=\frac{1}{2} \big( \mathsf{b}^{\hat{\imath}} \mathsf{b}^{\hat{\imath}} -k^{2} \mathsf{e}^{2} \big) 
\approx 0 \,,
\label{3.11d}
\\
\chi^{(h)} &:=\epsilon^{ij} \varpi_{i \dot{\alpha}} \varpi_{j}^{\dot{\alpha}} -\sqrt{2}\:\! m e^{i\varphi} 
\approx 0 \,,
\label{3.11e}
\\
\chi^{(\bar{h})} &:=\epsilon_{ij} \bar{\varpi}^{i}_{\alpha} \bar{\varpi}^{j \alpha} -\sqrt{2}\:\! m e^{-i\varphi} 
\approx 0 \,,
\label{3.11f}
\\
\chi^{(\varphi)} &:= i\big( h e^{i\varphi} -\bar{h} e^{-i\varphi} \big) 
\approx 0 \,.
\label{3.11g}
\end{align}
\end{subequations}
All the Poisson brackets between $H_{\rm C}$ and the secondary constraint functions $\chi$'s vanish. 
The Poisson brackets between the primary and secondary constraint functions  
are found to be   
\begin{alignat}{3}
\left\{ \chi^{(a)}, \phi^{i}_{A} \right\} &=\bar{\mathsf{Z}}_{A}^{i} \,, 
& \!\!\!\!\!\!\!\!\!\!\!\!\!\!\!
\Big\{ \chi^{(a)}, \bar{\phi}_{i}^{A} \Big\} &=\mathsf{Z}^{A}_{i} , 
\notag
\\
\left\{ \chi^{(\mathsf{b})}_{r}, \phi^{i}_{A} \right\} &=\sigma_{r j}{}^{i} \bar{\mathsf{Z}}_{A}^{j}\,, 
& \!\!\!\!\!\!\!\!\!\!\!\!\!\!\!
\Big\{ \chi^{(\mathsf{b})}_{r}, \bar{\phi}_{i}^{A} \Big\} &=\sigma_{r i}{}^{j} \mathsf{Z}^{A}_{j}, 
\notag
\\
\left\{ \chi^{(\mathsf{b})}_{\hat{\imath}}, \phi^{(\mathsf{b})}_{\hat{\jmath}} \right\} 
&= -\frac{1}{\mathsf{e}} \delta_{{\hat{\imath}}{\hat{\jmath}}} \,, 
& \!\!\!\!\!\!\!\!\!\!\!\!\!\!\! 
\left\{ \chi^{(\mathsf{b})}_{\hat{\imath}}, \phi^{(\mathsf{e})} \right\} 
&= \frac{1}{\mathsf{e}^{2}} \mathsf{b}_{\hat{\imath}} \,,
\notag
\\
\left\{ \chi^{(\mathsf{e})}, \phi^{(\mathsf{b})}_{\hat{\jmath}} \right\} 
&= \mathsf{b}_{\hat{\jmath}} \,, 
& \!\!\!\!\!\!\!\!\!\!\!\!\!\!\! 
\left\{ \chi^{(\mathsf{e})}, \phi^{(\mathsf{e})} \right\} 
&= -k^{2} \mathsf{e} \,,
\notag
\\
\left\{ \chi^{(h)}, \phi^{i\dot{\alpha}} \right\} &=2\epsilon^{ij} \varpi_{j}^{\dot{\alpha}} , 
& \!\!\!\!\!\!\!\!\!\!\!\!\!\!\!
\left\{ \chi^{(h)}, \phi^{(\varphi)} \right\} &=-i\sqrt{2} \:\! me^{i\varphi} , 
\notag
\\
\left\{ \chi^{(\bar{h})}, \bar{\phi}_{i}^{\alpha} \right\} &=2\epsilon_{ij} \bar{\varpi}^{j \alpha} , 
& \!\!\!\!\!\!\!\!\!\!\!\!\!\!\!
\left\{ \chi^{(\bar{h})}, \phi^{(\varphi)} \right\} &=i\sqrt{2} \:\! me^{-i\varphi}, 
\notag
\\
\left\{ \chi^{(\varphi)}, \phi^{(h)} \right\} &= ie^{i\varphi} , 
& \!\!\!\!\!\!\!\!\!\!\!\!\!\!\!
\left\{ \chi^{(\varphi)}, \phi^{(\bar{h})} \right\} &= -ie^{-i\varphi} , 
\notag
\\
\left\{ \chi^{(\varphi)}, \phi^{(\varphi)} \right\} &= -\big( h e^{i\varphi} +\bar{h} e^{-i\varphi} \big)   \,, 
~
\notag
\\
\mbox{all others} &=0 \,, 
\label{3.12}
\end{alignat}
where $\phi^{i\dot{\alpha}}$ and $\bar{\phi}{}_{i}^{\alpha}$ are spinor components 
of $\phi^{i}_{A}=\big(\phi^{i}_{\alpha}, \phi^{i\dot{\alpha}} \big)$ and 
$\bar{\phi}_{i}^{A}=\big( \bar{\phi}{}_{i}^{\alpha}, \bar{\phi}_{i\dot{\alpha}} \big)$, respectively. 
All the Poisson brackets between the secondary constraint functions vanish.

Next we investigate the time evolution of the secondary constraint functions using Eqs. (\ref{3.8}) 
and (\ref{3.12}). The time evolution of $\chi^{(a)}$ is evaluated as 
\begin{align}
\dot{\chi}^{(a)} 
=\left\{ \chi^{(a)}, H_{\rm{T}} \right\}
\approx
u_{i}^{A} \bar{\mathsf{Z}}_{A}^{i} +\bar{u}^{i}_{A} \mathsf{Z}^{A}_{i}.  
\label{3.13}
\end{align}
The condition $\dot{\chi}^{(a)}\approx 0$ is identically fulfilled  
with the aid of Eqs. (\ref{3.10a}), (\ref{3.10b}), (\ref{3.11e}), (\ref{3.11f}), and (\ref{3.11g}), 
and hence no new constraints are obtained from $\dot{\chi}^{(a)}\approx 0$. 
The time evolution of $\chi^{(\mathsf{b})}_{r}$ is evaluated as  
\begin{align}
\dot{\chi}{}^{(\mathsf{b})}_{r}
&=\left\{ \chi^{(\mathsf{b})}_{r}, H_{\rm{T}} \right\}
\notag 
\\
& \approx
u_{i}^{A} \sigma_{r j}{}^{i} \bar{\mathsf{Z}}_{A}^{j} 
+\bar{u}^{i}_{A} \sigma_{r i}{}^{j} \mathsf{Z}^{A}_{j} 
\notag
\\
& \quad \;\! +u^{s}_{(\mathsf{b})} \Big\{ \chi^{(\mathsf{b})}_{r}, \phi^{(\mathsf{b})}_{s} \Big\}  
+u_{(\mathsf{e})} \Big \{ \chi^{(\mathsf{b})}_{r}, \phi^{(\mathsf{e})} \Big\}  
\notag 
\\
& =-2\epsilon_{rst} \mathsf{b}^{s} \bar{\mathsf{Z}}_{A}^{j} \sigma_{tj}{}^{k} \mathsf{Z}^{A}_{k} 
\notag 
\\
& \quad \;\! +u^{s}_{(\mathsf{b})} \Big\{ \chi^{(\mathsf{b})}_{r}, \phi^{(\mathsf{b})}_{s} \Big\}  
+u_{(\mathsf{e})} \Big \{ \chi^{(\mathsf{b})}_{r}, \phi^{(\mathsf{e})} \Big\}  
\notag
\\
& \approx
-\frac{2}{\mathsf{e}} \epsilon_{rs \hat{\jmath}} \mathsf{b}^{s} \mathsf{b}^{\hat{\jmath}} 
-4t \epsilon_{rs3} \mathsf{b}^{s} 
\notag 
\\
& \quad \;\! +u^{s}_{(\mathsf{b})} \Big\{ \chi^{(\mathsf{b})}_{r}, \phi^{(\mathsf{b})}_{s} \Big\}  
+u_{(\mathsf{e})} \Big \{ \chi^{(\mathsf{b})}_{r}, \phi^{(\mathsf{e})} \Big\}  
\label{3.14}
\end{align}
by using Eqs. (\ref{3.10a}), (\ref{3.10b}), (\ref{3.11b}), and (\ref{3.11c}), 
together with the formulas  
$\sigma_{rk}{}^{i} \epsilon^{kj}=\sigma_{rk}{}^{j} \epsilon^{ki}$ and 
$\sigma_{ri}{}^{k} \epsilon_{kj}=\sigma_{rj}{}^{k} \epsilon_{ki}$. 
Then we see  that the condition $\dot{\chi}{}^{(\mathsf{b})}_{\hat{\imath}} \approx 0$ 
determines $u^{\hat{\imath}}_{(\mathsf{b})}$ as follows: 
\begin{align}
u^{\hat{\imath}}_{(\mathsf{b})}
=2\epsilon^{\hat{\imath} \hat{\jmath}} \mathsf{b}_{\hat{\jmath}} 
\!\left( \mathsf{b}^{3} -2t\mathsf{e} \right) 
+\frac{1}{\mathsf{e}} \mathsf{b}^{\hat{\imath}} u_{(\mathsf{e})} \,, 
\label{3.15}
\end{align}
while $\dot{\chi}{}^{(\mathsf{b})}_{3}\approx 0$ is identically satisfied. 
The time evolution of $\chi^{(\mathsf{e})}$ is calculated as 
\begin{align}
\dot{\chi}{}^{(\mathsf{e})}
&=\left\{ \chi^{(\mathsf{e})}, H_{\rm{T}} \right\} 
\notag 
\\
& \approx 
\mathsf{b}_{\hat{\imath}} u^{\hat{\imath}}_{(\mathsf{b})} -k^{2} \mathsf{e} u_{(\mathsf{e})} 
=\frac{2}{\mathsf{e}} \chi^{(\mathsf{e})} u_{(\mathsf{e})} \approx 0
\label{(3.16)}
\end{align}
by using Eqs. (\ref{3.15}) and (\ref{3.11d}). 
Hence $\dot{\chi}{}^{(\mathsf{e})} \approx 0$ is identically satisfied. 
The time evolution of $\chi^{(h)}$ is evaluated as  
\begin{align}
\dot{\chi}{}^{(h)}
&=\left\{ \chi^{(h)}, H_{\rm{T}} \right\}
\notag 
\\
& \approx
2\epsilon^{ij} u_{i\dot{\alpha}} \varpi_{j}^{\dot{\alpha}} -i\sqrt{2}\:\! m e^{i\varphi} u_{(\varphi)}
\notag 
\\
&=2ia\epsilon^{ij} \varpi_{i \dot{\alpha}} \varpi_{j}^{\dot{\alpha}} 
-i\sqrt{2}\:\! m e^{i\varphi} u_{(\varphi)}
\notag 
\\
& \approx 
i\sqrt{2}\:\! m e^{i\varphi} (2a-u_{(\varphi)})
\label{3.17}
\end{align}
by using Eqs. (\ref{3.10b}), (\ref{3.11e}) and the formula 
$\sigma_{rk}{}^{i} \epsilon^{kj}=\sigma_{rk}{}^{j} \epsilon^{ki}$. 
From the condition $\dot{\chi}{}^{(h)} \approx 0$, 
the Lagrange multiplier $u_{(\varphi)}$ is determined to be $u_{(\varphi)}=2a$. 
Similarly,  $\dot{\chi}{}^{(\bar{h})} \approx -i\sqrt{2}^{\:\!} m e^{-i\varphi} (2a-u_{(\varphi)})\approx 0$ 
leads to $u_{(\varphi)}=2a$. 
The time evolution of $\chi^{(\varphi)}$ is found to be   
\begin{align}
\dot{\chi}{}^{(\varphi)}
&=\left\{ \chi^{(\varphi)}, H_{\rm{T}} \right\}
\notag 
\\
& \approx 
i \big(u_{(h)}-u_{(\bar{h})} \big)-2a \big(he^{i\varphi}+\bar{h}e^{-i\varphi} \big) \,, 
\label{3.18}
\end{align}
so that the condition $\dot{\chi}{}^{(\varphi)} \approx 0$ gives 
$u_{(h)}-u_{(\bar{h})}=-2ia \big(he^{i\varphi}+\bar{h}e^{-i\varphi} \big)$. 
From the above analysis, we see that no further constraints can be derived; 
thus, the procedure for deriving constraints is now completed.  
We also see that $u_{i}^{A}$, $\bar{u}^{i}_{A}$, $u_{(\mathsf{b})}^{\hat{\imath}}$, 
$u_{(h)}-u_{(\bar{h})}$, and $u_{(\varphi)}$ are determined to be what are 
written in terms of other variables such as the canonical coordinates, 
while $u_{(a)}$, $u_{(\mathsf{b})}^{3}$, $u_{(\mathsf{e})}$, and $u_{(h)}+u_{(\bar{h})}$ 
still remain as arbitrary functions of $\tau$.

We have obtained all the Poisson brackets between 
the constraint functions, as in Eqs. (\ref{3.5}) and (\ref{3.12}). 
However, it is difficult to classify the constraints in Eqs. (\ref{3.4}) and (\ref{3.11}) 
into first and second classes on the basis of Eqs. (\ref{3.5}) and (\ref{3.12}) 
together with the vanishing Poisson brackets between the secondary constraint functions. 
To find simpler forms of the relevant Poisson brackets, we first define  
\begin{subequations}
\label{3.19}
\begin{align}
\tilde{\phi}^{(\mathsf{e})}&:=\phi^{(\mathsf{e})}
+\frac{1}{\mathsf{e}} \mathsf{b}^{\hat{\imath}} \phi^{(\mathsf{b})}_{\hat{\imath}} , 
\label{3.19a}
\\
\tilde{\phi}{}^{(\varphi)}&:=\phi^{(\varphi)}-ih\phi^{(h)}+i\bar{h}\phi^{(\bar{h})} , 
\label{3.19b}
\\
\tilde{\chi}^{(a)}&:=\chi^{(a)}+i\bar{\mathsf{Z}}{}^{j}_{A} \bar{\phi}_{j}^{A}
-i\phi^{j}_{A} \mathsf{Z}_{j}^{A} ,
\label{3.19c}
\\
\tilde{\chi}^{(\mathsf{b})}_{\hat{\imath}}&:=\chi^{(\mathsf{b})}_{\hat{\imath}}
+i\bar{\mathsf{Z}}{}^{j}_{A} \sigma_{\hat{\imath} j}{}^{k} \bar{\phi}_{k}^{A}
-i\phi^{j}_{A} \sigma_{\hat{\imath} j}{}^{k} \mathsf{Z}_{k}^{A} 
\notag 
\\
& \quad \,\,
-2t \mathsf{e} \:\! \epsilon_{\hat{\imath} \hat{\jmath}} \phi^{(\mathsf{b})}_{\hat{\jmath}} , 
\label{3.19d}
\\
\tilde{\chi}^{(\mathsf{b})}_{3}&:=\chi^{(\mathsf{b})}_{3}
+i\bar{\mathsf{Z}}{}^{j}_{A} \sigma_{3j}{}^{k} \bar{\phi}_{k}^{A}
-i\phi^{j}_{A} \sigma_{3j}{}^{k} \mathsf{Z}_{k}^{A} 
\notag 
\\
& \quad \,\,
+2 \epsilon_{\hat{\imath} \hat{\jmath}} \mathsf{b}^{\hat{\imath}} \phi^{(\mathsf{b})}_{\hat{\jmath}} , 
\label{3.19e}
\\
\tilde{\chi}^{(\mathsf{e})}&:=\chi^{(\mathsf{e})} +\mathsf{e} \:\! \mathsf{b}^{\hat{\imath}} 
\Big( \tilde{\chi}^{(\mathsf{b})}_{\hat{\imath}}
-2t \mathsf{e} \:\! \epsilon_{\hat{\imath} \hat{\jmath}} \phi^{(\mathsf{b})}_{\hat{\jmath}} \Big) \,, 
\label{3.19f}
\\
\tilde{\chi}{}^{(h)}&:=\chi^{(h)}+2i\epsilon^{jk} \bar{\phi}_{j\dot{\alpha}} 
\varpi^{\dot{\alpha}}_{k} \,,
\label{3.19g}
\\
\tilde{\chi}{}^{(\bar{h})}&:=\chi^{(\bar{h})}-2i\epsilon_{jk} \phi^{j}_{\alpha} 
\bar{\varpi}^{k \alpha} ,
\label{3.19h}
\end{align}
\end{subequations}
where $\bar{\phi}_{i\dot{\alpha}}$ and $\phi^{i}_{\alpha}$ are spinor components 
of $\bar{\phi}_{i}^{A}$ and $\phi^{i}_{A}$, respectively. 
Furthermore, it is convenient to define 
\begin{subequations}
\label{3.20}
\begin{align}
\upsilon^{(\pm)}&:=\frac{1}{2\sqrt{2} \:\! m} \bigg(\tilde{\phi}{}^{(\varphi)} 
\pm \frac{1}{2} \tilde{\chi}^{(a)} \bigg) \,, 
\label{3.20a}
\\
\phi^{(+)}&:=\frac{1}{2} \Big( e^{-i\varphi} \phi^{(h)} +e^{i\varphi} \phi^{(\bar{h})} \Big) \,, 
\label{3.20b}
\\
\phi^{(-)}&:=\frac{1}{2i} \Big( e^{-i\varphi} \phi^{(h)} -e^{i\varphi} \phi^{(\bar{h})} \Big) \,, 
\label{3.20c}
\\
\tilde{\chi}{}^{(+)}&:=\frac{1}{2} \Big( e^{-i\varphi} \tilde{\chi}{}^{(h)} +e^{i\varphi} \tilde{\chi}{}^{(\bar{h})} \Big) \,, 
\label{3.20d}
\\
\tilde{\chi}{}^{(-)}&:=\frac{1}{2i} \Big( e^{-i\varphi} \tilde{\chi}{}^{(h)} -e^{i\varphi} \tilde{\chi}{}^{(\bar{h})} \Big) \,. 
\label{3.20e}
\end{align}
\end{subequations}
It can readily be seen that the set of all the constraints given in Eqs. (\ref{3.4}) and (\ref{3.11}), i.e., 
\begin{widetext}
\begin{align}
\Big( &\phi_{A}^{i}, \bar{\phi}^{A}_{i}, \phi^{(a)}, \phi^{(\mathsf{b})}_{r}, \phi^{(\mathsf{e})}, 
\phi^{(h)}, \phi^{(\bar{h})}, \phi^{(\varphi)},  
\chi^{(a)}, 
\chi{}^{(\mathsf{b})}_{\hat{\imath}}, \chi{}^{(\mathsf{b})}_{3}, \chi{}^{(\mathsf{e})}, 
\chi^{(h)}, \chi^{(\bar{h})},  \chi^{(\varphi)} \Big)  
\approx 0 \,, 
\label{3.21}
\end{align}
is equivalent to the new set of constraints 
\begin{align}
\Big( &\phi_{A}^{i}, \bar{\phi}^{A}_{i}, \phi^{(a)}, \phi^{(\mathsf{b})}_{r}, \tilde{\phi}{}^{(\mathsf{e})},
\phi^{(+)}, \phi^{(-)}, \upsilon^{(+)}, \upsilon^{(-)}, 
\tilde{\chi}{}^{(\mathsf{b})}_{\hat{\imath}}, \tilde{\chi}{}^{(\mathsf{b})}_{3}, 
\tilde{\chi}{}^{(\mathsf{e})}, \tilde{\chi}^{(+)}, \tilde{\chi}^{(-)},  \chi^{(\varphi)} \Big)  
\approx 0 \,.
\label{3.22}
\end{align}
\end{widetext}
We can show that except for 
\begin{alignat}{3}
\left\{\phi_{A}^{i}, \bar{\phi}^{B}_{j} \right\} &=-i \delta_{j}^{i} \delta_{A}^{B}  \,, 
&\quad \;\; 
\left\{ \phi^{(\mathsf{b})}_{\hat{\imath}}, \tilde{\chi}{}^{(\mathsf{b})}_{\hat{\jmath}} \right\} &
= \frac{1}{\mathsf{e}} \delta_{\hat{\imath} \hat{\jmath}}\,,
\notag
\\
\left\{ \upsilon^{(+)},  \tilde{\chi}^{(-)} \right\} &= 1 \,, 
&\quad \;\;
\left\{ \chi^{(\varphi)}, \phi^{(-)} \right\} &= 1 \,, 
\label{3.23}
\end{alignat}
all other Poisson brackets between the constraint functions in Eq. (\ref{3.22}) vanish. 
In this way, the relevant Poisson brackets are simplified with the aid of 
the new constraint functions. 
The Poisson brackets between the constraint functions are summarized in a matrix form as 
\begin{widetext}
\begin{align}
\! \bordermatrix{ 
& \;\: \phi_{B}^{j} & \bar{\phi}^{B}_{j} & \phi^{(a)} & \phi^{(\mathsf{b})}_{\hat{\jmath}} &\phi^{(\mathsf{b})}_{3} & 
\tilde{\phi}{}^{(\mathsf{e})} & \phi^{(+)} & \phi^{(-)} & \upsilon^{(+)} & \upsilon^{(-)} & \tilde{\chi}{}^{(\mathsf{b})}_{\hat{\jmath}} & 
\tilde{\chi}{}^{(\mathsf{b})}_{3} & \tilde{\chi}{}^{(\mathsf{e})} & \tilde{\chi}^{(+)} & \tilde{\chi}^{(-)} &  \chi^{(\varphi)} \cr
\phi_{A}^{i} & 0 & -i \delta_{j}^{i} \delta_{A}^{B}  & 0 & 0 & 0 & 0 & 0 & 0 & 0 & 0 & 0 & 0 & 0 & 0 & 0 & 0 \cr 
\bar{\phi}^{A}_{i} & i \delta_{i}^{j} \delta_{B}^{A} & 0 & 0 & 0 & 0 & 0 & 0 & 0 & 0 & 0 & 0 & 0 & 0 & 0 & 0 & 0 \cr 
\phi^{(a)} & 0 & 0 & 0 & 0 & 0 & 0 & 0 & 0 & 0 & 0 & 0 & 0 & 0 & 0 & 0 & 0 \cr 
\phi^{(\mathsf{b})}_{\hat{\imath}} & 0 & 0 & 0 & 0 & 0 & 0 & 0 & 0 & 0 & 0& \dfrac{1}{\mathsf{e}} \delta_{\hat{\imath} \hat{\jmath}} & 0 & 0 & 0 & 0 & 0 \cr 
\phi^{(\mathsf{b})}_{3} & 0 & 0 & 0 & 0 & 0 & 0 & 0 & 0 & 0 & 0 & 0 & 0 & 0 & 0 & 0 & 0 \cr 
\tilde{\phi}^{(\mathsf{e})} & 0 & 0 & 0 & 0 & 0 & 0 & 0 & 0 & 0 & 0 & 0 & 0 & 0 & 0 & 0 & 0 \cr 
\phi^{(+)} & 0 & 0 & 0 & 0 & 0 & 0 & 0 & 0 & 0 & 0 & 0 & 0 & 0 & 0 & 0 & 0 \cr 
\phi^{(-)} & 0 & 0 & 0 & 0 & 0 & 0 & 0 & 0 & 0 & 0 & 0 & 0 & 0 & 0 & 0 & -1 \cr 
\upsilon^{(+)} & 0 & 0 & 0 & 0 & 0 & 0 & 0 & 0 & 0 & 0 & 0 & 0 & 0 & 0 & 1 & 0 \cr 
\vspace{1mm}
\upsilon^{(-)} & 0 & 0 & 0 & 0 & 0 & 0 & 0 & 0 & 0 & 0 & 0 & 0 & 0 & 0 & 0 & 0 \cr 
\tilde{\chi}{}^{(\mathsf{b})}_{\hat{\imath}} & 0 & 0 & 0 & -\dfrac{1}{\mathsf{e}} \delta_{\hat{\imath} \hat{\jmath}} & 0 & 0 & 0 & 0 & 0 & 0 & 0 & 0 & 0 & 0 & 0 & 0 \cr 
\tilde{\chi}{}^{(\mathsf{b})}_{3} & 0 & 0 & 0 & 0 & 0 & 0 & 0 & 0 & 0 & 0 & 0 & 0 & 0 & 0 & 0 & 0 \cr 
\tilde{\chi}^{(\mathsf{e})} & 0 & 0 & 0 & 0 & 0 & 0 & 0 & 0 & 0 & 0 & 0 & 0 & 0 & 0 & 0 & 0 \cr 
\tilde{\chi}^{(+)} & 0 & 0 & 0 & 0 & 0 & 0 & 0 & 0 & 0 & 0 & 0 & 0 & 0 & 0 & 0 & 0 \cr 
\tilde{\chi}^{(-)} &  0 & 0 & 0 & 0 & 0 & 0 & 0 & 0 & -1 & 0 & 0 & 0 & 0 & 0 & 0 & 0 \cr 
\chi^{(\varphi)} & 0 & 0 & 0 & 0 & 0 & 0 & 0 & 1 & 0 & 0 & 0 & 0 & 0 & 0 & 0 & 0 
 }  . 
\label{3.24}
\end{align}
\end{widetext}

We can immediately see from this matrix that 
$\phi^{(a)} \approx 0$, 
$\phi^{(\mathsf{b})}_{3} \approx 0$, 
$\tilde{\phi}^{(\mathsf{e})} \approx 0$, 
$\phi^{(+)} \approx 0$, 
$\upsilon^{(-)} \approx 0$, 
$\tilde{\chi}{}^{(\mathsf{b})}_{3} \approx 0$, 
$\tilde{\chi}^{(\mathsf{e})} \approx 0$, and 
$\tilde{\chi}^{(+)} \approx 0$
are first-class constraints, while 
$\phi_{A}^{i} \approx 0$, 
$\bar{\phi}^{A}_{i} \approx 0$, 
$\phi^{(\mathsf{b})}_{\hat{\imath}} \approx 0$, 
$\phi^{(-)} \approx 0$, 
$\upsilon^{(+)} \approx 0$, 
$\tilde{\chi}{}^{(\mathsf{b})}_{\hat{\imath}} \approx 0$,  
$\tilde{\chi}^{(-)} \approx 0$, and 
$\chi^{(\varphi)} \approx 0$ are second-class constraints. 
Following Dirac's approach to second-class constraints, 
we define the Dirac bracket by using the largest invertible submatrix of the matrix (\ref{3.24}). 
For arbitrary smooth functions $f$ and $g$ of the canonical variables, the Dirac bracket is defined by 
\begin{align}
&\left\{ f , g \right\}_{\rm D} 
\notag
\\ 
&:=\left\{ f , g \right\} 
+i \left\{ f , \phi_{A}^{i} \right\} \! \left\{ \bar{\phi}^{A}_{i} , g \right\} 
-i \left\{ f , \bar{\phi}^{A}_{i} \right\} \! \left\{ \phi_{A}^{i} , g \right\} 
\notag
\\
&\;\,\quad -\mathsf{e} \left\{ f , \tilde{\chi}{}^{(\mathsf{b})}_{\hat{\imath}} \right\} 
\! \left\{\phi^{(\mathsf{b})}_{\hat{\imath}}, g \right\} 
+\mathsf{e} \left\{ f , \phi^{(\mathsf{b})}_{\hat{\imath}} \right\}  
\! \left\{\tilde{\chi}{}^{(\mathsf{b})}_{\hat{\imath}} , g \right\} 
\notag
\\
&\;\,\quad +\left\{ f , \chi^{(\varphi)} \right\} \!\left\{ \phi^{(-)} , g \right\}
-\left\{ f , \phi^{(-)} \right\} \!\left\{ \chi^{(\varphi)} , g \right\}
\notag
\\
&\;\,\quad +\left\{ f , \upsilon^{(+)} \right\} \!\left\{ \tilde{\chi}^{(-)} , g \right\}
-\left\{ f ,\tilde{\chi}^{(-)} \right\} \!\left\{ \upsilon^{(+)} , g \right\} . 
\label{3.25} 
\end{align}
The Dirac bracket between $f$ and each of the constraint functions  
$\phi_{A}^{i}$, $\bar{\phi}^{A}_{i}$, $\phi^{(\mathsf{b})}_{\hat{\imath}}$, 
$\phi^{(-)}$, $\upsilon^{(+)}$, $\tilde{\chi}{}^{(\mathsf{b})}_{\hat{\imath}}$,  
$\tilde{\chi}^{(-)}$, and $\chi^{(\varphi)}$ vanishes identically.  
For this reason, the second-class constraints 
can be set strongly equal to zero and may be expressed as  
$\phi_{A}^{i}=0$, $\bar{\phi}^{A}_{i}=0$, $\phi^{(\mathsf{b})}_{\hat{\imath}}=0$, 
$\phi^{(-)}=0$, $\upsilon^{(+)}=0$, $\tilde{\chi}{}^{(\mathsf{b})}_{\hat{\imath}}=0$,  
$\tilde{\chi}^{(-)}=0$, and $\chi^{(\varphi)}=0$,   
as long as the Dirac bracket $\left\{ f , g \right\}_{\rm D}$ is adopted. 
We see that the second-class constraints lead to 
\begin{subequations}
\label{3.26}
\begin{alignat}{3}
P_{A}^{i} &=\frac{i}{2} \bar{\mathsf{Z}}_{A}^{i} \,, 
&\quad 
\bar{P}^{A}_{i} &=-\frac{i}{2} \mathsf{Z}^{A}_{i} \,, 
\label{3.26a}
\\
\mathsf{b}_{\hat{\imath}} &=\mathsf{e}\:\! \bar{\mathsf{Z}}_{A}^{j} \sigma_{\hat{\imath} j}{}^{k} \mathsf{Z}^{A}_{k} \,,  
&\quad 
P^{(\mathsf{b})}_{\hat{\imath}} &=0 \, 
\label{3.26b}
\\
h &=\mathsf{h} e^{-i\varphi} \,, 
&\quad 
P^{(h)} &=e^{i\varphi} P^{(\mathsf{h})} \,,
\label{3.26c}
\\
\bar{h} &=\mathsf{h} e^{i\varphi} \,, 
&\quad 
P^{(\bar{h})} &=e^{-i\varphi} P^{(\mathsf{h})} \,,
\label{3.26d}
\\
\varphi &=-\frac{i}{2} 
\ln \! \left( \frac{\epsilon^{ij} \varpi_{i \dot{\alpha}} \varpi_{j}^{\dot{\alpha}}}
{\epsilon_{ij} \bar{\varpi}^{i}_{\alpha} \bar{\varpi}^{j \alpha}} \right), 
&\quad 
P^{(\varphi)} &=-\frac{1}{2} \chi^{(a)} \,, 
\label{3.26e}
\end{alignat}
\end{subequations}
where $\mathsf{h}=\mathsf{h}(\tau)$ is a real scalar-density field of weight 1 on $\mathcal{T}$,  
and $P^{(\mathsf{h})}$ its associated momentum variable. 
At this stage, $P_{A}^{i}$, $\bar{P}^{A}_{i}$,  $\mathsf{b}^{\hat{\imath}}$, 
$P^{(\mathsf{b})}_{\hat{\imath}}$, $h$, $P^{(h)}$, $\bar{h}$, $P^{(\bar{h})}$, 
$\varphi$, and $P^{(\varphi)}$  
are treated as dependent variables specified by Eq. (\ref{3.26}), 
while the other canonical variables 
$\mathsf{Z}^{A}_{i}$, $\bar{\mathsf{Z}}_{A}^{i}$, $a$, $P^{(a)}$, $\mathsf{b}^{3}$, 
$P^{(\mathsf{b})}_{3}$, $\mathsf{e}$, $P^{(\mathsf{e})}$, $\mathsf{h}$, 
and $P^{(\mathsf{h})}$ are treated as independent variables. 
By virtue of the strong equalities of the second-class constraints, 
the set of all the first-class constraints, i.e, 
\begin{align}
\left( \phi^{(a)}, \phi^{(\mathsf{b})}_{3}, \tilde{\phi}{}^{(\mathsf{e})},
\phi^{(+)}, \upsilon^{(-)},  \tilde{\chi}{}^{(\mathsf{b})}_{3}, 
\tilde{\chi}{}^{(\mathsf{e})}, \tilde{\chi}^{(+)} \right)  
\approx 0 \,, 
\label{3.27}
\end{align}
turns out to be equivalent to the set consisting of 
\begin{subequations}
\label{3.28}
\begin{align}
\phi^{(a)}	& \approx 0 \,, 
\label{3.28a}
\\
\phi^{(\mathsf{b})}_{3} & \approx 0 \,, 
\label{3.28b}
\\
\phi^{(\mathsf{e})} & \approx 0 \,,
\label{3.28c}
\\
\phi^{(\mathsf{h})} &:=P^{(\mathsf{h})} \approx 0 \,, 
\label{3.28d}
\\
\chi^{(a)} & \approx 0 \,, 
\label{3.28e}
\\
\chi^{(\mathsf{b})}_{3} & \approx 0 \,, 
\label{3.28f}
\\
\chi^{(\mathsf{e})} & \approx 0 \,, 
\label{3.28g}
\\
\chi^{(h)} & \approx 0 \,,
\label{3.28h}
\\
\chi^{(\bar{h})} & \approx 0 \,. 
\label{3.28i}
\end{align}
\end{subequations}
Here we have taken into account both Eqs. (\ref{3.28h}) and (\ref{3.28i}) for later convenience, 
although it is sufficient to consider one of them in actuality.

The Dirac brackets between the spinor components of $\mathsf{Z}^{A}_{i}$ and $\bar{\mathsf{Z}}_{A}^{i}$ 
are found from Eq. (\ref{3.25}) to be 
\begin{align}
\left\{ \varrho_{i}^{\alpha}, \varrho_{j}^{\beta} \right\}_{\rm D}
&=\frac{i}{4\sqrt{2} \:\! m} e^{i\varphi} \Big( \varrho_{i}^{\alpha} \epsilon_{jk} \bar{\varpi}^{k\beta} 
-\varrho_{j}^{\beta} \epsilon_{ik} \bar{\varpi}^{k\alpha} \Big) \,, 
\notag 
\\
\left\{ \varrho_{i}^{\alpha}, \varpi_{j \dot{\beta}} \right\}_{\rm D}
&=-\frac{i}{4\sqrt{2} \:\! m} e^{i\varphi} \epsilon_{ik} \bar{\varpi}^{k\alpha} \varpi_{j \dot{\beta}} \,, 
\notag 
\\
\left\{ \varpi_{i \dot{\alpha}}, \varpi_{j \dot{\beta}} \right\}_{\rm D}
&=0 \,, 
\notag 
\\
\left\{ \varrho_{i}^{\alpha}, \bar{\varrho}{}^{\:\! j \dot{\beta}} \right\}_{\rm D}
&=\frac{i}{4\sqrt{2} \:\! m} \Big( e^{i\varphi} \epsilon_{ik} \bar{\varpi}^{k\alpha} \bar{\varrho}{}^{\:\! j \dot{\beta}} 
+ e^{-i\varphi} \varrho_{i}^{\alpha} \epsilon^{jk} \varpi{}_{k}^{\dot{\beta}} \Big) \,, 
\notag 
\\
\left\{ \varrho_{i}^{\alpha}, \bar{\varpi}^{j}_{\beta} \right\}_{\rm D}
&=-i \delta_{i}^{j} \delta^{\alpha}_{\beta}  
+\frac{i}{4\sqrt{2} \:\! m} e^{i\varphi} \epsilon_{ik} \bar{\varpi}^{k\alpha} \bar{\varpi}^{j}_{\beta} \,, 
\notag 
\\
\left\{ \varpi_{i \dot{\alpha}}, \bar{\varrho}{}^{\:\! j \dot{\beta}} \right\}_{\rm D}
&=-i \delta_{i}^{j} \delta_{\dot{\alpha}}^{\dot{\beta}}   
+\frac{i}{4\sqrt{2} \:\! m} e^{-i\varphi} \varpi_{i \dot{\alpha}} \epsilon^{jk} \varpi_{k}^{\dot{\beta}}  \,, 
\notag 
\\
\left\{ \varpi_{i \dot{\alpha}}, \bar{\varpi}^{j}_{\beta} \right\}_{\rm D}
&=0 \,, 
\notag 
\\
\left\{ \bar{\varrho}^{i\dot{\alpha}}, \bar{\varrho}{}^{\:\! j \dot{\beta}} \right\}_{\rm D}
&=-\frac{i}{4\sqrt{2} \:\! m} e^{-i\varphi} \Big( \bar{\varrho}^{i\dot{\alpha}} \epsilon^{jk} \varpi_{k}^{\dot{\beta}} 
-\bar{\varrho}{}^{\:\! j \dot{\beta}} \epsilon^{ik} \varpi_{k}^{\dot{\alpha}} \Big) \,, 
\notag 
\\
\left\{ \bar{\varrho}^{i\dot{\alpha}}, \bar{\varpi}^{j}_{\beta} \right\}_{\rm D}
&=\frac{i}{4\sqrt{2} \:\! m} e^{-i\varphi} \epsilon^{ik} \varpi_{k}^{\dot{\alpha}} \bar{\varpi}^{j}_{\beta} \,, 
\notag 
\\
\left\{ \bar{\varpi}^{i}_{\alpha}, \bar{\varpi}^{j}_{\beta} \right\}_{\rm D}
&=0 \,.
\label{3.29}
\end{align}
Using Eq. (\ref{3.29}) and taking into account Eqs. (\ref{3.28h}) and (\ref{3.28i}), we can show that
\begin{alignat}{3}
\left\{ \chi^{(a)},  \varrho_{i}^{\alpha} \right\}_{\rm D}
&=\frac{i}{2} \varrho_{i}^{\alpha} \,,
&\quad  
\left\{ \chi^{(a)},  \varpi_{i \dot{\alpha}} \right\}_{\rm D}
&=\frac{i}{2}  \varpi_{i \dot{\alpha}} \,,
\notag 
\\
\left\{ \chi^{(a)},  \bar{\varpi}^{i}_{\alpha} \right\}_{\rm D}
&=-\frac{i}{2}  \bar{\varpi}^{i}_{\alpha} \,,
&\quad  
\left\{\chi^{(a)},  \bar{\varrho}^{i \dot{\alpha}} \right\}_{\rm D}
&=-\frac{i}{2}  \bar{\varrho}^{i \dot{\alpha}} \,.
\label{3.30}
\end{alignat}
Many of the Dirac brackets in Eq. (\ref{3.29}) are rather complicated. 
Fortunately, however, Eq. (\ref{3.29}) can be expressed in the form of simple canonical brackets as  
\begin{align}
\left\{ \rho_{i}^{\alpha}, \bar{\varpi}^{j}_{\beta} \right\}_{\rm D}
&=-i \delta_{i}^{j} \delta^{\alpha}_{\beta}  \,,
\quad 
\left\{ \varpi_{i \dot{\alpha}}, \bar{\rho}{}^{\:\! j \dot{\beta}} \right\}_{\rm D}
=-i \delta_{i}^{j} \delta_{\dot{\alpha}}^{\dot{\beta}} \,, 
\notag
\\
\mbox{all others} &=0 \,, 
\label{3.31}
\end{align}
in terms of $\varpi_{i \dot{\alpha}}$, $\bar{\varpi}^{i}_{\alpha}$, and 
\begin{subequations}
\label{3.32}
\begin{align}
\rho_{i}^{\alpha}&:=\varrho_{i}^{\alpha}
+\frac{1}{2\sqrt{2} \:\! m} e^{i\varphi} \epsilon_{ij} \bar{\varpi}^{j\alpha} \chi^{(a)} ,
\label{3.32a}
\\
\bar{\rho}^{i\dot{\alpha}}&:=\bar{\varrho}^{i\dot{\alpha}}
+\frac{1}{2\sqrt{2} \:\! m} e^{-i\varphi} \epsilon^{ij} \varpi_{j}^{\dot{\alpha}}\chi^{(a)} .
\label{3.32b}
\end{align}
\end{subequations}
In showing this fact, it is convenient to use Eqs. (\ref{3.28e}) and (\ref{3.30}). 
Note here that the weak equalities $\rho_{i}^{\alpha}\approx \varrho_{i}^{\alpha}$,   
$\bar{\rho}^{i\dot{\alpha}}\approx \bar{\varrho}^{i\dot{\alpha}}$ hold owing to Eq. (\ref{3.28e}). 
Now we define the new twistors 
$\mathsf{W}_{i}^{A} :=(\rho_{i}^{\alpha}, \varpi_{i \dot{\alpha}})$ and 
$\bar{\mathsf{W}}^{i}_{A} :=(\bar{\varpi}^{i}_{\alpha}, \bar{\rho}{}^{i\dot{\alpha}})$, 
with which Eq. (\ref{3.31}) can concisely be written as 
\begin{align}
\left\{ \mathsf{W}_{i}^{A},  \bar{\mathsf{W}}^{j}_{B} \right\}_{\rm D}
&=-i \delta_{i}^{j} \delta^{A}_{B}  \,, 
\notag
\\ 
\Big\{ \mathsf{W}_{i}^{A},  \mathsf{W}_{j}^{B} \Big\}_{\rm D}
&=0 \,, 
\quad 
\left\{ \bar{\mathsf{W}}^{i}_{A},  \bar{\mathsf{W}}^{j}_{B} \right\}_{\rm D}
=0 \,.
\label{3.33}
\end{align}

Using Eqs. (\ref{3.28h}), (\ref{3.28i}), and the formulas given under Eq. (\ref{3.17}), 
we can show for 
\begin{subequations}
\label{3.34}
\begin{align}
\breve{\chi}^{(a)} &:= \bar{\mathsf{W}}_{A}^{i} \mathsf{W}^{A}_{i} -2s \,,
\label{3.34a}
\\
\breve{\chi}^{(\mathsf{b})}_{3} &:= \bar{\mathsf{W}}_{A}^{j} \sigma_{3j}{}^{k} \mathsf{W}^{A}_{k} -2t 
\label{3.34b}
\end{align}
\end{subequations}
that 
\begin{subequations}
\label{3.35}
\begin{align}
\breve{\chi}^{(a)} &=2\chi^{(a)} , 
\label{3.35a}
\\
\breve{\chi}_{3}^{(\mathsf{b})} &=\chi_{3}^{(\mathsf{b})} .
\label{3.35b}
\end{align}
\end{subequations}
Accordingly, the first-class constraints (\ref{3.28e}) and (\ref{3.28f}) read 
\begin{subequations}
\label{3.36}
\begin{align}
\breve{\chi}^{(a)} & \approx 0 \,, 
\label{3.36a}
\\
\breve{\chi}^{(\mathsf{b})}_{3} & \approx 0 \,. 
\label{3.36b}
\end{align}
\end{subequations}
With Eq. (\ref{3.35a}), Eqs. (\ref{3.32a}) and (\ref{3.32b}) can be solved inversely as 
\begin{subequations}
\label{3.37}
\begin{align}
\varrho_{i}^{\alpha}&=\rho_{i}^{\alpha}
-\frac{1}{4\sqrt{2} \:\! m} e^{i\varphi} \epsilon_{ij} \bar{\varpi}^{j\alpha} \breve{\chi}^{(a)} ,
\label{3.37a}
\\
\bar{\varrho}^{i\dot{\alpha}}&=\bar{\rho}^{i\dot{\alpha}}
-\frac{1}{4\sqrt{2} \:\! m} e^{-i\varphi} \epsilon^{ij} \varpi_{j}^{\dot{\alpha}} \breve{\chi}^{(a)} .  
\label{3.37b}
\end{align}
\end{subequations}
Hence it follows that there is a one-to-one correspondence between $(\mathsf{Z}_{i}^{A}, \bar{\mathsf{Z}}^{i}_{A})$  
and $(\mathsf{W}_{i}^{A}, \bar{\mathsf{W}}^{i}_{A})$. 
Taking into account this fact, 
we hereafter adopt $\mathsf{W}_{i}^{A}$ and $\bar{\mathsf{W}}^{i}_{A}$ as canonical variables 
instead of $\mathsf{Z}_{i}^{A}$ and $\bar{\mathsf{Z}}^{i}_{A}$. 
The first equation in Eq. (\ref{3.26b}) can be written as 
$\mathsf{b}_{\hat{\imath}}=\mathsf{e}_{\:\!}
\bar{\mathsf{W}}_{A}^{j} \sigma_{\hat{\imath} j}{}^{k} \mathsf{W}^{A}_{k}$.  
Substituting this into Eq. (\ref{3.11d}), we see that the first-class constraint $\chi^{(\mathsf{e})} \approx 0$ 
can be expressed as   
\begin{align}
\breve{\chi}^{(\mathsf{e})} := \mathsf{T}_{\hat{\imath}} \mathsf{T}_{\hat{\imath}} -\frac{1}{4} k^2 \approx 0 \,, 
\label{3.38}
\end{align}
where $\mathsf{T}_{\hat{\imath}}$ $(\hat{\imath}=1,2)$ are defined in 
\begin{align}
\mathsf{T}_{0}:=\frac{1}{2} \bar{\mathsf{W}}_{A}^{i} \mathsf{W}^{A}_{i} \,, \quad 
\mathsf{T}_{r}:=\frac{1}{2} \bar{\mathsf{W}}_{A}^{j} \sigma_{rj}{}^{k} \mathsf{W}^{A}_{k} \,.
\label{3.39}
\end{align}
Using Eq. (\ref{3.33}), we can readily verify that $\mathsf{T}_{0}$ and $\mathsf{T}_{r}$ constitute 
a bases of the $U(1)_{a} \times SU(2)$ Lie algebra 
in the following sense: 
\begin{align}
\left\{ \mathsf{T}_{0}, \mathsf{T}_{r} \right\}_{\rm D}=0\,, 
\quad 
\left\{ \mathsf{T}_{r}, \mathsf{T}_{s} \right\}_{\rm D}= \epsilon_{rst} \mathsf{T}_{t} \,. 
\label{3.40}
\end{align}

The canonical variables that we need to consider at the present stage are 
$\mathsf{W}^{A}_{i}$, $\bar{\mathsf{W}}_{A}^{i}$, $a$, $P^{(a)}$, $\mathsf{b}^{3}$, 
$P^{(\mathsf{b})}_{3}$, $\mathsf{e}$, $P^{(\mathsf{e})}$, $\mathsf{h}$, and $P^{(\mathsf{h})}$.  
All the Dirac brackets between these variables are given in Eq. (\ref{3.33}) and  
\begin{alignat}{3}
\left\{ a,  P^{(a)} \right\}_{\rm D} &=1  \,, 
&
\quad \left\{ \mathsf{b}^{3},  P^{(\mathsf{b})}_{3} \right\}_{\rm D} &=1  \,, 
\notag
\\
\left\{ \mathsf{e},  P^{(\mathsf{e})} \right\}_{\rm D} &=1  \,, 
& 
\left\{ \mathsf{h},  P^{(\mathsf{h})} \right\}_{\rm D} &=\frac{1}{2}  \,, 
\notag
\\
\mbox{all others} &=0 \,. 
\label{3.41}
\end{alignat}
We also need to consider the first class constraints (\ref{3.28a})--(\ref{3.28d}),  
(\ref{3.36a}), (\ref{3.36b}), (\ref{3.38}), (\ref{3.28h}), and (\ref{3.28i}).

\section{Canonical quantization}

In this section, we perform the canonical quantization of the Hamiltonian system studied in Sec. III. 
To this end, in accordance with Dirac's method of quantization, we introduce the operators 
$\hat{f}$ and $\hat{g}$ corresponding to the functions $f$ and $g$, respectively, 
and set the commutation relation 
\begin{align}
\left[\:\! \hat{f}, \hat{g} \:\! \right] \!=i \;\! \widehat {\left\{ f , g \right\} }_{\rm D} 
\label{4.1}
\end{align}
in units such that $\hbar=1$.  
Here, $\widehat {\left\{ f , g \right\} }_{\rm D}$ denotes the operator corresponding to 
the Dirac bracket $\left\{ f , g \right\}_{\rm D}$. 
From Eqs. (\ref{3.33}), (\ref{3.41}), and (\ref{4.1}), we have the canonical commutation relations 
\begin{subequations}
\label{4.2}
\begin{alignat}{3}
\left[ \;\! \hat{\mathsf{W}}_{i}^{A},  \Hat{\Bar{\mathsf{W}}}{}^{j}_{B} \right] &=\delta_{i}^{j} \delta^{A}_{B}  \,, 
&
\quad &~
\label{4.2a}
\\
\left[ \;\! \hat{\mathsf{W}}_{i}^{A},  \hat{\mathsf{W}}_{j}^{B} \right] &=0 \,, 
&
\left[ \;\! \Hat{\Bar{\mathsf{W}}}{}^{i}_{A},  \Hat{\Bar{\mathsf{W}}}^{j}_{B} \right] &= 0 \,, 
\label{4.2b}
\\
\left[ \;\! \hat{a},  \hat{P}^{(a)} \right] &=i  \,,  
&
\left[ \;\! \hat{\mathsf{b}}{}^{3},  \hat{P}{}^{(\mathsf{b})}_{3} \right] &=i  \,, 
\label{4.2c}
\\
\left[ \;\! \hat{\mathsf{e}},  \hat{P}{}^{(\mathsf{e})} \right] &=i  \,,   
&
\left[ \;\! \hat{\mathsf{h}},  \hat{P}{}^{(\mathsf{h})} \right] &=\frac{i}{2}  \,, 
\label{4.2d}
\\
\mbox{all others} &=0 \,. 
\label{4.2e}
\end{alignat}
\end{subequations}
The commutation relations (\ref{4.2a}) and (\ref{4.2b}) govern together so-called twistor quantization 
\cite{PenMac, PenRin}.

In the procedure of canonical quantization, 
the first-class constraints are treated as conditions imposed on the physical states,  
after the replacement of the first-class constraint functions by the corresponding operators. 
In the present model, the physical state conditions are found from 
Eqs. (\ref{3.28a})--(\ref{3.28d}), (\ref{3.36a}), (\ref{3.36b}), (\ref{3.38}), (\ref{3.28h}), and (\ref{3.28i}) to be  
\begin{subequations}
\label{4.3}
\begin{align}
\hat{\phi}{}^{(a)} |F \:\! \rangle & =\hat{P}{}^{(a)} |F \:\! \rangle =0 \,, 
\label{4.3a}
\\
\hat{\phi}{}^{(\mathsf{b})}_{3} |F \:\! \rangle & =\hat{P}{}^{(\mathsf{b})}_{3} |F \:\! \rangle =0 \,, 
\label{4.3b}
\\
\hat{\phi}{}^{(\mathsf{e})} |F \:\! \rangle & =\hat{P}{}^{(\mathsf{e})} |F \:\! \rangle =0 \,,
\label{4.3c}
\\
\hat{\phi}{}^{(\mathsf{h})} |F \:\! \rangle &=\hat{P}{}^{(\mathsf{h})} |F \:\! \rangle =0 \,, 
\label{4.3d}
\\
\Hat{\Breve{\chi}}{}^{(a)} |F \:\! \rangle 
& =\bigg[ \;\! \frac{1}{2}\Big( \Hat{\Bar{\mathsf{W}}}_{A}^{i} \Hat{\mathsf{W}}{}^{A}_{i}
+\Hat{\mathsf{W}}{}^{A}_{i} \Hat{\Bar{\mathsf{W}}}_{A}^{i} \Big) -2s \bigg]  |F \:\! \rangle 
\notag
\\
& = 2 \Big( \hat{\mathsf{T}}_{0} -s-2 \Big) |F \:\! \rangle=0 \,, 
\label{4.3e}
\\
\Hat{\Breve{\chi}}{}^{(\mathsf{b})}_{3} |F \:\! \rangle 
& =\bigg[ \;\! \frac{1}{2}\Big( \Hat{\Bar{\mathsf{W}}}_{A}^{j} \sigma_{3j}{}^{k} \Hat{\mathsf{W}}{}^{A}_{k} 
+\Hat{\mathsf{W}}{}^{A}_{k} \sigma_{3j}{}^{k} \Hat{\Bar{\mathsf{W}}}_{A}^{j} \Big) -2t \bigg]  |F \:\! \rangle 
\notag
\\
& =2 \Big( \hat{\mathsf{T}}_{3} -t \Big) |F \:\! \rangle=0 \,, 
\label{4.3f}
\\
\Hat{\Breve{\chi}}{}^{(\mathsf{e})} |F \:\! \rangle & 
=\bigg( \hat{\mathsf{T}}_{\hat{\imath}} \hat{\mathsf{T}}_{\hat{\imath}} -\frac{1}{4} k^2 \bigg) |F \:\! \rangle=0 \,, 
\label{4.3g}
\\
\Hat{\chi}{}^{(h)} |F \:\! \rangle & 
=\Big( \epsilon^{ij} \hat{\varpi}_{i \dot{\alpha}} \hat{\varpi}_{j}^{\dot{\alpha}} 
-\sqrt{2}\:\! m e^{i\hat{\varphi}} \Big) |F \:\! \rangle=0 \,, 
\label{4.3h}
\\
\Hat{\chi}{}^{(\bar{h})} |F \:\! \rangle &
=\Big( \epsilon_{ij} \Hat{\Bar{\varpi}}{}^{i}_{\alpha} \Hat{\Bar{\varpi}}^{j \alpha} 
-\sqrt{2}\:\! m e^{-i\hat{\varphi}} \Big) |F \:\! \rangle=0 \,. 
\label{4.3i}
\end{align}
\end{subequations}
Here, $|F^{\:\!} \rangle$ denotes a physical state, 
$\hat{\mathsf{T}}_{0}$ and $\hat{\mathsf{T}}_{r}$ $(r={\hat{\imath}}, 3)$ are defined by 
\begin{align}
\hat{\mathsf{T}}_{0}:=\frac{1}{2} \Hat{\mathsf{W}}{}^{A}_{i} \Hat{\Bar{\mathsf{W}}}{}_{A}^{i}\,, \quad 
\hat{\mathsf{T}}_{r}:=\frac{1}{2} \sigma_{rj}{}^{k} \Hat{\mathsf{W}}{}^{A}_{k} \Hat{\Bar{\mathsf{W}}}_{A}^{j}\,, 
\label{4.4}
\end{align}
and $\hat{\varphi}$ is defined according to the first equation in Eq. (\ref{3.26e}) as 
\begin{align}
\hat{\varphi} :=-\frac{i}{2} \!\left[ \;\!
\ln \! \left( \epsilon^{ij} \hat{\varpi}_{i \dot{\alpha}} \hat{\varpi}_{j}^{\dot{\alpha}} \right) \!
-\ln \! \left( {\epsilon_{ij} \Hat{\Bar{\varpi}}{}^{i}_{\alpha} \Hat{\Bar{\varpi}}{}^{j \alpha}} \right) \right] . 
\label{4.5}
\end{align}
In defining the operators $\Hat{\Breve{\chi}}{}^{(a)}$, $\Hat{\Breve{\chi}}{}^{(\mathsf{b})}_{3}$, and 
$\Hat{\Breve{\chi}}{}^{(\mathsf{e})}$, we have obeyed the Weyl ordering rule and have used the 
commutation relation (\ref{4.2a}) to simplify the Weyl ordered operators. 
Using Eqs. (\ref{4.2a}) and (\ref{4.2b}), we can easily show that 
\begin{align}
\left[ \;\! \hat{\mathsf{T}}_{0}, \hat{\mathsf{T}}_{r} \right]=0\,, 
\quad 
\left[ \;\! \hat{\mathsf{T}}_{r}, \hat{\mathsf{T}}_{s} \right]= i\epsilon_{rst} \hat{\mathsf{T}}_{t} \,,
\label{4.6}
\end{align}
which is precisely the quantum mechanical counterpart of Eq. (\ref{3.40}). 
It is evident that $\hat{\mathsf{T}}_{0}$ is the generator of $U(1)_{a}$ 
and $\hat{\mathsf{T}}_{r}$ $(r=1,2,3)$ are the generators of $SU(2)$.  
In particular, $\hat{\mathsf{T}}_{3}$ is the generator of $U(1)_{\mathsf{b}}$.

Now we introduce the bra-vector 
\begin{align}
\langle \;\! \mathsf{W}, a, \mathsf{b}^{3}, \mathsf{e}, \mathsf{h} \;\!|
& :=\langle 0 | \exp \!\Big(\!  -\mathsf{W}^{A}_{i} \Hat{\Bar{\mathsf{W}}}{}^{i}_{A} 
+ia \hat{P}^{(a)} 
+i\mathsf{b}^{3} \hat{P}{}^{(\mathsf{b})}_{3} 
\notag
\\
& \quad \;\:
+i\mathsf{e} \hat{P}{}^{(\mathsf{e})} +2i\mathsf{h} \hat{P}{}^{(\mathsf{h})} \Big)
\label{4.7}
\end{align}
with a reference bra-vector $\langle 0 |$ satisfying 
\begin{align}
\langle 0 |\:\! \hat{\mathsf{W}}_{i}^{A} =\langle 0 |\:\!\hat{a} 
=\langle 0 |\:\! \hat{\mathsf{b}}{}^{3} =\langle 0 |\:\! \hat{\mathsf{e}} 
=\langle 0 |\:\! \hat{\mathsf{h}} =0\,.
\label{4.8}
\end{align}
Using the commutation relations (\ref{4.2a})--(\ref{4.2e}), we can show that 
\begin{subequations}
\label{4.9}
\begin{align}
\langle \;\! \mathsf{W}, a, \mathsf{b}^{3}, \mathsf{e}, \mathsf{h} \;\!| \:\! \hat{\mathsf{W}}_{i}^{A}
&=\mathsf{W}^{A}_{i} \langle \;\! \mathsf{W}, a, \mathsf{b}^{3}, \mathsf{e}, \mathsf{h} \;\!| \,,
\label{4.9a}
\\
\langle \;\! \mathsf{W}, a, \mathsf{b}^{3}, \mathsf{e}, \mathsf{h} \;\!| \:\! \hat{a}
&=a \:\! \langle \;\! \mathsf{W}, a, \mathsf{b}^{3}, \mathsf{e}, \mathsf{h} \;\!| \,,
\label{4.9b}
\\
\langle \;\! \mathsf{W}, a, \mathsf{b}^{3}, \mathsf{e}, \mathsf{h} \;\!| \:\! \hat{\mathsf{b}}{}^{3}
&=\mathsf{b}^{3} \langle \;\! \mathsf{W}, a, \mathsf{b}^{3}, \mathsf{e}, \mathsf{h} \;\!| \,,
\label{4.9c}
\\
\langle \;\! \mathsf{W}, a, \mathsf{b}^{3}, \mathsf{e}, \mathsf{h} \;\!| \:\! \hat{\mathsf{e}} 
&=\mathsf{e} \:\! \langle \;\! \mathsf{W}, a, \mathsf{b}^{3}, \mathsf{e}, \mathsf{h} \;\!| \,,
\label{4.9d}
\\
\langle \;\! \mathsf{W}, a, \mathsf{b}^{3}, \mathsf{e}, \mathsf{h} \;\!| \:\! \hat{\mathsf{h}} 
&=\mathsf{h} \:\! \langle \;\! \mathsf{W}, a, \mathsf{b}^{3}, \mathsf{e}, \mathsf{h} \;\!| \,.
\label{4.9e}
\end{align}
\end{subequations}
Equation (\ref{4.9a}) can be decomposed into two parts, 
\begin{subequations}
\label{4.10}
\begin{align}
\langle \;\! \mathsf{W}, a, \mathsf{b}^{3}, \mathsf{e}, \mathsf{h} \;\!| \:\! \hat{\rho}{}_{i}^{\alpha}
&=\rho_{i}^{\alpha} \langle \;\! \mathsf{W}, a, \mathsf{b}^{3}, \mathsf{e}, \mathsf{h} \;\!| \,,
\label{4.10a}
\\
\langle \;\! \mathsf{W}, a, \mathsf{b}^{3}, \mathsf{e}, \mathsf{h} \;\!| \:\! \hat{\varpi}_{i \dot{\alpha}}
&=\varpi_{i \dot{\alpha}} \langle \;\! \mathsf{W}, a, \mathsf{b}^{3}, \mathsf{e}, \mathsf{h} \;\!| \,.
\label{4.10b}
\end{align}
\end{subequations}
Also, it is easy to see that 
\begin{subequations}
\label{4.11}
\begin{align}
\langle \;\! \mathsf{W}, a, \mathsf{b}^{3}, \mathsf{e}, \mathsf{h} \;\!| \:\! \Hat{\Bar{\mathsf{W}}}{}^{i}_{A}
&=-\frac{\partial}{\partial \mathsf{W}^{A}_{i}} 
\langle \;\! \mathsf{W}, a, \mathsf{b}^{3}, \mathsf{e}, \mathsf{h} \;\!| \,,
\label{4.11a}
\\
\langle \;\! \mathsf{W}, a, \mathsf{b}^{3}, \mathsf{e}, \mathsf{h} \;\!|  \hat{P}^{(a)} 
&=-i \frac{\partial}{\partial a}  
\langle \;\! \mathsf{W}, a, \mathsf{b}^{3}, \mathsf{e}, \mathsf{h} \;\!| \,,
\label{4.11b}
\\
\langle \;\! \mathsf{W}, a, \mathsf{b}^{3}, \mathsf{e}, \mathsf{h} \;\!|  \hat{P}{}^{(\mathsf{b})}_{3} 
&=-i \frac{\partial}{\partial \mathsf{b}^{3}} 
\langle \;\! \mathsf{W}, a, \mathsf{b}^{3}, \mathsf{e}, \mathsf{h} \;\!| \,,
\label{4.11c}
\\
\langle \;\! \mathsf{W}, a, \mathsf{b}^{3}, \mathsf{e}, \mathsf{h} \;\!|  \hat{P}{}^{(\mathsf{e})} 
&=-i \frac{\partial}{\partial \mathsf{e}} 
\langle \;\! \mathsf{W}, a, \mathsf{b}^{3}, \mathsf{e}, \mathsf{h} \;\!| \,,
\label{4.11d}
\\
\langle \;\! \mathsf{W}, a, \mathsf{b}^{3}, \mathsf{e}, \mathsf{h} \;\!|  \hat{P}{}^{(\mathsf{h})}
&=-\frac{i}{2} \frac{\partial}{\partial \mathsf{h}} 
\langle \;\! \mathsf{W}, a, \mathsf{b}^{3}, \mathsf{e}, \mathsf{h} \;\!| \,.
\label{4.11e}
\end{align}
\end{subequations}
Equation (\ref{4.11a}) can be decomposed into two parts, 
\begin{subequations}
\label{4.12}
\begin{align}
\langle \;\! \mathsf{W}, a, \mathsf{b}^{3}, \mathsf{e}, \mathsf{h} \;\!| \:\! 
\Hat{\Bar{\varpi}}{}^{i}_{\alpha}  
&=-\frac{\partial}{\partial \rho_{i}^{\alpha}} 
\langle \;\! \mathsf{W}, a, \mathsf{b}^{3}, \mathsf{e}, \mathsf{h} \;\!| \,,
\label{4.12a}
\\
\langle \;\! \mathsf{W}, a, \mathsf{b}^{3}, \mathsf{e}, \mathsf{h} \;\!| \:\! 
\Hat{\Bar{\rho}}{}^{i\dot{\alpha}} 
&=-\frac{\partial}{\partial \varpi_{i \dot{\alpha}}} 
\langle \;\! \mathsf{W}, a, \mathsf{b}^{3}, \mathsf{e}, \mathsf{h} \;\!| \,.
\label{4.12b}
\end{align}
\end{subequations}
Multiplying each of Eqs. (\ref{4.3a})--(\ref{4.3i}) by 
$\langle^{\;\!} \mathsf{W}, a, \mathsf{b}^{3}, \mathsf{e}, \mathsf{h}^{\;\!}|$ on the left 
and using Eqs. (\ref{4.9})--(\ref{4.12}), we obtain 
a set of simultaneous differential equations for  
$F(\mathsf{W}, a, \mathsf{b}^{3}, \mathsf{e}, \mathsf{h})
:=\langle^{\;\!} \mathsf{W}, a, \mathsf{b}^{3}, \mathsf{e}, \mathsf{h}^{\;\!} |F^{\:\!} \rangle$ 
as follows: 
\begin{subequations}
\label{4.13}
\begin{align}
\frac{\partial}{\partial a} F &=0 \,, 
\label{4.13a}
\\
\frac{\partial}{\partial \mathsf{b}^{3}} F &=0 \,, 
\label{4.13b}
\\
\frac{\partial}{\partial \mathsf{e}} F &=0 \,, 
\label{4.13c}
\\
\frac{\partial}{\partial \mathsf{h}} F &=0 \,, 
\label{4.13d}
\\
\check{\mathsf{T}}_{0} F &=(s+2)F \,, 
\label{4.13e}
\\
\check{\mathsf{T}}_{3} F &= tF \,, 
\label{4.13f}
\\
\check{\mathsf{T}}_{\hat{\imath}} \check{\mathsf{T}}_{\hat{\imath}} F& =\frac{1}{4} k^2 F \,, 
\label{4.13g}
\\
\epsilon^{ij} \varpi_{i \dot{\alpha}} \varpi_{j}^{\dot{\alpha}} F 
&=\sqrt{2}\:\! m e^{i\check{\varphi}} F \,, 
\label{4.13h}
\\
\epsilon_{ij} \epsilon^{\alpha \beta}
\frac{\partial}{\partial \rho_{i}^{\alpha}} \frac{\partial}{\partial \rho_{j}^{\beta}} F
&=\sqrt{2}\:\! m e^{-i\check{\varphi}} F \,.  
\label{4.13i}
\end{align}
\end{subequations}
Here, $\check{\mathsf{T}}_{0}$ and $\check{\mathsf{T}}_{r}$ $(r={\hat{\imath}}, 3)$ are defined by
\begin{align}
\check{\mathsf{T}}_{0}:=-\frac{1}{2} \mathsf{W}^{A}_{i} \frac{\partial}{\partial \mathsf{W}^{A}_{i}} \,,
\quad 
\check{\mathsf{T}}_{r}:=-\frac{1}{2} \sigma_{rj}{}^{k} \mathsf{W}^{A}_{k} \frac{\partial}{\partial \mathsf{W}^{A}_{j}} \,,
\label{4.14}
\end{align}
and $\check{\varphi}$ is defined by 
\begin{align}
\check{\varphi} :=-\frac{i}{2} \!\left[ \;\!
\ln \! \left( \epsilon^{ij} \varpi_{i \dot{\alpha}} \varpi_{j}^{\dot{\alpha}} \right) \!
-\ln \! \left( \epsilon_{ij} \epsilon^{\alpha \beta}
\frac{\partial}{\partial \rho_{i}^{\alpha}} \frac{\partial}{\partial \rho_{j}^{\beta}} 
\right) \right] . 
\label{4.15}
\end{align}
Equations (\ref{4.13a})--(\ref{4.13d}) imply that $F$ is actually independent of  
$a$, $\mathsf{b}^{3}$, $\mathsf{e}$, and $\mathsf{h}$. 
Hence it follows that $F$ is a function of the twistors $\mathsf{W}^{A}_{i}$ only. 
The holomorphic functions of $\mathsf{W}^{A}_{i}$, such as $F$,  
are often referred to as the twistor functions.   
As can be seen immediately, Eqs. (\ref{4.13h}) and (\ref{4.13i}) are respectively equivalent to
\begin{subequations}
\label{4.16}
\begin{align}
\varpi_{i \dot{\alpha}} \varpi_{j}^{\dot{\alpha}} F 
&=\frac{m}{\sqrt{2}} \epsilon_{ij} e^{i\check{\varphi}} F \,, 
\label{4.16a}
\\
\epsilon^{\alpha \beta}
\frac{\partial}{\partial \rho_{i}^{\alpha}} \frac{\partial}{\partial \rho_{j}^{\beta}} F 
&=\frac{m}{\sqrt{2}} \epsilon^{ij} e^{-i\check{\varphi}} F \,.    
\label{4.16b}
\end{align}
\end{subequations}

Combining Eqs. (\ref{4.13e}) and (\ref{4.13f}), we have 
\begin{subequations}
\label{4.17}
\begin{align}
\mathsf{W}^{A}_{1} \frac{\partial}{\partial \mathsf{W}^{A}_{1}} F &=-2(s_{1}+1)F \,, 
\label{4.17a}
\\
\mathsf{W}^{A}_{2} \frac{\partial}{\partial \mathsf{W}^{A}_{2}} F &=-2(s_{2}+1)F \,, 
\label{4.17b}
\end{align}
\end{subequations}
where 
\begin{align}
s_{1}:=\frac{1}{2}(s+t) \,,  \quad s_{2}:=\frac{1}{2}(s-t) \,.
\label{4.18}
\end{align}
The pair of Eqs. (\ref{4.13e}) and (\ref{4.13f}) is equivalent to the pair of Eqs. (\ref{4.17a}) and (\ref{4.17b}). 
Obviously, Eqs. (\ref{4.17a}) and (\ref{4.17b}) are simultaneously satisfied by a homogeneous
twistor function of degree $-2s_{1}-2$ with respect to $\mathsf{W}^{A}_{1}$ 
and degree $-2s_{2}-2$ with respect to $\mathsf{W}^{A}_{2}$.  
These degrees must be integers so that $F$ can be
a single-valued function of $\mathsf{W}^{A}_{i}$.  
In this way, the allowed values of $s_{1}$ and $s_{2}$ are restricted to arbitrary integer or half-integer values,   
and accordingly $s$ and $t$ are also restricted to arbitrary integer or half-integer values. 
We thus see that the Chern-Simons coefficients $2s$ and $2t$, which are coefficients of  
the one-dimensional Chern-Simons terms $S_{a}$ and $S_{\mathsf{b}3}$, respectively, are   
quantized to be arbitrary integer values.

The operators $\check{\mathsf{T}}_{r}$ fulfill the $SU(2)$ commutation relation  
\begin{align}
\left[ \;\! \check{\mathsf{T}}_{r}, \check{\mathsf{T}}_{s} \right]= i\epsilon_{rst} \check{\mathsf{T}}_{t} \,. 
\label{4.19}
\end{align}
Following the general method for solving the eigenvalue problem in  
the $SU(2)$ Lie algebra \cite{Sakurai},  
we can simultaneously solve the eigenvalue equation for the Casimir operator 
$\check{\mathsf{T}}_{r} \check{\mathsf{T}}_{r}
=\check{\mathsf{T}}_{\hat{\imath}}  \check{\mathsf{T}}_{\hat{\imath}} 
+\check{\mathsf{T}}_{3} \check{\mathsf{T}}_{3}$, i.e., 
\begin{align}
\check{\mathsf{T}}_{r} \check{\mathsf{T}}_{r} F=\varLambda F \,, 
\label{4.20}
\end{align}
and Eq. (\ref{4.13f}) to obtain  
\begin{subequations}
\label{4.21}
\begin{align}
\varLambda &=I(I+1) \,, \quad  I=0, \frac{1}{2}, 1, \frac{3}{2}, \ldots ,
\label{4.21a}
\\
t &=-I, \:\! -I+1, \ldots , I-1, I \,. 
\label{4.21b}
\end{align}
\end{subequations}
In deriving Eqs. (\ref{4.21a}) and (\ref{4.21b}), 
we assume the existence of a positive-definite inner product 
in the function space consisting of twistor functions.  
(As for the twistor formulation of a massless system, 
a twistor-function space with a positive-definite inner product has been established \cite{DegNot}.) 
Since $t$ takes integer or half-integer values as explained above, 
$I$ also takes integer or half-integer values accordingly. 
From Eqs. (\ref{4.13f}), (\ref{4.13g}), (\ref{4.20}), and (\ref{4.21a}), the allowed values of 
the positive constant $k$ are determined to be 
\begin{align}
k=2 \sqrt{ I(I+1)-t^2} \,.
\label{4.22}
\end{align}
In this way, the coefficient of $S_{\mathsf{b}12}$ is also quantized in addition to 
the Chern-Simons coefficients.  
It is now clear that the twistor function $F$ is characterized by 
the set of three quantum numbers $(s,I,t)$, or equivalently, by $(I,s_{1},s_{2})$;     
for this reason, it is convenient to label $F$ as $F_{s,I,t}$ or $F_{I,s_{1},s_{2}}$.

\section{Penrose transform and a massive spinor field of arbitrary rank}

In this section, we define a spinor field of arbitrary rank by the Penrose transform of $F_{I,s_{1},s_{2}}$. 
We also demonstrate that this spinor field satisfies generalized DFP equations with $SU(2)$ indices.

Let us consider the Penrose transform of $F_{I,s_{1},s_{2}}$ specified by 
\begin{align}
& \varPsi^{\:\! i_1\ldots i_p}_{\alpha_1 \ldots \alpha_p ; \;\! 
 j_1\ldots j_q, \:\! \dot{\alpha}_1 \ldots \dot{\alpha}_q} (z) 
\notag
\\
&=\frac{1}{(2\pi i)^{4}} \oint_{\varSigma} 
e^{ip \check{\varphi}}  
\varpi_{j_{1} \dot{\alpha}_{1}} \cdots \varpi_{j_{q} \dot{\alpha}_{q}} 
\notag
\\
& \quad\, \times \frac{\partial}{\partial \rho_{i_{1}}^{\alpha_{1}}}
\cdots  \frac{\partial}{\partial \rho_{i_{p}}^{\alpha_{p}}} 
F_{I,s_{1},s_{2}}(\mathsf{W}) 
\:\! d^4 \varpi 
\label{5.1}
\end{align} 
with 
\begin{align}
d^4 \varpi :=d\varpi_{1\dot{0}} \wedge d\varpi_{1\dot{1}} 
\wedge d\varpi_{2\dot{0}} \wedge d\varpi_{2\dot{1}}
\label{5.2}
\end{align} 
to define the rank-$(p+q)$ spinor field 
$\varPsi^{\:\! i_1\ldots i_p}_{\alpha_1 \ldots \alpha_p ; \;\! 
 j_1\ldots j_q, \:\! \dot{\alpha}_1 \ldots \dot{\alpha}_q}$ (occasionally abbreviated as $\varPsi$)   
on complexified Minkowski space $\Bbb{C}\mathbf{M}$. 
Here, $\varSigma$ denotes a suitable four-dimensional contour.  
Equation (\ref{5.1}) is identified as a nonprojective form of the Penrose transform in 
the massive case \cite{PenRin}.\footnote{The two-dimensional projective form of the Penrose transform (\ref{5.1}) 
is given by 
\begin{align*}
& \varPsi^{\:\! i_1\ldots i_p}_{\alpha_1 \ldots \alpha_p ; \;\! 
 j_1\ldots j_q, \:\! \dot{\alpha}_1 \ldots \dot{\alpha}_q} (z) 
\notag
\\
&=\frac{1}{(2\pi i)^{2}} \oint_{\varGamma} 
e^{ip \check{\varphi}}  
\varpi_{j_{1} \dot{\alpha}_{1}} \cdots \varpi_{j_{q} \dot{\alpha}_{q}} 
\notag
\\
& \quad\, \times \frac{\partial}{\partial \rho_{i_{1}}^{\alpha_{1}}}
\cdots  \frac{\partial}{\partial \rho_{i_{p}}^{\alpha_{p}}} 
F_{I,s_{1},s_{2}}(\mathsf{W}) 
\:\! \varpi_{1\dot{\beta}} d\varpi_{1}^{\dot{\beta}} \wedge 
\varpi_{2\dot{\gamma}} d\varpi_{2}^{\dot{\gamma}}  \,, 
\end{align*}
where $\varGamma$ denotes a suitable two-dimensional contour \cite{HugTod}. 
We can also find the three-dimensional projective form of the Penrose transform (\ref{5.1}) \cite{Hughston}. 
}  
It should be noted that $\varPsi$ has the upper and lower $SU(2)$ indices  
in addition to the dotted and undotted spinor indices.  
Because of the structure of Eq. (\ref{5.1}),  
the number of upper (lower) $SU(2)$ indices is equal to the number of undotted (dotted) spinor indices.  
It is obvious that $\varPsi$ has the symmetric properties 
\begin{subequations}
\label{5.3}
\begin{align}
&\varPsi^{\:\! i_1\ldots i_m \ldots i_n \ldots i_p}_{\alpha_1 \ldots \alpha_m \ldots \alpha_n \ldots \alpha_p ; \;\! 
 j_1\ldots j_q, \:\! \dot{\alpha}_1 \ldots \dot{\alpha}_q} 
\notag 
\\
&= \varPsi^{\:\! i_1\ldots i_n \ldots i_m \ldots i_p}_{\alpha_1 \ldots \alpha_n \ldots \alpha_m \ldots \alpha_p ; \;\! 
 j_1\ldots j_q, \:\! \dot{\alpha}_1 \ldots \dot{\alpha}_q} \,, 
\label{5.3a}
\\
&\varPsi^{\:\! i_1\ldots i_p}_{\alpha_1 \ldots \alpha_p ; \;\! 
 j_1\ldots j_a \ldots j_b \ldots j_q, \:\! \dot{\alpha}_1 \ldots \dot{\alpha}_a \ldots \dot{\alpha}_b \ldots \dot{\alpha}_q} 
\notag
\\
&= \varPsi^{\:\! i_1\ldots i_p}_{\alpha_1 \ldots \alpha_p ; \;\! 
 j_1\ldots j_b \ldots j_a \ldots j_q, \:\! \dot{\alpha}_1 \ldots \dot{\alpha}_b \ldots \dot{\alpha}_a \ldots \dot{\alpha}_q} \,.
 \label{5.3b}
\end{align}
\end{subequations}
Suppose that among $i_1,\ldots, i_p$, the number of 1's is $p_{1}$ and the number of 2's is 
$p_{2}(=p-p_{1})$. 
Similarly, suppose that among $j_1,\ldots, j_q$, the number of 1's is $q_{1}$ and the number of 2's is 
$q_{2}(=q-q_{1})$. 
The integral in Eq. (\ref{5.1}) can remain nonvanishing  if 
\begin{align}
s_{1}=\frac{1}{2}(q_{1}-p_{1})\,, \quad s_{2}=\frac{1}{2}(q_{2}-p_{2}) \,. 
\label{5.4}
\end{align}
Combining Eqs. (\ref{4.18}) and (\ref{5.4}), we have 
\begin{subequations}
\label{5.5}
\begin{align}
s=\frac{1}{2} (q_{1}-p_{1}+q_{2}-p_{2}) \,, 
\label{5.5a}
\\
t=\frac{1}{2} (q_{1}-p_{1}-q_{2}+p_{2}) \,.  
\label{5.5b}
\end{align}
\end{subequations}

Now we can show that 
\begin{align}
& \frac{\partial}{\partial z_{\beta\dot{\beta}}} F(\mathsf{W})
=\frac{\partial \rho^{\gamma}_{k}}{\partial z_{\beta\dot{\beta}}}
\frac{\partial}{\partial \rho^{\gamma}_{k}} F(\mathsf{W}) 
=\frac{\partial \varrho^{\gamma}_{k}}{\partial z_{\beta\dot{\beta}}}
\frac{\partial}{\partial \rho^{\gamma}_{k}} F(\mathsf{W}) 
\notag
\\
&=\frac{\partial \!\left( i z^{\gamma \dot{\gamma}} \varpi_{k \dot{\gamma}} \right)}
{\partial z_{\beta \dot{\beta}}}
\frac{\partial}{\partial \rho^{\gamma}_{k}} F(\mathsf{W}) 
=i \varpi_{k}^{\dot{\beta}} \epsilon^{\beta \gamma}
\frac{\partial}{\partial \rho_{k}^{\gamma}} F(\mathsf{W}) \,.  
\label{5.6}
\end{align}
Here the weak equality $\rho^{\gamma}_{j} \approx \varrho^{\gamma}_{j}$, Eq. (\ref{2.22}),  
and the formula  
$\partial/\partial z_{\beta\dot{\beta}} =\epsilon^{\beta\alpha} 
\epsilon^{\dot{\beta} \dot{\alpha}} \partial/\partial z^{\alpha\dot{\alpha}}$ 
have been used. 
The derivative of $\varPsi$ with respect to $z_{\beta\dot{\beta}}$ can be calculated  
by using Eq. (\ref{5.6}) as follows: 
\begin{align}
& \frac{\partial}{\partial z_{\beta\dot{\beta}}} 
\varPsi^{\:\! i_1\ldots i_p}_{\alpha_1 \ldots \alpha_p ; \;\! 
 j_1\ldots j_q, \:\! \dot{\alpha}_1 \ldots \dot{\alpha}_q} (z) 
\notag
\\
&=\frac{1}{(2\pi i)^{4}} \oint_{\varSigma} 
e^{ip \check{\varphi}}  
\varpi_{j_{1} \dot{\alpha}_{1}} \cdots \varpi_{j_{q} \dot{\alpha}_{q}} 
\notag
\\
& \quad\, \times 
\frac{\partial}{\partial \rho_{i_{1}}^{\alpha_{1}}}
\cdots  \frac{\partial}{\partial \rho_{i_{p}}^{\alpha_{p}}} 
\frac{\partial}{\partial z_{\beta\dot{\beta}}} F_{I,s_{1},s_{2}}(\mathsf{W}) \:\! d^4 \varpi 
\notag
\\
&=\frac{i}{(2\pi i)^{4}} \oint_{\varSigma} 
e^{ip \check{\varphi}}  
\varpi_{j_{1} \dot{\alpha}_{1}} \varpi_{k}^{\dot{\beta}} 
\varpi_{j_{2} \dot{\alpha}_{2}} \cdots \varpi_{j_{q} \dot{\alpha}_{q}} 
\notag
\\
& \quad\, \times 
\frac{\partial}{\partial \rho_{i_{2}}^{\alpha_{2}}}
\cdots  \frac{\partial}{\partial \rho_{i_{p}}^{\alpha_{p}}} 
\epsilon^{\beta \gamma} 
\frac{\partial}{\partial \rho_{i_{1}}^{\alpha_{1}}} \frac{\partial}{\partial \rho_{k}^{\gamma}} 
F_{I,s_{1},s_{2}}(\mathsf{W}) \:\! d^4 \varpi \,. 
\label{5.7}
\end{align}
Contracting over the indices $\dot{\beta}$ and $\dot{\alpha}_{1}$ in Eq. (\ref{5.7}) and using Eq. (\ref{4.16a}), 
we obtain 
\begin{align}
&  \frac{\partial}{\partial z_{\beta\dot{\beta}}} 
\varPsi^{\:\! i_1\ldots i_p}_{\alpha_1 \ldots \alpha_p ; \;\! 
 j_1\ldots j_q, \:\! \dot{\beta} \dot{\alpha}_2 \ldots \dot{\alpha}_q} (z) 
\notag
\\
&= \frac{m}{\sqrt{2}} \epsilon^{\beta\gamma} \epsilon_{j_{1} k}
\frac{i}{(2\pi i)^{4}} \oint_{\varSigma} 
e^{i(p+1) \check{\varphi}}  
\varpi_{j_{2} \dot{\alpha}_{2}} \cdots \varpi_{j_{q} \dot{\alpha}_{q}} 
\notag
\\
& \quad\, \times \frac{\partial}{\partial \rho_{k}^{\gamma}}
\frac{\partial}{\partial \rho_{i_{1}}^{\alpha_{1}}}
\cdots  \frac{\partial}{\partial \rho_{i_{p}}^{\alpha_{p}}} 
F_{I,s_{1},s_{2}}(\mathsf{W}) \:\! d^4 \varpi 
\notag
\\
&= \frac{im}{\sqrt{2}} \epsilon^{\beta\gamma} \epsilon_{j_{1} k}  
\varPsi^{\:\! k i_1\ldots i_p}_{\gamma \alpha_1 \ldots \alpha_p ; \;\! 
 j_2\ldots j_q, \:\! \dot{\alpha}_2 \ldots \dot{\alpha}_q} (z) \,. 
\label{5.8}
\end{align}
Similarly, contracting over the indices $\beta$ and $\alpha_{1}$ in Eq. (\ref{5.7}) and using Eq. (\ref{4.16b}), 
we obtain 
\begin{align}
&  \frac{\partial}{\partial z_{\beta\dot{\beta}}} 
\varPsi^{\:\! i_1\ldots i_p}_{\beta \alpha_2 \ldots \alpha_p ; \;\! 
 j_1\ldots j_q, \:\! \dot{\alpha}_1 \ldots \dot{\alpha}_q} (z) 
\notag
\\
&= \frac{m}{\sqrt{2}} \epsilon^{\dot{\beta} \dot{\gamma}} \epsilon^{i_{1} k}
\frac{i}{(2\pi i)^{4}} \oint_{\varSigma} 
e^{i(p-1) \check{\varphi}}  
\varpi_{k \dot{\gamma}} \varpi_{j_{1} \dot{\alpha}_{1}} \cdots \varpi_{j_{q} \dot{\alpha}_{q}} 
\notag
\\
& \quad\, \times 
\frac{\partial}{\partial \rho_{i_{2}}^{\alpha_{2}}}
\cdots  \frac{\partial}{\partial \rho_{i_{p}}^{\alpha_{p}}} 
F_{I,s_{1},s_{2}}(\mathsf{W}) \:\! d^4 \varpi 
\notag
\\
&= \frac{im}{\sqrt{2}} \epsilon^{\dot{\beta} \dot{\gamma}} \epsilon^{i_{1} k}
\varPsi^{\:\! i_2 \ldots i_p}_{\alpha_2 \ldots \alpha_p ; \;\! 
k j_1\ldots j_q, \:\! \dot{\gamma} \dot{\alpha}_1 \ldots \dot{\alpha}_q} (z) \,. 
\label{5.9}
\end{align}
In this way, it has been shown that the spinor field $\varPsi$ satisfies 
the generalized DFP equations with $SU(2)$ indices 
\begin{subequations}
\label{5.10}
\begin{align}
& i \sqrt{2}\:\! \frac{\partial}{\partial z_{\beta\dot{\beta}}} 
\varPsi^{\:\! i_{1}\ldots i_{p}}_{\alpha_1 \ldots \alpha_p; \;\! 
j_1\ldots j_q, \;\! \dot{\beta} \dot{\alpha}_2 \ldots \dot{\alpha}_q} 
\notag
\\
&+m \epsilon^{\beta\gamma} \epsilon_{j_{1} k}
\varPsi^{\:\! k i_1\ldots i_p}_{\gamma \alpha_1 \ldots \alpha_p ; \;\! 
 j_2\ldots j_q, \;\! \dot{\alpha}_2 \ldots \dot{\alpha}_q}\! =0 \,, 
 \label{5.10a}
 \\
& i \sqrt{2}\:\! \frac{\partial}{\partial z_{\beta\dot{\beta}}} 
\varPsi^{\:\! i_{1}\ldots i_{p}}_{\beta \alpha_2 \ldots \alpha_p ; \;\! 
 j_1\ldots j_q, \:\! \dot{\alpha}_1 \ldots \dot{\alpha}_q} 
 \notag
 \\
&+m \epsilon^{\dot{\beta} \dot{\gamma}} \epsilon^{i_{1} k}
\varPsi^{\:\! i_2\ldots i_p}_{\alpha_2 \ldots \alpha_p ; \;\! 
k j_1\ldots j_q, \:\! \dot{\gamma} \dot{\alpha}_1 \ldots \dot{\alpha}_q}\! =0 \,. 
\label{5.10b}
\end{align}
\end{subequations}
Using Eqs. (\ref{5.10a}) and (\ref{5.10b}) and noting 
\begin{align}
\frac{\partial}{\partial z^{\alpha\dot{\beta}}} \frac{\partial}{\partial z_{\beta\dot{\beta}}} 
=\frac{1}{2} \delta_{\alpha}^{\beta} 
\frac{\partial}{\partial z^{\gamma\dot{\gamma}}} \frac{\partial}{\partial z_{\gamma\dot{\gamma}}} \,, 
\label{5.11}
\end{align}
we can derive the Klein-Gordon equation  
\begin{align}
\left( \frac{\partial}{\partial z^{\beta\dot{\beta}}} \frac{\partial}{\partial z_{\beta\dot{\beta}}} 
+m^2 \right) \! 
\varPsi^{\:\! i_1\ldots i_p}_{\alpha_1 \ldots \alpha_p ; \;\! 
 j_1\ldots j_q, \:\! \dot{\alpha}_1 \ldots \dot{\alpha}_q} =0 \,.  
\label{5.12}
\end{align}
This makes it clear that $\varPsi$ is a field of mass $m$. 
Thus, we obtain a spinor field of arbitrary rank with mass $m$ by means of the Penrose transform (\ref{5.1}).

\section{Rank-one spinor fields and physical meanings of the gauge symmetries}

In this section, we investigate the rank-one spinor fields in detail to clarify the physical meanings of 
the $U(1)_{a}$, $U(1)_{\mathsf{b}}$, and $SU(2)$ symmetries 
as well as those of the constants $s$ and $t$.

Now we particularly consider Eq. (\ref{5.10a}) in the case $(p,q)=(0,1)$ and Eq. (\ref{5.10b}) in the case $(p,q)=(1,0)$,    
which respectively read  
\begin{subequations}
\label{6.1}
\begin{align}
& i \sqrt{2}\:\! \frac{\partial}{\partial z^{\alpha\dot{\beta}}} 
\varPsi^{\:\! \dot{\beta}}_{i} (z)
-m \epsilon_{ij} \varPsi^{\:\! j}_{\alpha} (z) =0 \,, 
\label{6.1a}
\\
& i \sqrt{2}\:\! \frac{\partial}{\partial z_{\beta\dot{\alpha}}} 
\varPsi^{\:\! i}_{\beta} (z)
+m \epsilon^{ij} \varPsi^{\:\! \dot{\alpha}}_{j} (z)=0 \,, 
\label{6.1b}
\end{align}
\end{subequations}
with $\varPsi^{\:\! \dot{\beta}}_{i} :=\epsilon^{\dot{\beta} \dot{\gamma}} \varPsi_{i \dot{\gamma}}$.  
Equation (\ref{6.1a}) with $i=1$ and Eq. (\ref{6.1b}) with $i=2$ can be combined  
in the form of the ordinary Dirac equation 
\begin{align}
D\psi_{1}(z)=0 \,, 
\quad 
\psi_{1}(z):=\! \left( \begin{array}{c} \varPsi^{2}_{\beta}(z) \\ \varPsi_{1}^{\:\! \dot{\beta}}(z) \end{array} \right) , 
\label{6.2}
\end{align}
while Eq. (\ref{6.1a}) with $i=2$ and Eq. (\ref{6.1b}) with $i=1$ can be combined, 
after replacing $z^{\alpha\dot{\alpha}}$ by $-z^{\alpha\dot{\alpha}}$, as  
\begin{align}
D\psi_{2}(z)=0 \,, 
\quad 
\psi_{2}(z):=\! \left( \begin{array}{c} \varPsi^{1}_{\beta}(-z) \\ \varPsi_{2}^{\:\! \dot{\beta}}(-z) \end{array} \right) .
\label{6.3}
\end{align}
In Eqs. (\ref{6.2}) and (\ref{6.3}),  $D$ denotes the Dirac operator 
\begin{align}
D:=\left(
\begin{array}{cc}
-m\delta_{\alpha}^{\beta} & \; i \sqrt{2}\:\! \dfrac{\partial}{\partial z^{\alpha\dot{\beta}}}  \\
i \sqrt{2}\:\! \dfrac{\partial}{\partial z_{\beta\dot{\alpha}}}  &\;  -m\delta_{\dot{\beta}}^{\dot{\alpha}}   
\end{array}
\right).  
\label{6.4}
\end{align}
The charge conjugate of $\psi_{1}(z)$ is found to be 
\begin{align}
\psi_{1}^{\mathrm{c}}(z)
&:=\left(
\begin{array}{cc}
0 &  -\epsilon_{\beta \gamma}  \\
\epsilon^{\dot{\beta} \dot{\gamma}}  &  0
\end{array}
\right) \overline{\psi_{1}(\bar{z})} 
\notag
\\
&\; =\left(
\begin{array}{cc}
0 &  -\epsilon_{\beta \gamma}  \\
\epsilon^{\dot{\beta} \dot{\gamma}}  &  0
\end{array}
\right) \!
\left( \begin{array}{c} \bar{\varPsi}_{2\dot{\gamma}}(z) 
\\ \bar{\varPsi}^{1\gamma}(z) \end{array} \right)
=
\left( \begin{array}{c} \bar{\varPsi}^{1}_{\beta}(z) 
\\ \bar{\varPsi}_{2}^{\:\! \dot{\beta}}(z) \end{array} \right) , 
\label{6.5}
\end{align}
where the arguments of $\psi_{1}$, namely $z^{\alpha\dot{\alpha}}$, have been replaced by 
their complex conjugates $\bar{z}^{\alpha\dot{\alpha}}:=\overline{z^{\alpha\dot{\alpha}}}$ so that 
$\psi_{1}^{\mathrm{c}}$ can be a holomorphic function of $z^{\alpha\dot{\alpha}}$. 
Using the complex conjugates of Eqs. (\ref{6.1a}) and (\ref{6.1b}),  
we can show that $D\psi_{1}^{\mathrm{c}}(z)=0$. 
Since $\psi_{2}$ and $\psi_{1}^{\mathrm{c}}$ satisfy the same Dirac equation and 
have the same spinor and $SU(2)$ indices, 
they can be identified with each other up to an overall constant.\footnote{The 
plane wave solution of Eq. (\ref{6.1}) given by 
\begin{align*}
\varPsi^{\:\! i}_{\alpha} (z) &=-Ce^{i\varphi/2} \bar{\varpi}{}^{i}_{\alpha}  
\exp\! \left(-iz^{\gamma\dot{\gamma}} \bar{\varpi}{}^{k}_{\gamma} \varpi_{k\dot{\gamma}} \right) ,
\\
\varPsi^{\:\! \dot{\alpha}}_{i} (z) &=Ce^{-i\varphi/2} \varpi_{i}^{\dot{\alpha}} 
\exp\! \left(-iz^{\gamma\dot{\gamma}} \bar{\varpi}{}^{k}_{\gamma} \varpi_{k\dot{\gamma}} \right) 
\end{align*}
fulfills the conditions 
$\varPsi^{\:\! i}_{\alpha} (-z)= -\big(C/\bar{C}\big) \bar{\varPsi}^{\:\! i}_{\alpha}(z)$ and 
$\varPsi^{\:\! \dot{\alpha}}_{i} (-z)= -\big(C/\bar{C}\big) \bar{\varPsi}^{\:\! \dot{\alpha}}_{i}(z)$.  
Here, $C$ is a complex constant and $\varphi$ is given in Eq. (\ref{3.26e}). 
These conditions lead to $\psi_{2}(z)= -\big(C/\bar{C}\big) \psi_{1}^{\mathrm{c}}(z)$, and hence, 
in this case, $\psi_{2}$ and $\psi_{1}^{\mathrm{c}}$ can indeed be identified with each other.
For verifying that the plane wave solution satisfies Eq. (\ref{6.1}),  
it is convenient to use the classical counterparts of Eqs. (\ref{4.16a}) and (\ref{4.16b}): 
\begin{align*}
\varpi_{i \dot{\alpha}} \varpi_{j}^{\dot{\alpha}}  
=\frac{m}{\sqrt{2}} \epsilon_{ij} e^{i\varphi} , 
\quad \;
\bar{\varpi}{}^{i}_{\alpha} \bar{\varpi}{}^{j \alpha}  
=\frac{m}{\sqrt{2}} \epsilon^{ij} e^{-i\varphi} . 
\end{align*}
}
(This identification may be confirmed by the $CPT$ symmetry.) 
If $\psi_{1}(z)$ is a spinor field of a particle with four-momentum $(E, \bm{p})$, 
then $\psi_{1}^{\mathrm{c}}(z) \big( \simeq \psi_{2}(z) \big)$ is regarded as a spinor field of 
a corresponding antiparticle with four-momentum $(-E, -\bm{p})$.  
Accordingly, $\psi_{2}(-z) =\big( \varPsi^{1}_{\alpha}(z) , \varPsi_{2}^{\:\! \dot{\alpha}}(z) \big){}^{\mathrm{T}}$ 
is considered a spinor field of the antiparticle with four-momentum $(E, \bm{p})$.  
In the light of this fact, 
it is clear that $\varPsi^{2}_{\alpha}(z)$ and $\varPsi^{1}_{\alpha}(z)$ represent 
a left-handed particle and a corresponding left-handed antiparticle, respectively,  
while $\varPsi_{1}^{\:\! \dot{\alpha}}(z)$ and $\varPsi_{2}^{\:\! \dot{\alpha}}(z)$ represent  
a right-handed particle and a corresponding right-handed antiparticle, respectively, 
as summarized in Table I. 
\begin{table}
\caption{A classification of the rank-one spinor fields.}
\begin{center}
\begin{tabular}{lcc}
\hline
\hline \\[-9pt] 
&  \hspace{0.3pt} Particle  \hspace{0.8pt} & \hspace{1.1pt}  Antiparticle  \\
\hline \\[-6pt]
\hspace{0.1pt}  Left-handed  \hspace{1.3pt} & \hspace{0.10pt} $\varPsi^{2}_{\alpha}$ & \hspace{0.10pt} $\varPsi^{1}_{\alpha} $  \\[4pt]
\hspace{0.1pt}  Right-handed \hspace{1.3pt} & \hspace{0.10pt} $\varPsi_{1}^{\:\! \dot{\alpha}}$ & \hspace{0.10pt} $\varPsi_{2}^{\:\! \dot{\alpha}}$ \\[3pt]
\hline
\hline
\end{tabular}
\end{center}
\end{table}
We thus see that the index $i$ of $\varPsi^{\:\! i}_{\alpha} (z)$ and $\varPsi^{\:\! \dot{\alpha}}_{i} (z)$  
distinguishes between a particle and its antiparticle.

Using Eq. (\ref{5.5}), we can obtain the possible values of $s$ and $t$ for each of  
the rank-one spinor fields as in Table II. 
\begin{table}
\caption{The values of $s$ and $t$ of the rank-one spinor fields.}
\begin{center}
\begin{tabular}{lcc| lcc}
\hline
\hline & & & & & \\[-9pt] 
&  \hspace{6pt} $s$ \hspace{4pt}  & \hspace{8pt} $t$ \hspace{10pt} & & \hspace{9pt} $s$ \hspace{4pt} & \hspace{8pt} $t$ \hspace{4pt}  \\
\hline & & & & &  \\[-2pt]
\hspace{0.1pt} $\varPsi^{2}_{\alpha}$  \hspace{1pt} & $- \dfrac{1}{2}$ & $ \hspace{6pt} \dfrac{1}{2} $ \hspace{5pt} & \hspace{6pt} 
$\varPsi^{1}_{\alpha}$ & $  \hspace{2pt}  - \dfrac{1}{2}$ & $  \hspace{2pt} - \dfrac{1}{2}$ \\[13pt]
\hspace{0.1pt} $ \varPsi_{1}^{\:\! \dot{\alpha}}$ \hspace{1pt} & $  \hspace{2pt} \dfrac{1}{2} $ & $ \hspace{6pt} \dfrac{1}{2}$ \hspace{5pt} & \hspace{6pt} 
$\varPsi_{2}^{\:\! \dot{\alpha}}$ & $ \hspace{5pt} \dfrac{1}{2}$ &  $  \hspace{2pt} - \dfrac{1}{2} $ \\[11pt]
\hline 
\hline
\end{tabular}
\end{center}
\end{table}
We observe that the left-handed spinor fields $\varPsi^{\:\! i}_{\alpha} (z)$ ($i=1,2$) have $s=-1/2$, while 
the right-handed spinor fields $\varPsi^{\:\! \dot{\alpha}}_{i} (z)$  ($i=1,2$) have $s=1/2$. 
Hence, $s$ turns out to be a quantum number specifying the chirality of a spinor field. 
Since $s$ is an eigenvalue of $\check{\mathsf{T}}_{0}$ up to the additive constant $2$,   
as can be seen from (\ref{4.13e}), 
$\check{\mathsf{T}}_{0}$ can be interpreted as the operator of chirality. 
Accordingly, $U(1)_{a}$ can be identified as the gauge group of chirality, and 
the $U(1)_{a}$ symmetry is physically understood as a gauge symmetry leading to chirality conservation. 
We also observe that the particle spinor fields $\varPsi^{2}_{\alpha}(z)$ and $\varPsi_{1}^{\:\! \dot{\alpha}}(z)$ have 
$t=1/2$, while the antiparticle spinor fields $\varPsi^{1}_{\alpha}(z)$ and $\varPsi_{2}^{\:\! \dot{\alpha}}(z)$ 
have $t=-1/2$. Hence, $t$ turns out to be a quantum number distinguishing between a particle and its antiparticle. 
Then it follows that $t$ is proportional to the electric charge of the particle/antiparticle. 
Since $t$ is an eigenvalue of $\check{\mathsf{T}}_{3}$ as can be seen from (\ref{4.13f}), 
$\check{\mathsf{T}}_{3}$ can be interpreted as the operator of electric charge up to a constant of proportionality. 
Accordingly, $U(1)_{\mathsf{b}}$ can be identified with the gauge group of electric charge, 
and the $U(1)_{\mathsf{b}}$ symmetry is physically understood as a gauge symmetry  
leading to electric charge conservation.

Now we recall that our study has been performed in the unitary gauge in which 
the GGS action takes the form of Eq. (\ref{2.20}) or Eq. (\ref{2.24}). 
In the unitary gauge, the local $SU(2)$ symmetry is hidden and the $U(1)_{\mathsf{b}}$ 
symmetry is linearly realized in accordance with Eq. (2.12). 
The manifestly $SU(2)$ covariant formulation can be developed on the basis of the action (\ref{2.23}). 
The rank-one spinor fields found in this formulation, denoted by  
$\varOmega^{\:\! \dot{\alpha}}_{i}$ and $\varOmega^{\:\! i}_{\alpha}$, 
are related to $\varPsi^{\:\! \dot{\alpha}}_{i}$ and $\varPsi^{\:\! i}_{\alpha}$ by\footnote{The 
rank-$(p+q)$ spinor field in the manifestly $SU(2)$ covariant formulation is given by 
\begin{align*}
& \varOmega^{\:\! i_1\ldots i_p}_{\alpha_1 \ldots \alpha_p ; \;\! 
 j_1\ldots j_q, \:\! \dot{\alpha}_1 \ldots \dot{\alpha}_q} (z) 
\notag
\\
&=\frac{1}{(2\pi i)^{4}} \oint_{\varSigma} 
e^{ip \check{\varphi}}  
\pi_{j_{1} \dot{\alpha}_{1}} \cdots \pi_{j_{q} \dot{\alpha}_{q}} 
\notag
\\
& \quad\, \times 
\frac{\partial}{\partial \mu_{i_{1}}^{\alpha_{1}}}
\cdots  \frac{\partial}{\partial \mu_{i_{p}}^{\alpha_{p}}} 
F_{I,s_{1},s_{2}}(\mu, \pi) 
\:\! d^4 \pi \,, 
\end{align*} 
where $\mu_{i}^{\alpha}$ is a spinor related to $\omega_{i}^{\alpha}$ by the weak equality 
$\mu_{i}^{\alpha} \approx \omega_{i}^{\alpha}$. }
\begin{align}
\varOmega^{\:\! \dot{\alpha}}_{i} (z)=V_{i}{}^{j} \varPsi^{\:\! \dot{\alpha}}_{j} (z) \,,  
\quad \;
\varOmega^{\:\! i}_{\alpha} (z)=\varPsi^{\:\! j}_{\alpha} (z) V^{\dagger}{}_{j}{}^{i} \,. 
\label{6.6}
\end{align}
Because $V$ is independent of $z^{\alpha\dot{\alpha}}$, we can readily verify 
by using Eqs. (\ref{6.1a}) and (\ref{6.1b}) that 
\begin{subequations}
\label{6.7}
\begin{align}
& i \sqrt{2}\:\! \frac{\partial}{\partial z^{\alpha\dot{\beta}}} 
\varOmega^{\:\! \dot{\beta}}_{i} (z)
-m \epsilon_{ij} \varOmega^{\:\! j}_{\alpha} (z) =0 \,, 
\label{6.7a}
\\
& i \sqrt{2}\:\! \frac{\partial}{\partial z_{\beta\dot{\alpha}}} 
\varOmega^{\:\! i}_{\beta} (z)
+m \epsilon^{ij} \varOmega^{\:\! \dot{\alpha}}_{j} (z)=0 \,.  
\label{6.7b}
\end{align}
\end{subequations}
Following the above consideration for 
$\varPsi^{\:\! \dot{\alpha}}_{i} (z)$ and $\varPsi^{\:\! i}_{\alpha} (z)$, we see that 
$\varOmega^{2}_{\alpha}(z)$ and $\varOmega^{1}_{\alpha}(z)$ constitute a doublet of 
left-handed particle and antiparticle spinor fields, 
while $\varOmega^{\:\! \dot{\alpha}}_{1}(z)$ and $\varOmega^{\:\! \dot{\alpha}}_{2}(z)$ 
constitute a doublet of right-handed particle and antiparticle spinor fields.  
Under the $SU(2)$ transformation, $\varOmega^{\:\! \dot{\alpha}}_{i}$ and $\varOmega^{\:\! i}_{\alpha}$ 
transform linearly as 
\begin{align}
\varOmega^{\:\! \dot{\alpha}}_{i}
\rightarrow \varOmega^{\prime \:\! \dot{\alpha}}_{i} =U_{i}{}^{j} \varOmega^{\:\! \dot{\alpha}}_{j} \,, 
\quad \;
\varOmega^{\:\! i}_{\alpha}
\rightarrow \varOmega^{\prime \:\! i}_{\alpha}=\varOmega^{\:\! j}_{\alpha} U^{\dagger}{}_{j}{}^{i} \,, 
\label{6.8}
\end{align}
whereas $\varPsi^{\:\! \dot{\alpha}}_{i}$ and $\varPsi^{\:\! i}_{\alpha}$ transform according to 
the $U(1)_{\mathsf{b}}$ transformation 
\begin{align}
\varPsi^{\:\! \dot{\alpha}}_{i}
\rightarrow \varPsi^{\prime \:\! \dot{\alpha}}_{i} =\varTheta_{i}{}^{j} \varPsi^{\:\! \dot{\alpha}}_{j} \,, 
\quad \;
\varPsi^{\:\! i}_{\alpha}
\rightarrow \varPsi^{\prime \:\! i}_{\alpha}=\varPsi^{\:\! j}_{\alpha} \varTheta^{\dagger}{}_{j}{}^{i} \,.  
\label{6.9}
\end{align}
As seen from Eq. (\ref{6.8}), the $SU(2)$ transformation causes   
a continuous transformation between the particle spinor field 
$\varOmega^{2}_{\alpha}$ $\left( \varOmega^{\:\! \dot{\alpha}}_{1} \right)$  
and the antiparticle spinor field 
$\varOmega^{1}_{\alpha}$ $\left( \varOmega^{\:\! \dot{\alpha}}_{2} \right)$. 
The $SU(2)$ symmetry therefore turns out to be a gauge symmetry realized in  
the particle-antiparticle doublets 
$\left( \varOmega^{2}_{\alpha}, \varOmega^{1}_{\alpha} \right)$ and 
$\left( \varOmega^{\:\! \dot{\alpha}}_{1}, \varOmega^{\:\! \dot{\alpha}}_{2} \right)$. 
Such a symmetry, however, is not observed in nature; hence, it should be considered  
that the $SU(2)$ symmetry is hidden or broken. 
The formulation in the unitary gauge is appropriate for this situation, 
because, in the unitary gauge, the $SU(2)$ symmetry is hidden  
and the $U(1)_{\mathsf{b}}$ symmetry is manifestly exhibited instead.

\section{Summary and discussion}

We have presented a gauged twistor model of a free {\em massive} spinning particle in four dimensions. 
This model is a non-Abelian extension of the gauged twistor model of a free {\em massless} spinning particle 
in four dimensions, presented in Refs. \cite{BarPic, DEN, DNOS}.  
The extended model is governed by the GGS action that was elaborated by 
adding the 1D Chen-Simons terms $S_{a}$ and $S_{\mathsf{b}3}$ and the novel term $S_{\mathsf{be}}$ 
to the gauged twistorial action $S_{m \mathrm{g}}$ $[^{\!\:}$see Eq. (2.20)]. 
The GGS action remains invariant under the reparametrization and the $U(1)_{a}$  
and local $SU(2)$ transformations, 
although the $SU(2)$ symmetry is nonlinearly realized in the action. 
In the unitary gauge, the $U(1)_{\mathsf{b}}$ symmetry is manifestly exhibited, 
while the $SU(2)$ symmetry is hidden.

We have studied the canonical Hamiltonian formalism based on the GGS action in the unitary gauge 
by following Dirac's recipe for constrained Hamiltonian systems. 
The classification of the constraints into first and second classes was carried out strictly,   
and the Dirac brackets between the canonical variables were obtained concretely. 
It was demonstrated that just sufficient constraints for the twistor variables are consistently derived as 
the secondary first-class constraints $[^{\!\:}$see Eqs. (\ref{3.28e})--(\ref{3.28i})].

The subsequent canonical quantization of the system was performed 
in terms of the new twistor variables $\mathsf{W}_{i}^{A}$ and $\bar{\mathsf{W}}^{i}_{A}$, 
because they satisfy the simple Dirac brackets given in Eq. (\ref{3.33}).  
We have shown that the Chern-Simons coefficients $2s$ and $2t$ are quantized to be 
arbitrary integer values as a result of the canonical quantization based on the commutation relations 
(\ref{4.2a})--(\ref{4.2e}). 
In general, the quantization of Chern-Simons coefficient is a common consequence in certain theories 
in which the Chern-Simons terms play crucial roles (see e.g. Refs. \cite{DJT, Polychronakos, Witten, TonWon}). 
Our gauged twistor model can be regarded as a specific example of such theories.  
Intriguingly, the coefficient $k$ of $S_{\mathsf{b}12}$ is also quantized via solving 
the eigenvalue problem of the $SU(2)$ Lie algebra. 
We found that the twistor functions in our model are eigenfunctions of the relevant  
differential operators governed by the $U(1)_{a} \times SU(2)$ Lie algebra  
$[^{\!\:}$see Eqs. (\ref{4.13e})--(\ref{4.13g})].  
Each twistor function $F$ is then labeled by a set of three quantum numbers 
associated with the $U(1)_{a} \times SU(2)$ Lie algebra.

We have carried out the Penrose transform of the twistor function $F$ 
to obtain a massive spinor field of arbitrary rank defined on complexified Minkowski space $[^{\!\:}$see Eq. (\ref{5.1})]. 
As emphasized earlier, this spinor field has the upper and lower $SU(2)$ indices  
in addition to the dotted and undotted spinor indices. 
In fact, we observed that the number of upper (lower) $SU(2)$ indices is equal 
to the number of undotted (dotted) spinor indices.  
We also demonstrated that the spinor field satisfies the generalized DFP equations with $SU(2)$ indices,  
given in Eq. (\ref{5.10}).

We have investigated the rank-one spinor fields in detail to clarify the physical meanings of 
the gauge symmetries as well as those of the constants $s$ and $t$.  
It turned out that $s$ is a quantum number specifying the chirality of a spinor field and 
that the $U(1)_{a}$ symmetry is a gauge symmetry leading to chirality conservation. 
It also turned out that $t$ is a quantum number proportional to the electric charge of a spinor field and 
that the $U(1)_{\mathsf{b}}$ symmetry is a gauge symmetry leading to electric charge conservation. 
The $SU(2)$ symmetry was shown to be a gauge symmetry realized in the particle-antiparticle doublets.  
Such a symmetry, however, is not observed in nature, so that it should be considered to be hidden or broken. 
Fortunately our twistor formulation in the unitary gauge is appropriate for describing this situation. 
Since the $SU(2)$ symmetry is a symmetry realized in the particle-antiparticle doublets, 
it cannot be identified with the weak isospin symmetry. 
We thus conclude that the idea proposed by Penrose, Perj\'{e}s, and Hughston 
\cite{Penrose, Perjes1, Perjes2, Perjes3, Hughston}
is not valid in our gauged twistor model.

The observation that $s$ is a quantum number specifying the chirality of a spinor field 
is supported for the following reason: 
The gauged Shirafuji action for a massless spinning particle enjoys the $U(1)_{a}$ symmetry and 
contains its associated constant $s$ \cite{BarPic, DEN, DNOS}. 
This constant is indeed shown to be the helicity of a massless spinning particle. 
As is well known,  the chirality is an analog of the helicity, while the chirality is 
a Lorentz invariant quantity valid for massive particles as well as massless particles.  
(For massless particles, chirality is the same as helicity.) 
For this reason, in the present twistor model, it is quite natural to identify the Lorentz invariant 
quantity $s$ as the chirality quantum number.

We have seen that each eigenstate of $\check{\mathsf{T}}_{3}$ corresponds 
(via the Penrose transform) to a particle or antiparticle state represented by its own spinor field. 
Remarkably, we encounter a similar situation in studying the rigid body model \cite{HTY, HarGot}.   
In this model, the rigid body rotation leads to an intrinsic $SU(2)$ symmetry in addition to the spin $SU(2)$ symmetry.  
Hara {\it et al.} showed that the eigenstates of the third generator of the intrinsic $SU(2)$ group 
are assigned to particle and antiparticle spinor fields. 
They also pointed out that this generator cannot be identified with the third component of the isospin generators. 
(Accordingly, it turns out that the intrinsic $SU(2)$ symmetry cannot be regarded as the isospin symmetry. 
This result contradicts the earlier idea concerning isospin proposed in Refs. \cite{Nakano, Takabayasi}.)  
We thus see that the gauged twistor model and the rigid body model share common aspects.

Now we recall that the secondary first-class constraints (\ref{3.28e})--(\ref{3.28g}), or equivalently, 
Eqs. (\ref{3.36a}), (\ref{3.36b}), and (\ref{3.38}), have been derived systematically on the basis of 
the $U(1)_{a}$, $U(1)_{\mathsf{b}}$, and reparametrization symmetries of the GGS action. 
By contrast, the remaining secondary first-class constraints 
(\ref{3.28h}) and (\ref{3.28i}) have been derived as a result of incorporating the mass-shell condition $(\ref{2.3})$ 
into the GGS action by hand. 
Considering this fact, we can never say that the present approach for constructing the GGS action 
is satisfactory from the gauge-theoretical point of view. 
To make our gauged twistor formulation complete,  
we need to establish an approach in which the mass-shell condition $(\ref{2.3})$ is supplied  
as an inevitable outcome of an extra gauge symmetry.

In this paper, we have not presented precise definitions of the chirality and charge conjugation 
for a massive spinor field of arbitrary rank.  
The chirality may be defined on the basis of the type of spinor indices of the field. 
For clarifying the definition of charge conjugation and its associated concept of particle-antiparticle, 
it is necessary to examine coupling of a massive spinor field of arbitrary rank to the electromagnetic field.  
The precise definitions of chirality and charge conjugation should confirm our observation on 
the physical meanings of the constants $s$ and $t$. 
We hope to address the aforementioned issues in the near future.

\section*{Acknowledgments}

We would like to thank Shigefumi Naka for useful comments.  
The work of S.D. is supported in part by  
Grant-in-Aid for Fundamental Scientific Research from   
College of Science and Technology, Nihon University.

\appendix*
\section
{POINCAR\'{E} SYMMETRY AND PAULI-LUBANSKI PSEUDOVECTOR}

In this appendix, we consider the Poincar\'{e} symmetry and 
the Pauli-Lubanski pseudovector 
within the framework of the gauged twistor formulation.  

We can easily show that the GGS action (\ref{2.20}) remains invariant under 
the infinitesimal Poincar\'{e} transformation 
$[^{\!\:}$or more accurately, the infinitesimal $SL(2, \Bbb{C}) \ltimes \Bbb{R}^{1,3}$ transformation],    
\begin{subequations}
\label{A1}
\begin{align}
\varrho_{i}^{\alpha} & \rightarrow 
\varrho_{i}^{\prime \alpha} =\varrho_{i}^{\alpha} -\varepsilon^{\alpha}{}_{\beta} \varrho_{i}^{\beta}
-i \varepsilon^{\alpha \dot{\beta}} \varpi_{i\dot{\beta}} \,, 
\label{A1a}
\\
\bar{\varrho}^{i \dot{\alpha}} & \rightarrow 
\bar{\varrho}^{\prime\:\! i \dot{\alpha}} =\bar{\varrho}^{i \dot{\alpha}}
-\bar{\varepsilon}^{\dot{\alpha}}{}_{\dot{\beta}} \bar{\varrho}^{i\dot{\beta}}
+i \varepsilon^{\beta \dot{\alpha}} \bar{\varpi}^{i}_{\beta} \,, 
\label{A1b}
\\
\varpi_{i \dot{\alpha}} & \rightarrow 
\varpi^{\prime}_{i \dot{\alpha}} =\varpi_{i \dot{\alpha}}+\bar{\varepsilon}_{\dot{\alpha}}{}^{\dot{\beta}} \varpi_{i \dot{\beta}} \,, 
\label{A1c}
\\
\bar{\varpi}^{i}_{\alpha} & \rightarrow 
\bar{\varpi}^{\prime\:\! i}_{\alpha} =\bar{\varpi}^{i}_{\alpha}+\varepsilon_{\alpha}{}^{\beta} \bar{\varpi}^{i}_{\beta} \,. 
\label{A1d}
\end{align}
\end{subequations}
Here, $\varepsilon^{\alpha\beta}$ and 
$\bar{\varepsilon}^{\dot{\alpha} \dot{\beta}} \big(:=\overline{\varepsilon^{\alpha\beta}} \:\! \big)$ are parameters 
of the infinitesimal Lorentz transformation 
$[^{\!\:}$or more accurately, the infinitesimal $SL(2, \Bbb{C})$ transformation], 
satisfying the symmetric properties $\varepsilon^{\alpha\beta}=\varepsilon^{\beta\alpha}$ and 
$\bar{\varepsilon}^{\dot{\alpha} \dot{\beta}}=\bar{\varepsilon}^{\dot{\beta} \dot{\alpha}}$, 
while $\varepsilon^{\alpha \dot{\beta}}$ is a parameter of the infinitesimal translation, satisfying 
the Hermiticity $\overline{\varepsilon^{\alpha\dot{\beta}}}=\varepsilon^{\beta\dot{\alpha}}$. 
The fields $h$, $\bar{h}$, $a$, and $\mathsf{b}^{r}$ are assumed to be Poincar\'{e} invariant. 
Since the GGS action is Poincar\'{e} invariant, 
we can derive conserved quantities 
by applying Noether's theorem. 
The conserved quantities corresponding to $\varepsilon^{\alpha\beta}$, 
$\bar{\varepsilon}^{\dot{\alpha} \dot{\beta}}$, and $\varepsilon^{\alpha\dot{\beta}}$ 
are found to be 
\begin{subequations}
\label{A2}
\begin{align}
\mu_{\alpha\beta} &:=\frac{i}{2} \big( \varrho_{i\alpha} \bar{\varpi}^{i}_{\beta} 
+\varrho_{i\beta} \bar{\varpi}^{i}_{\alpha} \big) \,, 
\label{A2a}
\\
\bar{\mu}_{\dot{\alpha} \dot{\beta}} &:=-\frac{i}{2} 
\big( \bar{\varrho}{}^{i}_{\dot{\alpha}} \varpi_{i\dot{\beta}}  
+\bar{\varrho}{}^{i}_{\dot{\beta}} \varpi_{i\dot{\alpha}} \big) \,, 
\label{A2b}
\\
p_{\alpha\dot{\beta}} &:=\bar{\varpi}^{i}_{\alpha} \varpi_{i\dot{\beta}} \,. 
\label{A2c}
\end{align}
\end{subequations}
Substituting Eqs. (\ref{3.37a}) and (\ref{3.37b}) into Eqs. (\ref{A2a}) and (\ref{A2b}), respectively, 
we can rewrite $\mu_{\alpha\beta}$ and $\bar{\mu}_{\dot{\alpha} \dot{\beta}}$ as 
\begin{subequations}
\label{A3}
\begin{align}
\mu_{\alpha\beta} &=\frac{i}{2} \big( \rho_{i\alpha} \bar{\varpi}^{i}_{\beta} 
+\rho_{i\beta} \bar{\varpi}^{i}_{\alpha} \big) \,, 
\label{A3a}
\\
\bar{\mu}_{\dot{\alpha} \dot{\beta}} &=-\frac{i}{2} 
\big( \bar{\rho}{}^{i}_{\dot{\alpha}} \varpi_{i\dot{\beta}}  
+\bar{\rho}{}^{i}_{\dot{\beta}} \varpi_{i\dot{\alpha}} \big) \,. 
\label{A3b}
\end{align}
\end{subequations}
The angular momentum tensor is given by 
\begin{align}
M_{\alpha\dot{\alpha}\beta\dot{\beta}} 
:=\mu_{\alpha\beta} \epsilon_{\dot{\alpha}\dot{\beta}} +\bar{\mu}_{\dot{\alpha} \dot{\beta}} 
\epsilon_{\alpha\beta} \,, 
\label{A4}
\end{align}
and the four-momentum vector is given by Eq. (\ref{A2c}).

The Pauli-Lubanski pseudovector is defined by \cite{PenRin, BarWig, Ryder} 
\begin{align}
W^{\alpha\dot{\alpha}}:=\frac{1}{2} \epsilon^{\alpha\dot{\alpha} \beta\dot{\beta} \gamma\dot{\gamma} \delta\dot{\delta}} 
p_{\beta\dot{\beta}} M_{\gamma\dot{\gamma} \delta\dot{\delta}} \,, 
\label{A5}
\end{align}
which can be written as 
\begin{align}
W^{\alpha\dot{\alpha}}=-i\mu^{\alpha\beta} p_{\beta}{}^{\dot{\alpha}} +i\bar{\mu}^{\dot{\alpha} \dot{\beta}} 
p^{\alpha}{}_{\dot{\beta}} 
\label{A6}
\end{align}
by using the formula 
\begin{align}
\epsilon^{\alpha\dot{\alpha}\beta\dot{\beta}\gamma\dot{\gamma}\delta\dot{\delta}}
&=i \Big( \epsilon^{\alpha\gamma} \epsilon^{\beta\delta} \epsilon^{\dot{\alpha}\dot{\delta}} 
\epsilon^{\dot{\beta}\dot{\gamma}} 
-\epsilon^{\alpha\delta} \epsilon^{\beta\gamma} \epsilon^{\dot{\alpha}\dot{\gamma}} 
\epsilon^{\dot{\beta}\dot{\delta}} \Big)\,. 
\label{A7}
\end{align}
Using the identity 
\begin{align}
\epsilon^{\alpha\beta} \rho^{\gamma}_{i} +\epsilon^{\beta\gamma} \rho^{\alpha}_{i} 
+\epsilon^{\gamma\alpha} \rho^{\beta}_{i} =0 
\label{A8}
\end{align}
and its complex conjugate, we can express Eq. (\ref{A6}) as 
\begin{align}
W^{\alpha\dot{\alpha}} &=\Big( \rho_{i}^{\beta} \bar{\varpi}{}^{j}_{\beta} 
+\varpi_{i\dot{\beta}} \bar{\rho}{}^{j \dot{\beta}} \Big) 
\bar{\varpi}{}^{i \alpha} \varpi{}_{j}^{\dot{\alpha}} 
\notag 
\\
& \quad \, 
-\frac{1}{2} \Big( \rho_{i}^{\beta} \bar{\varpi}{}^{i}_{\beta} 
+\varpi_{i\dot{\beta}} \bar{\rho}{}^{i \dot{\beta}} \Big) 
\bar{\varpi}{}^{j \alpha} \varpi{}_{j}^{\dot{\alpha}} \,, 
\label{A9}
\end{align}
or more concisely, 
\begin{align}
W^{\alpha\dot{\alpha}} &=
\bigg(\delta_{i}^{l} \delta_{k}^{j} -\frac{1}{2} \delta_{i}^{j} \delta_{k}^{l} \bigg) 
 \bar{\mathsf{W}}_{B}^{k} \mathsf{W}^{B}_{l}
\bar{\varpi}{}^{i \alpha} \varpi{}_{j}^{\dot{\alpha}} \,. 
\label{A10}
\end{align}
Here, $\mathsf{W}_{k}^{B}$ and $\bar{\mathsf{W}}{}^{k}_{B}$ are the twistors defined by 
$\mathsf{W}_{k}^{B} :=\big(\rho_{k}^{\beta}, \varpi_{k \dot{\beta}} \big)$ and 
$\bar{\mathsf{W}}^{k}_{B} :=\big(\bar{\varpi}^{k}_{\beta}, \bar{\rho}{}^{k\dot{\beta}} \big)$ 
$[^{\!\:}$see the text above Eq. (\ref{3.33})]. 
Applying the formula 
\begin{align}
\frac{1}{2} \sigma_{ri}{}^{j} \sigma_{rk}{}^{l} 
=\delta_{i}^{l} \delta_{k}^{j} -\frac{1}{2} \delta_{i}^{j} \delta_{k}^{l}
\label{A11}
\end{align}
valid for the Pauli matrices $\sigma_{r}$ to Eq. (\ref{A10}), we obtain 
\begin{align}
W^{\alpha\dot{\alpha}}=\mathsf{T}_{r} \sigma_{ri}{}^{j} \bar{\varpi}{}^{i \alpha} \varpi{}_{j}^{\dot{\alpha}}  \,,
\label{A12}
\end{align}
with
\begin{align}
\mathsf{T}_{r} :=\frac{1}{2} \bar{\mathsf{W}}_{B}^{k} \sigma_{rk}{}^{l} \mathsf{W}^{B}_{l} 
\label{A13}
\end{align} 
$[^{\!\:}$see Eq. (\ref{3.39})]. 
Equation (\ref{A12}) can be written in terms of the (original) twistors $Z_{k}^{B}$ and $\bar{Z}^{k}_{B}$ as 
\begin{align}
W^{\alpha\dot{\alpha}}=T_{r} \sigma_{ri}{}^{j} \bar{\pi}{}^{i \alpha} \pi{}_{j}^{\dot{\alpha}} \,,  
\label{A14}
\end{align}
with
\begin{align}
T_{r} :=\frac{1}{2} \bar{Z}_{B}^{k} \sigma_{rk}{}^{l} Z^{B}_{l} \,.
\label{A15}
\end{align} 
Using the mass-shell constraints 
\begin{subequations}
\label{A16}
\begin{align}
\varpi_{i \dot{\alpha}} \varpi_{j}^{\dot{\alpha}}  
& \approx \frac{m}{\sqrt{2}} \epsilon_{ij} e^{i\varphi} , 
\label{A16a}
\\
\bar{\varpi}{}^{i}_{\alpha} \bar{\varpi}{}^{j \alpha}  
&\approx \frac{m}{\sqrt{2}} \epsilon^{ij} e^{-i\varphi} 
\label{A16b}
\end{align}
\end{subequations}
equivalent, respectively, to Eqs. (\ref{3.11e}) and (\ref{3.11f}), 
and utilizing the formula $\sigma_{2} \sigma_{r} \sigma_{2}=-\sigma_{r}^{\mathrm{T}}$,   
we can show for Eq. (\ref{A12}) that 
\begin{align}
W_{\alpha\dot{\alpha}} W^{\alpha\dot{\alpha}} 
\approx -m^2 \mathsf{T}_{r} \mathsf{T}_{r} \,. 
\label{A17}
\end{align}

In our model, twistor quantization is performed with the commutation relations 
(\ref{4.2a}) and (\ref{4.2b}), or equivalently, 
\begin{align}
\left[\:\! \hat{\rho}_{i \alpha}, \Hat{\Bar{\varpi}}{}^{j}_{\beta} \right]
&=-\delta_{i}^{j} \epsilon_{\alpha\beta}  \,,
\quad 
\left[\:\! \Hat{\Bar{\rho}}{}^{i}_{\dot{\alpha}},  \hat{\varpi}_{j \dot{\beta}} \right] 
=\delta_{j}^{i} \epsilon_{\dot{\alpha} \dot{\beta}} \,, 
\notag
\\
\mbox{all others} &=0 \,.  
\label{A18}
\end{align}
The operators corresponding to $\mu_{\alpha\beta}$ and $\bar{\mu}_{\dot{\alpha} \dot{\beta}}$ 
are defined by replacing the twistor variables in Eq. (\ref{A3}) with their corresponding operators 
and by obeying the Weyl ordering rule. 
After using the commutation relations in Eq. (\ref{A18}), we have 
\begin{subequations}
\label{A19}
\begin{align}
\hat{\mu}_{\alpha\beta} &=\frac{i}{2} \Big( \hat{\rho}_{i\alpha} \Hat{\Bar{\varpi}}{}^{i}_{\beta} 
+\hat{\rho}_{i\beta} \Hat{\Bar{\varpi}}{}^{i}_{\alpha} \Big) \,, 
\label{A19a}
\\
\Hat{\Bar{\mu}}_{\dot{\alpha} \dot{\beta}} &=-\frac{i}{2} 
\Big( \Hat{\Bar{\rho}}{}^{i}_{\dot{\alpha}} \hat{\varpi}_{i\dot{\beta}}  
+\Hat{\Bar{\rho}}{}^{i}_{\dot{\beta}} \hat{\varpi}_{i\dot{\alpha}} \Big) \,. 
\label{A19b}
\end{align}
\end{subequations}
The operator corresponding to $p_{\alpha\dot{\beta}}$ is found immediately from 
Eq. (\ref{A2c}) to be 
\begin{align}
\hat{p}_{\alpha\dot{\beta}} =\Hat{\Bar{\varpi}}{}^{i}_{\alpha} \Hat{\varpi}_{i\dot{\beta}} \,. 
\label{A20}
\end{align}
Using Eq. (\ref{A18}), we can calculate the commutation relations between 
$\hat{\mu}_{\alpha\beta}$, $\Hat{\Bar{\mu}}_{\dot{\alpha} \dot{\beta}}$, 
and $\hat{p}_{\alpha\dot{\beta}}$ to obtain 
\begin{align}
\Big[\;\! \hat{\mu}_{\alpha\beta} , \:\! \hat{\mu}_{\gamma\delta} \Big] 
&=-\frac{i}{2} \Big( 
\epsilon_{\alpha\gamma} \hat{\mu}_{\beta\delta} 
+\epsilon_{\alpha\delta} \hat{\mu}_{\beta\gamma} 
+\epsilon_{\beta\gamma} \hat{\mu}_{\alpha\delta} 
+\epsilon_{\beta\delta} \hat{\mu}_{\alpha\gamma} \Big) \:\! ,
\notag
\\
\Big[\;\! \Hat{\Bar{\mu}}_{\dot{\alpha} \dot{\beta}} ,  \:\! \Hat{\Bar{\mu}}_{\dot{\gamma} \dot{\delta}} \Big] 
&=-\frac{i}{2} \Big( 
\epsilon_{\dot{\alpha} \dot{\gamma}} \Hat{\Bar{\mu}}_{\dot{\beta} \dot{\delta}}  
+\epsilon_{\dot{\alpha} \dot{\delta}} \Hat{\Bar{\mu}}_{\dot{\beta} \dot{\gamma}}  
+\epsilon_{\dot{\beta} \dot{\gamma}} \Hat{\Bar{\mu}}_{\dot{\alpha} \dot{\delta}}  
+\epsilon_{\dot{\beta} \dot{\delta}} \Hat{\Bar{\mu}}_{\dot{\alpha} \dot{\gamma}} \Big) \:\! ,
\notag
\\
\Big[\;\! \hat{\mu}_{\alpha\beta} , \:\! \hat{p}_{\gamma\dot{\delta}} \Big] 
&=-\frac{i}{2} \Big( 
\epsilon_{\alpha\gamma} \hat{p}_{\beta\dot{\delta}}  
+\epsilon_{\beta\gamma} \hat{p}_{\alpha\dot{\delta}} \Big) \:\! ,
\notag
\\
\Big[\;\! \Hat{\Bar{\mu}}_{\dot{\alpha} \dot{\beta}} , \:\! \hat{p}_{\gamma\dot{\delta}} \Big] 
&=-\frac{i}{2} \Big( 
\epsilon_{\dot{\alpha} \dot{\delta}} \hat{p}_{\gamma\dot{\beta}}  
+\epsilon_{\dot{\beta} \dot{\delta}} \hat{p}_{\gamma\dot{\alpha}} \Big) \:\! , 
\notag
\\
\mbox{all others}&=0 \,.
\label{A21}
\end{align}
These commutation relations specify together a spinor representation of the Poincar\'{e} algebra. 
The operators $\hat{\mu}_{\alpha\beta}$, $\Hat{\Bar{\mu}}_{\dot{\alpha} \dot{\beta}}$, 
and $\hat{p}_{\alpha\dot{\beta}}$ are thus established as the generators of 
$SL(2, \Bbb{C}) \ltimes \Bbb{R}^{1,3}$. 
We can verify that $\hat{\mu}_{\alpha\beta}$, $\Hat{\Bar{\mu}}_{\dot{\alpha} \dot{\beta}}$, 
and $\hat{p}_{\alpha\dot{\beta}}$ commute with the generators $\hat{\mathsf{T}}_{0}$ and $\hat{\mathsf{T}}_{r}$ defined in 
Eq. (\ref{4.4}). This implies that the Poincar\'{e} symmetry and the $U(1)_{a} \times SU(2)$ 
internal symmetry are not combined, so that the result is consistent with 
the Coleman-Mandula theorem \cite{ColMan, Weinberg2}.

The Weyl ordered operator corresponding to the Pauli-Lubanski pseudovector $W^{\alpha \dot{\alpha}}$ 
can be simplified as 
\begin{align}
\hat{W}^{\alpha\dot{\alpha}}=\hat{\mathsf{T}}_{r} \sigma_{ri}{}^{j} \Hat{\Bar{\varpi}}{}^{i \alpha} 
\Hat{\varpi}{}_{j}^{\dot{\alpha}}  
\label{A22}
\end{align}
by using the commutation relation 
\begin{align}
\Big[\:\! \hat{\mathsf{T}}_{r}, \:\! \sigma_{si}{}^{j} \Hat{\Bar{\varpi}}{}^{i \alpha} \Hat{\varpi}{}_{j}^{\dot{\alpha}} \Big]
=i\epsilon_{rst} \sigma_{ti}{}^{j} \Hat{\Bar{\varpi}}{}^{i \alpha} \Hat{\varpi}{}_{j}^{\dot{\alpha}} \,. 
\label{A23}
\end{align}
Then, using the physical state conditions 
\begin{subequations}
\label{A24}
\begin{align}
\Hat{\varpi}_{i \dot{\alpha}} \Hat{\varpi}{}_{j}^{\dot{\alpha}} |F \:\! \rangle
&=\frac{m}{\sqrt{2}} \epsilon_{ij} e^{i\hat{\varphi}} |F \:\! \rangle , 
\label{A24a}
\\
\Hat{\bar{\varpi}}{}^{i}_{\alpha} \Hat{\bar{\varpi}}{}^{j \alpha} |F \:\! \rangle
&= \frac{m}{\sqrt{2}} \epsilon^{ij} e^{-i\hat{\varphi}} |F \:\! \rangle
\label{A24b}
\end{align}
\end{subequations}
equivalent, respectively, to Eqs. (\ref{4.3h}) and (\ref{4.3i}), 
we can show that  
\begin{align}
\hat{W}_{\alpha\dot{\alpha}} \hat{W}^{\alpha\dot{\alpha}} |F \:\! \rangle =
-m^2 \hat{\mathsf{T}}_{r} \hat{\mathsf{T}}_{r} |F \:\! \rangle \,. 
\label{A25}
\end{align}
This is precisely a quantum mechanical counterpart of Eq. (\ref{A17}). 
The Casimir operators of the Poincar\'{e} algebra are given by 
$\hat{p}_{\alpha\dot{\beta}} \hat{p}^{\alpha\dot{\beta}}$ and 
$\hat{W}_{\alpha\dot{\alpha}} \hat{W}^{\alpha\dot{\alpha}}$.  
From Eq. (\ref{A24}), it follows that 
\begin{align}
\hat{p}_{\alpha\dot{\beta}} \hat{p}^{\alpha\dot{\beta}} |F \:\! \rangle
=m^{2} |F \:\! \rangle \,.
\label{A26}
\end{align}
Then it can be shown that \cite{BarWig, Ryder} 
\begin{align}
\hat{W}_{\alpha\dot{\alpha}} \hat{W}^{\alpha\dot{\alpha}} |F \:\! \rangle
=-m^{2} J(J+1) |F \:\! \rangle \,, 
\label{A27}
\end{align}
where $J$ denotes the spin quantum number taking the values 
\begin{align}
 J=0, \frac{1}{2}, 1, \frac{3}{2}, \ldots .   
\label{A28}
\end{align}
Here, $|F^{\:\!} \rangle$ is assumed to be a simultaneous eigenvector of 
$\hat{W}_{\alpha\dot{\alpha}} \hat{W}^{\alpha\dot{\alpha}}$ 
and the other relevant operators $\hat{\mathsf{T}}_{0}$, $\hat{\mathsf{T}}_{3}$, 
$\hat{\mathsf{T}}_{\hat{\imath}} \hat{\mathsf{T}}_{\hat{\imath}}$, and 
$\hat{p}_{\alpha\dot{\beta}} \hat{p}^{\alpha\dot{\beta}}$ 
$[^{\!\:}$see Eqs. (\ref{4.3e}), (\ref{4.3f}), (\ref{4.3g}), and (\ref{A26})]. 
This assumption holds true, because the generators of $SL(2, \Bbb{C}) \ltimes \Bbb{R}^{1,3}$ commute with 
those of $U(1)_{a} \times SU(2)$. 
The vector $|F^{\:\!} \rangle$ turns out to be characterized by the set of quantum numbers $(s, I, t^{\:\!} ; m, J)$. 
In terms of $|F^{\:\!} \rangle$, Eq. (\ref{4.20}) reads 
\begin{align}
\hat{\mathsf{T}}_{r} \hat{\mathsf{T}}_{r} |F \:\! \rangle =\varLambda |F \:\! \rangle \,, 
\label{A29}
\end{align}
where $\varLambda$ is determined to be 
\begin{align}
\varLambda &=I(I+1) \,, \quad  I=0, \frac{1}{2}, 1, \frac{3}{2}, \ldots . 
\label{A30}
\end{align}
Applying Eqs. (\ref{A27}) and (\ref{A29}) to Eq. (\ref{A25}), we eventually have 
\begin{align}
I=J\,. 
\label{A31}
\end{align}
This result is consistent with the fact that 
the number of $SU(2)$ indices of the spinor field $\varPsi$ given in Eq. (\ref{5.1})  
is equal to the number of its spinor indices.

\end{document}